%% file: petreczky_jpgrev.tex
\documentclass[12pt]{iopart}
\usepackage{graphicx}

\newcommand{\beq}{\begin{equation}}
\newcommand{\eeq}{\end{equation}}
\newcommand{\be}{\begin{equation}} 
\newcommand{\ee}{\end{equation}}
\newcommand{\ber}{\begin{eqnarray}}
\newcommand{\eer}{\end{eqnarray}}
\newcommand{\bea}{\begin{eqnarray}}
\newcommand{\eea}{\end{eqnarray}}
\newcommand{\berr}{\begin{eqnarray*}}
\newcommand{\eerr}{\end{eqnarray*}}

\newcommand{\sgh}{\sigma_H(\omega,\vec{p})}

\newcommand{\non}{\nonumber}
\newcommand{\D}{{\cal D}}

\begin{document}

\title{Lattice QCD at non-zero temperature}

\author{P. Petreczky}

\address{Physics Department, Brookhaven National Laboratory, 
         Upton, NY 11973, USA}

\begin{abstract}
I review our current understanding of the properties of
strongly interacting matter at high temperatures, based upon numerical
calculations in lattice QCD. I discuss the chiral and deconfining aspects
of the QCD transition, the equation of state, fluctuations of conserved charges,
color screening,
meson correlation functions, and
the determination of some transport coefficients.
\end{abstract}

\input{intro.tex}

\input{lattice.tex}
\input{screening.tex}

\input{fluctuations.tex}

\input{chiral.tex}

\input{eos.tex}

\input{spf.tex}
\input{conclusion.tex}

\section*{Acknowledgments} 
This work was supported by U.S. Department of Energy under
Contract No. DE-AC02-98CH10886.  I would like to thank Rob Pisarski and Swagato Mukherjee for
useful discussions.

\bibliographystyle{h-physrev.bst}
\bibliography{HotQCD}{}

\end{document}

%% file: intro.tex
\section{Introduction}
\label{sec:intro}
It is expected that strongly interacting matter shows qualitatively
new behavior at temperatures and/or densities which are
comparable or larger than the typical hadronic scale.
It has been argued by Hagedorn that production of hadronic
resonances may lead to a limiting temperature above which
hadronic matter can no longer exist \cite{Hagedorn:1965st}.
This limiting temperature was re-interpreted by Cabibbo and Parisi
as the transition temperature to a new state of matter that consist of quarks \cite{Cabibbo:1975ig}.
Indeed, based of asymptotic freedom
one would expect that relevant degrees of freedom at very high
temperatures and/or densities are quarks and gluons which are
no longer subject to confinement \cite{Collins:1974ky}, i.e. deconfined.
Since the quarks and gluons interact weakly at high temperatures
the deconfined matter is analogous to the plasmas and is also
called the quark gluon plasma (QGP) \cite{Shuryak:1980tp}. In particular, 
it is characterized by color screening that is similar to the well-know
Debye screening \cite{Shuryak:1977ut}.

The existence of such deconfining transition was first shown using
strong coupling expansion in lattice QCD \cite{Polyakov:1978vu,Susskind:1979up},
followed by numerical
Monte-Carlo studies of the lattice $SU(2)$ gauge theory which confirmed it
\cite{Kuti:1980gh,McLerran:1981pb,Engels:1980ty}. 
The deconfinement of gluons was seen as onset of color
screening \cite{Kuti:1980gh,McLerran:1981pb}
and rapid increase in the energy density that indicate
liberation of many new degrees of freedom \cite{Engels:1980ty}.
The onset of color screening and rapid increase in number of
degrees of freedom remain the key signatures of deconfinement even today.

Since these pioneering studies QCD at finite temperature
became quite a large sub-field of lattice QCD. One of the
obvious reasons for this is that the transition from hadronic
to partonic matter cannot be described in perturbation theory. 
But even at high temperatures the physics is non-perturbative beyond 
the length scales $1/(g^2~T)$ ($g^2$ being the gauge coupling) \cite{Linde:1980ts}.  
Therefore lattice QCD
remains the only ab-initio tool for theoretical understanding of the
properties of strongly interacting matter under extreme condition
which is important for
the physics of the early universe \cite{Hindmarsh:2005ix,Asaka:2006rw,Laine:2006we}
and heavy ion collisions \cite{Muller:2006ee}.

Symmetries play an important role in understanding the transition to QGP.
For zero quark masses the QCD Lagrangian has an approximate chiral symmetry
$SU_A(3)$ as well as the axial symmetry $U_A(1)$. The later is broken by
quantum effects, while the former is spontaneously broken by the QCD vacuum 
\cite{Muta:1998vi}. At high temperatures the chiral symmetry is expected
to be restored \cite{Gross:1980br}. Therefore there should be a chiral
transition in QCD. The nature of the chiral transition was first discussed
in Ref. \cite{Pisarski:1983ms}, where it was argued that for three light quark
flavors the transition should be first order, while for two flavors it
could be second order belonging to the $O(4)$ universality class.
Furthermore, the axial symmetry is
also effectively restored at high temperatures as the breaking of the
axial symmetry is controlled by instanton density \cite{Pisarski:1983ms}
that vanishes at infinite temperature \cite{Gross:1980br}.
If the breaking of axial symmetry at the chiral transition temperature is
reduced the chiral transition in the two flavor theory could be 
first order \cite{Pisarski:1983ms}. I will comment on this issue in section \ref{sec:chiral}.

In the opposite limit of infinitely heavy quark masses the QCD partition
function has $Z(3)$ symmetry \cite{McLerran:1981pb}. This symmetry is broken
at high temperatures due to color screening. 
In this case we have a deconfining transition which is first order.
The $Z(3)$ symmetry is a good symmetry for sufficiently large quark mass.
For intermediate quark masses the transition is a crossover. 
The regions of the first order transitions are separated from the crossover
region by lines of second order transition that belong the $Z(2)$ universality
class. 
The region of the first order deconfinement
transition was mapped out in Refs. \cite{Alexandrou:1998wv,Karsch:2001ya,Saito:2011fs}. 
The endpoints of this first order
region correspond to a quark masses of the order of the charm quark mass, $m_q \sim 1.4$GeV 
\cite{Alexandrou:1998wv,Karsch:2001ya}. 

\begin{figure}
\includegraphics[width=8cm]{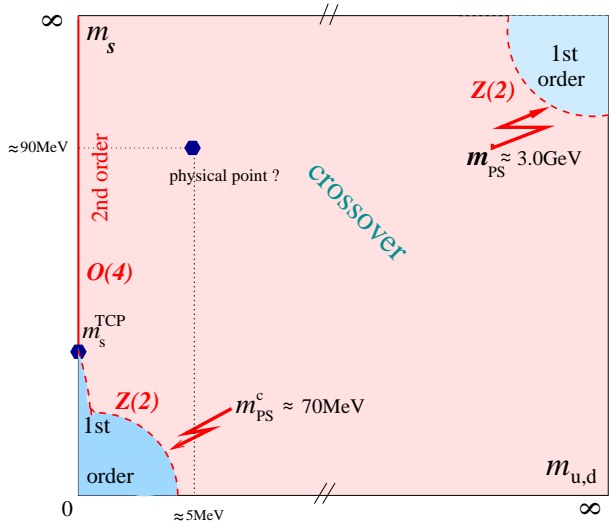}
\caption{The QCD transition as function of the quark masses.}
\label{fig:columbia}
\end{figure}

The boundary of the first order chiral transition was studied using effective models,
see Refs. \cite{Herpay:2005yr,Herpay:2006vc} for recent works (also references therein). 
It was found that
for three degenerate quark flavors the endpoint of the region 
where the transition is first order corresponds
to a pseudo-scalar meson mass of $(110 \pm 40)$MeV. 
There is no direct evidence for the first order chiral
transition for three very light quark flavors so far in lattice QCD. 
This maybe due to lattice artifacts.
The existing calculations that use
staggered fermion formulation give only an upper bound for the pseudo-scalar 
meson mass smaller than $70$MeV 
\cite{Karsch:2003va,Endrodi:2007gc,Ding:2011du},
which is not inconsistent with effective model analysis mentioned above. 
The end-point of the first order transition, or in more general case the line
that separates the first order transition from the crossover region
belongs to the $Z(2)$ universality class. This also means that for some value 
of the strange quark mass, $m_s^{TCP}$ we have a tricritical point that connects
$Z(2)$ and $O(4)$ lines as well as the first order region. Calculations in
linear sigma model suggest that $m_s^{TCP}$ corresponds to the kaon mass
of about $(1700-1850)$MeV \cite{Herpay:2006vc}. On the other hand, as we will see,
lattice QCD calculations indicate a much smaller value of $m_s^{TCP}$.
The above
picture of the finite temperature QCD transition is summarized in
Fig. \ref{fig:columbia} (also known as the Columbia plot). 

For the physical quark masses the transition is likely to be an analytic crossover 
\cite{Aoki:2006we} (for earlier works see Refs. \cite{Bernard:2004je,Cheng:2006qk} ). 
Therefore, the discussion of the deconfinement and the chiral aspects of the QCD transition at
finite temperature and the determination of the corresponding temperature requires 
a special care. In fact, it is only the chiral aspects of the transition 
that allow a determination of the corresponding transition
temperature. In sections \ref{sec:screening} and \ref{sec:fluct} I discuss the deconfinement aspects of
the QCD transition in terms of chromo-electric screening and fluctuations of conserved charges.
The chiral aspects of the QCD transition and the determination of the corresponding
transition temperature are then discussed in section \ref{sec:chiral}. 
Equation of state is one of the most important characteristic of hot strongly interacting medium. 
It will
be discussed
in section \ref{sec:chiral}. Further  insight into the properties of medium can be gained by
studying meson correlation functions and spectral functions. They are also important for heavy ion
phenomenology. Therefore section \ref{sec:spf} is dedicated to the discussion of meson correlation
functions and spectral functions. In the next section 
I will review some basic information on 
lattice QCD calculations which is relevant for the present discussions.

%% file: lattice.tex
\section{Basics of lattice QCD}
\label{sec_lgt_basics}

To study QCD non-perturbatively  we use lattice gauge theory
\cite{Wilson:1974sk}. In this formalism a field theory is defined in a
gauge-invariant way on a discrete space-time domain. This serves
at least two purposes: a) to provide an ultra-violet cut-off for the 
theory, restricting highest momentum to $\pi/a$ ($a$ being the lattice
spacing), and b) to evaluate the path integrals in the Euclidean
formulation stochastically using importance sampling.

On the lattice the fundamental degrees of freedom of a theory
with local $SU(3)$ gauge symmetry are fermion fields $\psi_x$ that
reside on the sites of the lattice and carry flavor, color and Dirac
indeces, which we suppress through the most of this paper, and
bosonic, gauge degrees of freedom that in the form of $SU(3)$ matrices 
$U_{x,\mu}$ reside on links. Sites on a four-dimensional lattice 
are labeled with $x\equiv(\tau,\vec{x})$. The link variables are related
to the usual gauge fields: $U_{x, \mu}=\exp(ig a A_{\mu}(x))$, 
where $a$ is the lattice spacing and $g$ is the gauge coupling. 

The theory is defined by the partition function
\begin{equation}\label{partfun}
  Z=\int DU D\bar\psi D\psi \exp(-S)
\end{equation}
where the action
\begin{equation}\label{Saction}
  S=S_g+S_f
\end{equation}
contains gauge, $S_g$ and fermionic, $S_f$ parts. The latter part is bi-linear
in fields and has the form
\begin{equation}\label{Sfaction}
  S_f=\sum_q \bar\psi_q D_q \psi_q
\end{equation}
where $D_q=D+m_q$ is the fermion matrix and the sum goes over the quark flavors. 
In the simplest formulation the lattice gauge action can
be written as 
\be
S_g=\beta  \sum_{x,\mu < \nu} \left( 1- \frac{1}{3}{\rm Re\; Tr}\; U_{x,\mu} U_{x+\hat{\mu},\nu} U^\dagger_{x+\hat{\nu},\mu}
U^\dagger_{x,\nu} \right),
\ee
$\beta=6/g^2$ with $g^2$ being the bare gauge coupling. This is the Wilson gauge
action \cite{Wilson:1974sk}. It is easy to see that expanding the link variables in $a$ the 
above expression gives the well known Yang-Mills action $\sum_x a^4 1/4 (F_{\mu \nu}^a)^2$
up to corrections that are suppressed as $a^2$. Thus calculations with Wilson gauge action have discretization
errors of order $a^2$. These discretization errors can be removed by the Symanzik improvement
program that relies on higher order difference scheme, i.e. considering combination of link variables
that  extend beyond the elementary square \cite{Luscher:1984xn}.

Th expectation value of an operator $\hat O$ is then given  by
\begin{equation}\label{expectO}
  \langle \hat O\rangle=\frac{1}{Z}
  \int DU D\bar\psi D\psi\hat O \exp(-S).
\end{equation}
Integration over the fermion fields (which are Grasmann variables)
can be carried out explicitly:
\begin{equation}\label{partfun_eff}
  Z=\int DU \left(\det D_q[U] \right)^{n_f} \exp(-S_g)\equiv
  \int DU \exp(-S_{eff}),
\end{equation}
where $S_{eff}=S_g-n_f \ln\det D_q[U]$ is the effective action.
The fermion determinant $\det D_q[U]$ describes the vacuum polarization effects due to the dynamical
quarks and makes the effective action non-local in gauge variables.
For this reason simulations with dynamical quarks are very resource demanding
and the quenched approximation is often employed, where $\det D_q[U]$ is set to 1.
This approximation will be relevant for the discussion of the meson spectral functions.

The most commonly used fermion formulations include : the Wilson fermion formulation,
staggered fermion formulation \cite{DeGrand:2006zz} 
and the domain wall fermion (DWF) formulation  \cite{Kaplan:1992bt,Furman:1994ky,Shamir:1993zy}.
The Wilson formulation
breaks the chiral symmetry of the QCD Lagrangian. As the result quark masses
acquire an additive renormalization which makes the calculations quite complicated and
numerically expensive. The domain wall fermions preserve all the symmetry of
the QCD Lagrangian at a price on introducing the extra fifth dimension.
Therefore the computational costs associated with DWF formulation are proportional
to the extent of the fifth dimension and are typically 30 to 100 times larger
than the ones with Wilson fermions. The staggered formulation describes four fermion
flavors (also called tastes) in the continuum limit. The full $U_L(4) \times U_R(4)$ symmetry
of the four flavor theory is reduced to $U_S(1) \times U_B(1)$, where $U_S(1)$ is the
subgroup of the $SU_A(4)$ symmetry. Thus the staggered formulation preserves a part
of the chiral symmetry. This makes finite temperature calculations with staggered fermions 
attractive for
two reasons. First, due to the remnant chiral symmetry there is no additive renormalization
of the quark mass. This makes staggered fermions computationally inexpensive. Second, the
remnant chiral symmetry makes it easy to study the chiral aspects of the finite
temperature QCD transitions. To study arbitrary number of flavors with the staggered fermion
formulation the rooting trick is used. This amounts to replacing $n_f$ in Eq. (\ref{partfun_eff})
with $n_f/4$. The validity of this procedure was studied in details in recent years
\cite{Shamir:2004zc,Shamir:2006nj,Sharpe:2006re,Adams:2008db,Bernard:2006vv,Bernard:2007ma,Bernard:2007eh}
though some open issues remain (see e.g. Ref. \cite{Sharpe:2006re,Creutz:2008nk} ). 

The standard staggered fermion formulations has discretization errors of order $a^2$.
They come either from the lattice effects in the fermion dispersion relation 
and formally start at order $g^0 a^2$, and from the breaking of taste symmetry,
which starts at order $g^2 a^2$. The former lead to large cutoff effects in the
thermodynamic quantities at high temperatures. The later lead to the distortion
of the hadron spectrum at non-zero lattice spacing. In particular, out of 16 pseudo-scalar
meson masses only one vanishes in the zero quark mass limit. The remaining 15 
pseudo-scalar meson have masses proportional to $a$ (
the squared mass of these mesons goes like $g^2 a^2$ for sufficiently small
lattice spacings) and thus vanish only in the continuum limit. 
These discretization errors can be eliminated by Symanzik improvement program.
Using 3-link terms in the lattice Dirac operator one can eliminate tree level
$a^2$ errors and largely reduce cutoff effects in thermodynamics quantities \cite{Heller:1999xz}.
Such improvement of the staggered fermion formulation is mandatory to
control discretization effects at high temperature \cite{Heller:1999xz}.
To reduce errors due to taste symmetry breaking so-called
fat links are used. These are combinations of the usual link 
$U_{x,\mu}$ and products of links along different paths.
One of the possibilities is 
to construct fat links by using APE smearing
\cite{Albanese:1987ds}: 
a link variable $U_{x,\mu}$ is replaced by a weighted average
of itself and a sum of the 3-link paths connecting the same sites as $U_{x,\mu}$:
\begin{equation}
  U_{x,\mu}\to U'_{x,\mu}=(1-6c)U_{x,\mu}+c\sum_{\nu\neq\mu}
  U_{x,\mu}U_{x+\hat\nu,\mu}U^\dagger_{x+\hat\mu,\nu}.
\end{equation}
Here $c$ is a weight factor, $0 < c < 1$.
More complicated paths can be used as well. In particular, using combination
of different paths up to length seven it is possible to eliminate taste breaking 
effects up to order $g^2 a^2$ \cite{Orginos:1999cr}. The fat links described above are not elements of
the gauge group $SU(3)$. While this is not a problem in principle, it tuns out that
projecting the fat links to elements of $SU(3)$ or $U(3)$ gauge group further reduces
the breaking of taste symmetry \cite{Hasenfratz:2001hp,Hasenfratz:2002vv,Hasenfratz:2007rf}.
The staggered fermion actions that used in large numerical calculations 
take certain combination of fat links  and improvement of the quark dispersion
relation and go under acronym $p4$, $asqtad$ and $HISQ/tree$. 
In $HISQ/tree$ action the fat links are projected to $U(3)$ \cite{Follana:2006rc}.
The $stout$ action used
by Budapest-Wuppertal collaboration only includes fat links 
\cite{Aoki:2005vt,Aoki:2006br,Aoki:2006we,Aoki:2009sc}.

To evaluate the path integral (\ref{expectO}) stochastically,
an ensemble of $N_U$ gauge configurations, weighted with $\exp(-S_{eff})$,
is generated using Monte Carlo methods. The state of the art Monte-Carlo algorithm
that can handle arbitrary number of quark flavors and most often used in QCD
thermodynamics goes under the name of the rational hybrid  Monte-Carlo (RHMC) 
algorithm \cite{Clark:2005sq}.
The expectation value of the operator is then approximated by the
ensemble average:
\begin{equation}\label{expectO_appr}
  \langle\hat O\rangle \simeq \frac{1}{N_U} \sum_{i=1}^{N_U}O_i(U),
\end{equation}
where $O_i(U)$ is the value of the operator $\hat O$, 
calculated on $i$-th configuration. This method works as long as $S_{eff}$ is
real. When quark chemical potentials are included, e.g. as in calculations at
non-zero net baryon density, $S_f$ and $S_{eff}$ become complex and Monte-Carlo
methods do not work. This is the well known sign problem. 

All the quantities calculated in lattice simulations are dimensionless, i.e. are
calculated in units of the lattice spacing $a$. The lattice spacing is varied by
varying the lattice gauge coupling $\beta \sim 1/g^2$. To set the scale in lattice
simulations we need to calculate a given quantity and assume  that it has the known
physical value. Different choices are possible and the lattice spacing determined
by different quantities should be the same up to discretization errors that vanish
when the continuum limit is taken. Common choices  used in the
literature are the scale parameters $r_0$ and $r_1$ determined by the static quark 
anti-quark potential $V(r)$:
\begin{equation}
r^2 \frac{d V}{d r}|_{r=r_0}=1.65,~~ r^2 \frac{d V}{d r}|_{r=r_1}=1.00
\end{equation}
Since the static quark anti-quark potential cannot be measured experimentally 
the value of the parameters $r_0$ and $r_1$ in physical units in the continuum
should be determined using some other experimentally measured quantity. The most
precise determination of $r_1$ comes from using the pion decay constant $f_{\pi}$
as an input, which gives $r_1=0.3106(20)$fm \cite{Bazavov:2010hj}. 
Furthermore, studying the shape of the static potential
from the above value of $r_1$ one can obtain the value of $r_0=0.468(4)$fm \cite{Bazavov:2011nk}. 
Alternatively one can use
$f_K$ to set the lattice spacing. While this procedure is more straightforward, for
the staggered fermion formulation it is also more problematic as one expects that
due to taste symmetry breaking $f_K$ has larger discretization errors than the static potential.

The above discussion referred to the zero temperature case. To consider the case of non-zero
temperature, $T$ the Euclidean time extent has to be $1/T$ and periodic and anti-periodic
boundary condition have to be imposed on the boson and fermion fields \cite{LeBellac:375551}.
Thus the temperature is related to the lattice spacing and the temporal extent of
the lattice, $T=1/(N_{\tau} a)$. Taking the continuum limit at fixed temperature implies
$N_{\tau} \rightarrow \infty$ while keeping the aspect ratio $N_{\sigma}/N_{\tau}$ fixed,
where $N_{\sigma}$ is the spatial extent of the lattice.

%% file: screening.tex
\section{Deconfinement : color screening}
\label{sec:screening}

\subsection{Color electric screening and order parameter for deconfinement}
As discussed in the introduction the deconfining transition in $SU(N)$
gauge theories is a true phase transition related to $Z(N)$ symmetry \footnote{The deconfinement
transition for other gauge groups have been discussed in Refs.\cite{Holland:2003mc,Holland:2003kg,Pepe:2006er}.}.
The order parameters of this phase transition are the expectation value of the Polyakov loop and 
the Polyakov loop correlator
\begin{eqnarray}
L(T)&=&\langle \frac1N {\rm Tr} W(\vec{x}) \rangle, ~W(\vec{x})=\prod_{\tau=0}^{N_{\tau}-1} U_0(\tau,\vec{x}),\\
C_{PL}(r,T)&=&\frac{1}{N^2} \langle {\rm Tr} W(r) {\rm Tr} W(0) \rangle.
\end{eqnarray}
The Polyakov loop transforms non-trivially under $Z(N)$ transformation and 
the expectation value of the Polyakov loop is zero in the confined phase and non-zero above
the transition temperature. The correlation function
of the Polyakov loop is related to the free energy of a static quark anti-quark pair \cite{McLerran:1981pb}.
The qualitative change in the behavior of the Polyakov loop and its correlator above the phase transition
temperature is related to color screening. As it was pointed out already in Ref. \cite{McLerran:1981pb} the Polyakov loop
and the Polyakov loop correlator require renormalization to be interpreted as the free energy of an isolated static quark
or the free energy of a static quark anti-quark $(Q \bar Q)$ pair. More precisely, they are related to the free energy 
difference of a system with static $Q\bar Q$ pair at some temperature and the system without $Q\bar Q$ pair at
the same temperature. 
Since in the zero temperature limit the free energy of a static
quark anti-quark pair should coincide with the static potential the Polyakov loop renormalization is determined
by the normalization constant of the static potential, namely

\begin{eqnarray}
L_{ren}(T)&=&\exp(-c/(2 T)) L(T)=\exp(-F_{\infty}(T)/(2 T))\\
C_{PL}(r,T)&=&\exp(-F(r,T)/T+c/T),~~F_{\infty}(T)=\lim_{r \rightarrow \infty} F(r,T),
\end{eqnarray}
where $c$ is an additive normalization that ensures that the static potential has a certain value at a given
distance. In the above expressions we made explicit that the free energy of an isolated static quark anti-quark pair
is half the free energy of $Q\bar Q$ pair at infinite separation. 
In the confined phase the free energy of a static quark anti-quark pair is proportional to $\sigma(T) r$ at
large distances $r$, as expected. The effective temperature dependent string tension $\sigma(T)$
is non-zero below the phase transition temperature \cite{Kaczmarek:1999mm,Digal:2003jc}. 
Consequently the free energy of an isolated static quark is infinity and 
$L_{ren}(T)=0$. In the deconfined phase $F_{\infty}(T)$ is finite due to color screening. In particular at
high temperatures $F_{\infty}=-4 \alpha_s m_D/3$ at leading order, where $m_D=g T$ is the leading order Debye mass.
Here and in what follows we consider only the case $N=3$.
The next-to-leading order correction for $F_{\infty}$ has
been calculated in Refs. \cite{Burnier:2009bk,Brambilla:2010xn} and was found to be small. 
For the free energy of a static $Q\bar Q$ pair at distance $r$ we have 
\cite{Nadkarni:1986cz}
\begin{equation}
F(r,T)=-\frac{1}{9} \frac{\alpha_s^2}{r^2} \exp(-2 m_D r).
\end{equation}
This result is in contrast to the free energy of static charges in QED which is $-\alpha \exp(-m_D r)/r$,
and is the consequence of non-Abelian nature of interactions. A static quark anti-quark pair could be either
in a singlet or in an octet state, and thus (in a fixed gauge) one can define the so-called color singlet and
octet free energy \cite{Nadkarni:1986cz,Nadkarni:1986as} \footnote{The terms color singlet free energy and octet free energy are misleading as only
$F(r,T)$ has the meaning of the free energy of static quark anti-quark pair. Furthermore, at low temperatures both $F_1$ and
$F_8$ are determined by color singlet asymptotic states \cite{Jahn:2004qr}. The color screening of static quark quark
interaction was discussed in Ref. \cite{Doring:2007uh}, while color screening of charges in higher representations was studied in Refs.
\cite{Gupta:2007ax,Mykkanen:2012ri}.}
\begin{eqnarray}
\displaystyle
\exp(-F_1(r,T)/T+c/T)&=&
\frac{1}{3} \langle {\rm Tr}[L^{\dagger}(x)  L(y)]\rangle,\label{F1def}\\
\displaystyle
\exp(-F_8(r,T)/T+c/T)&=&
\frac{1}{8}
\langle {\rm Tr} L^{\dagger}(x)  {\rm Tr} L(y) \rangle \nonumber\\
\displaystyle
&-&\frac{1}{24} \langle {\rm Tr}\left[L^{\dagger}(x)  L(y)\right]\rangle.\label{Fadef}
\end{eqnarray}
Then the Polyakov loop correlator can be written as the thermal average over 
the singlet and the octet contributions 
\cite{McLerran:1981pb,Nadkarni:1986cz,Nadkarni:1986as}
\begin{equation}
C_{PL}(r,T)=\frac19 \exp(-F_1(r,T)/T+c/T)+\frac89 \exp(-F_8(r,T)/T+c/T).
\label{decomp}
\end{equation}
The singlet and octet free energies can be calculated at high temperature in
leading order HTL approximation \cite{Petreczky:2005bd}, resulting in
\begin{eqnarray}
\displaystyle
F_1(r,T)&=&-\frac{4}{3} \frac{\alpha_s}{r} \exp(-m_D r)-\frac{4 \alpha_s m_D }{3},
\label{f1p}\\
\displaystyle
F_a(r,T)&=&\frac{1}{6} \frac{\alpha_s}{r} \exp(-m_D r)-\frac{4 \alpha_s m_D}{3}.
\label{f3p}
\end{eqnarray}
The singlet and octet free energies are gauge independent at this order. 
Inserting the above expressions into Eq. (\ref{decomp}) and expanding
in $\alpha_s$ one can see that the leading order contributions from the singlet and
octet channels cancel each other and one recovers the leading order
result for $F(r,T)$ in Eq. (\ref{decomp}). It was not clear how to generalize
the decomposition of the free energy into the singlet and octet contributions beyond
leading order. Recently using the effective theory approach, namely the potential
non-relativistic QCD (pNRQCD) at finite temperature \cite{Brambilla:1999xf,Brambilla:2008cx} 
it was shown that this decomposition indeed
holds at short distances \cite{Brambilla:2010xn}. As we will see below the singlet free energy turns out to be
useful when studying screening numerically in the lattice calculations.

The QCD partition function does not have the $Z(3)$ symmetry. This symmetry  is broken by the quark
contribution. In physical terms this means that a static quark anti-quark pair can be
screened already in the vacuum by light dynamical quarks. The free energy does not
rise linearly with distance but saturates at some distance, i.e. we see string breaking.
For light dynamical quarks the corresponding free energy $F_{\infty}$ is twice the binding energy 
of static-light meson
and thus is proportional to $\Lambda_{QCD}$. Therefore, the  Polyakov loop is expected to be ${\cal O}(1)$ 
in the transition region.
At very high temperatures, however, the screening in the QCD and purely gluonic plasma
is expected to be similar. The above discussion of the free energy of static quark anti-quark
pair still holds and the only difference is in the value of the Debye mass, which now becomes
$m_D=g T \sqrt{1+N_f/6}$. Here $N_f$ being the number of light quark flavors, i.e. $N_f=3$. 

\begin{figure}
\includegraphics[width=8cm]{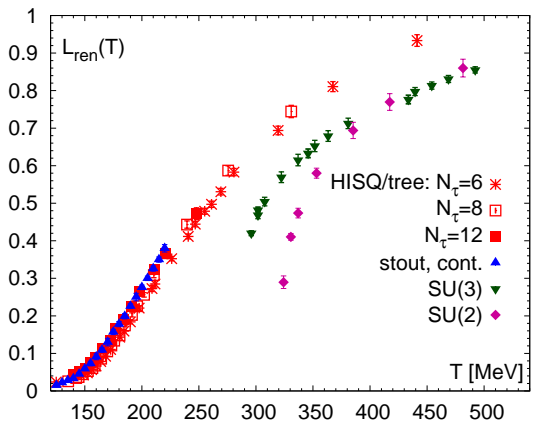}
\caption{The Polyakov loop as function of the temperature in 2+1 flavor QCD \cite{Borsanyi:2010bp,Bazavov:2011nk}
and in pure gauge theory \cite{Digal:2003jc,Kaczmarek:2002mc}.}
\label{fig:lren}
\end{figure}

In Fig. \ref{fig:lren} I show the recent lattice results for the renormalized Polyakov loop
in 2+1 flavor QCD with physical quark masses \cite{Borsanyi:2010bp} 
compared to the results in pure gluodynamics \cite{Digal:2003jc,Kaczmarek:2002mc}.
The temperature scale in pure gauge theory was set using the value $\sqrt{\sigma}=465.2$MeV for
the zero temperature string tension. We see a relative smooth increase in the Polyakov loop
in 2+1 flavor QCD starting at temperatures of about $160$MeV.
Earlier results for the renormalized Polyakov loop in 2 and 3 flavor QCD with significantly
heavier quark masses have been reported in Refs. \cite{Kaczmarek:2005ui,Petreczky:2004pz}
The behavior of the Polyakov loop in pure glue theory and full QCD is quite different in the
transition region, while at high temperatures it is qualitatively similar. 

The free energy of a static $Q\bar Q$ pair as well as the singlet free energy was calculated
using $p4$ action on $16^3 \times 4$ and $24^3 \times 6$ lattices for almost physical light quark masses,
namely $m_l=m_s/10$ \cite{Petreczky:2010yn,Kaczmarek:2007pb}.
There are also calculations with Wilson fermion formulation with much heavier quark masses \cite{Maezawa:2011aa}
as well as preliminary results for the $stout$ action \cite{Fodor:2007mi}. Qualitatively these results are similar
to the ones obtained with $p4$ action. A very detailed calculation of the free energy of $Q \bar Q$
pair in pure glue theory was presented in Refs. \cite{Kaczmarek:1999mm,Digal:2003jc,Petreczky:2001pd}, while  
the singlet free energy in Coulomb gauge was studied in Refs. \cite{Kaczmarek:2002mc,Kaczmarek:2004gv}.

\begin{figure}[htbp]
\begin{center} 
\includegraphics[width=7.7cm]{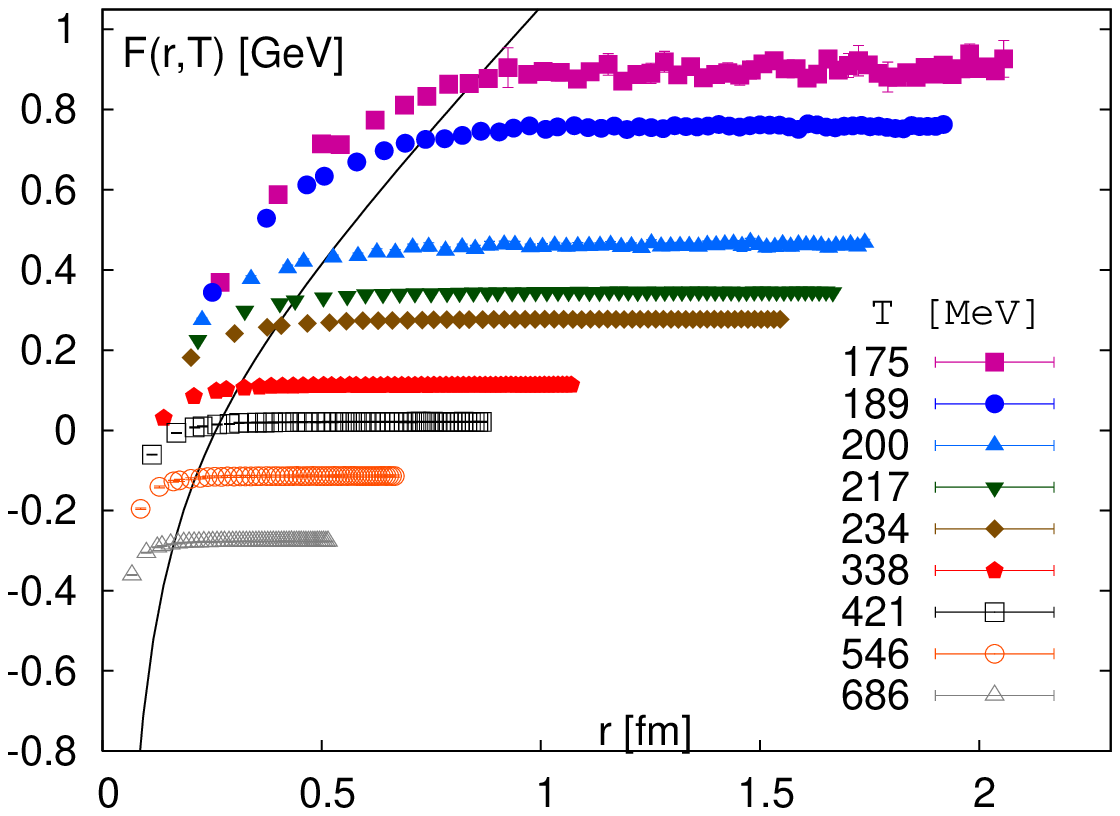}
\includegraphics[width=7.7cm]{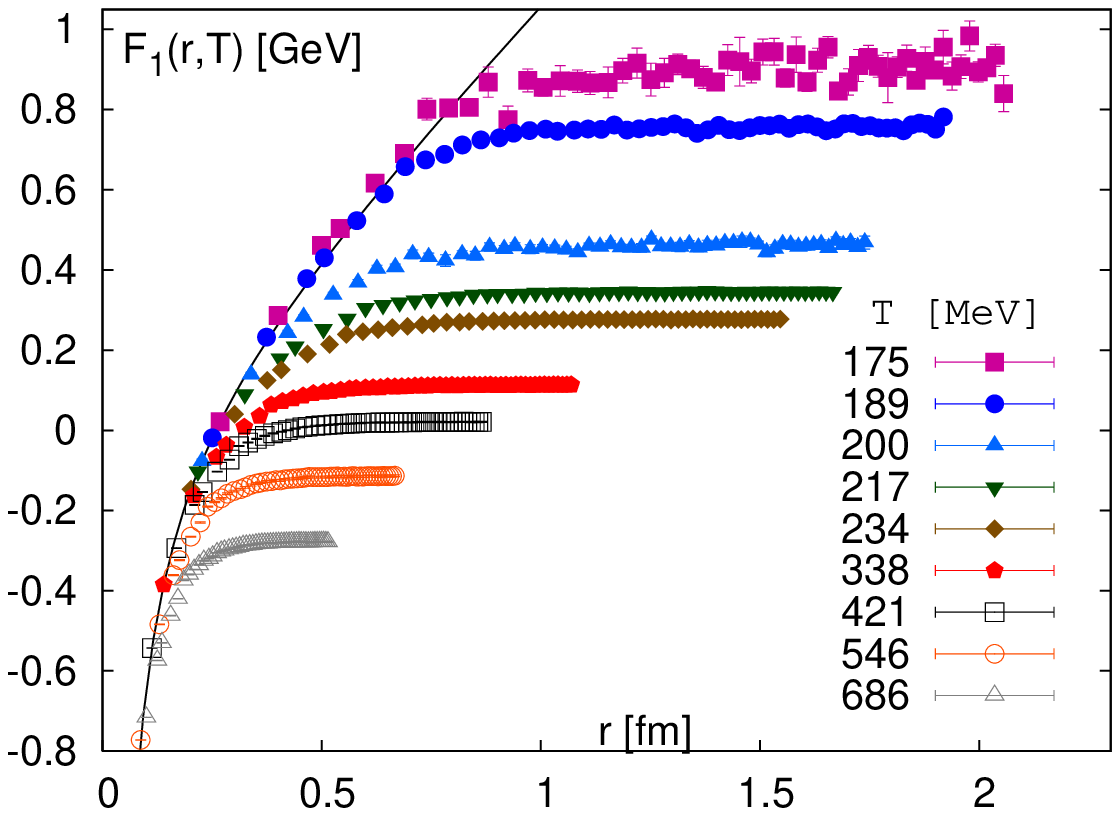}
\vspace*{-0.5cm}
\caption[]{The free energy (left) and the singlet free energy (right) of a 
static $Q\bar Q$ pair calculated in 2+1 flavor QCD
on $16^3 \times 4$ lattices at different temperatures \cite{Petreczky:2010yn}.
The solid black line  is the
parametrization of the zero temperature potential calculated in Ref. \cite{Cheng:2007jq}.
}
\label{fig:f}
\end{center}
\end{figure}

In Fig. \ref{fig:f} I show the numerical results for the free energy and the singlet free
energy obtained with $p4$ action on $16^3 \times 4$ lattice. At low temperatures
the qualitative behavior of the free energy and singlet free energy is similar. 
The singlet free energy agrees with the zero temperature potential shown in the figure
as the black line, while the free energy differs from it by the trivial color factor $T \ln 9$.
At high temperatures the free energy and the singlet free energy show very different behavior, in particular
the temperature dependence is much stronger in $F(r,T)$. The singlet free energy agrees with the zero
temperature potential at short distances, while this is not the case for the free energy even when the
factor  $T \ln 9$ is taken into account. This behavior of the $Q \bar Q$ free energy at high temperatures 
is consistent with the partial cancellation of the color singlet and color octet contributions as described above. 
Similar difference between the free energy and singlet
free energy have been observed in the pure glue theory \cite{Kaczmarek:1999mm,Kaczmarek:2002mc,Petreczky:2001pd}.

\begin{figure}[htbp]  
\begin{center}
\includegraphics[width=7.7cm]{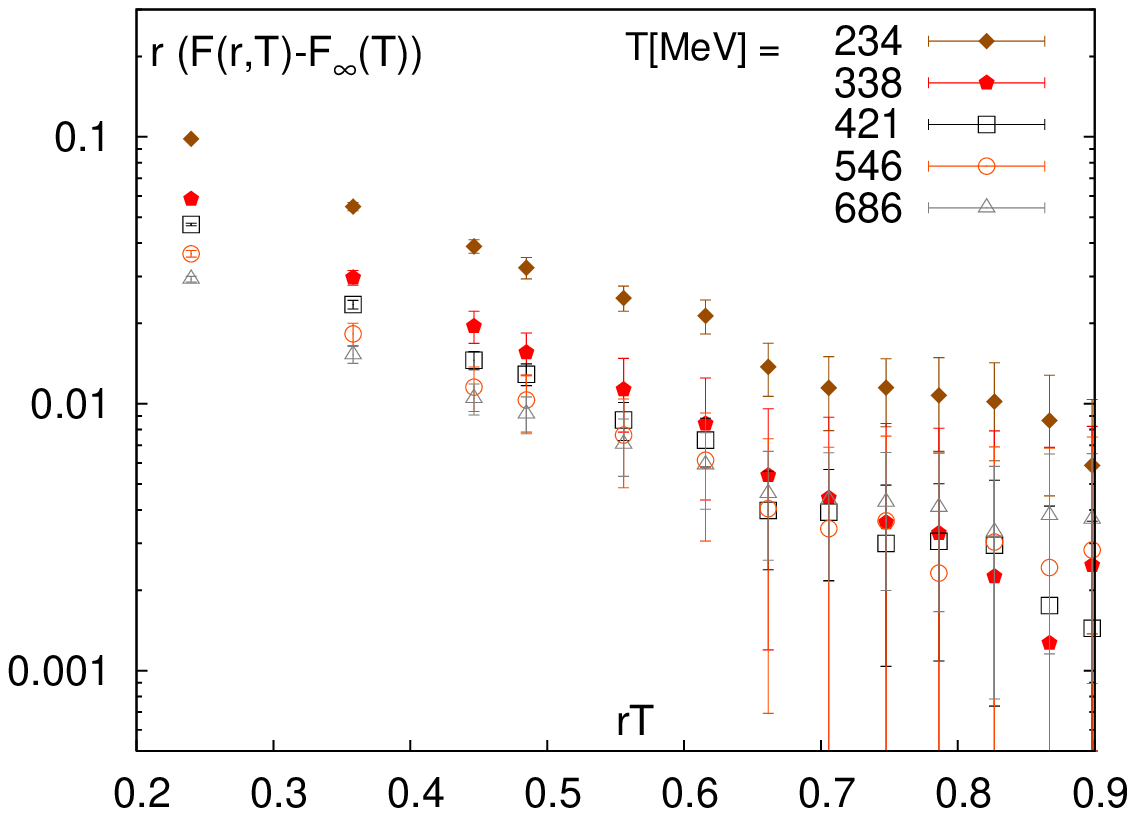}
\includegraphics[width=7.7cm]{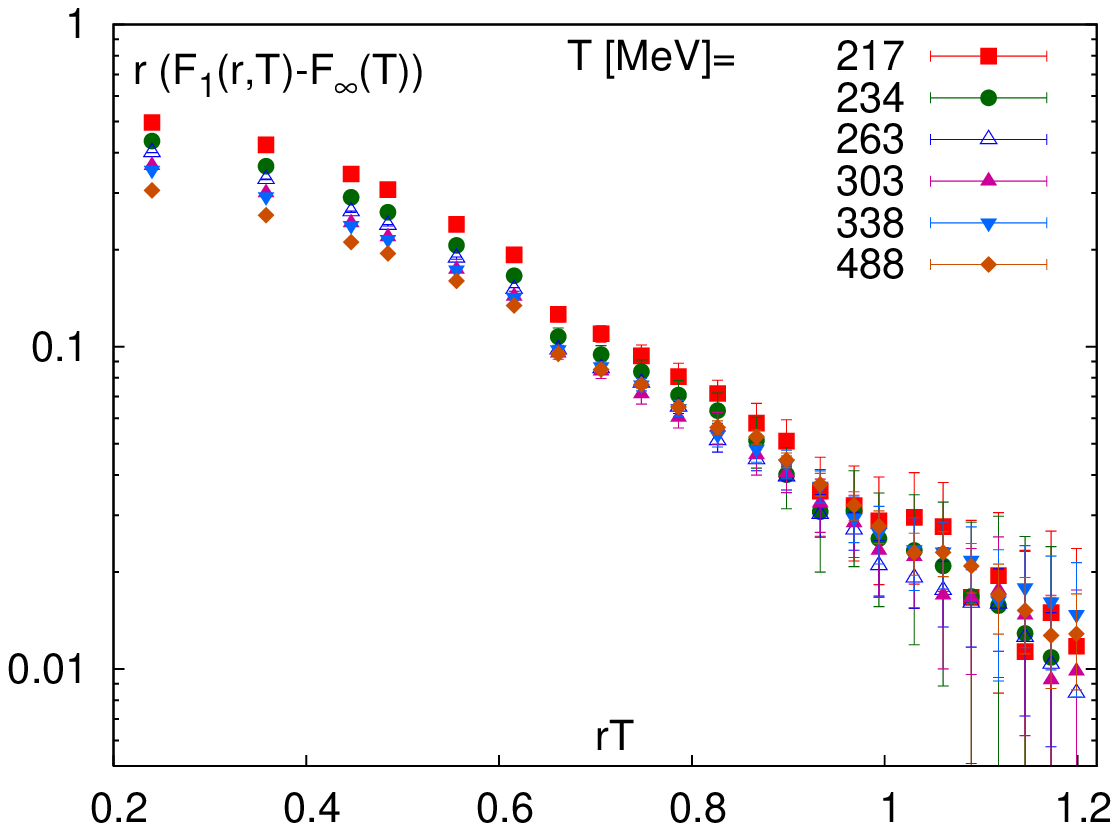}
\vspace*{-0.5cm}
\caption[]{
The combination $r (F(r,T)-F_{\infty}(T))$ (left) and $r (F_1(r,T)-F_{\infty}(T))$  (right)
for different temperatures as function of $r T$ \cite{Petreczky:2010yn}.
}
\label{fig:s}
\end{center}
\end{figure}

Color screening means that $r (F_1(r,T)-F_{\infty}(T))$
as well as $r (F(r,T)-F_{\infty}(T))$ should decay exponentially at large distances. Therefore in Fig. \ref{fig:s}
I show these combinations. At distance $rT >0.8$ we indeed see the expected exponential falloff.
Fitting the lattice data for $r (F_1(r,T)-F_{\infty}(T))$ to an exponential one can determine the Debye
mass non-perturbatively. This was done for pure glue theory and 2+1 flavor QCD. It turns out that 
the temperature dependence of the Debye mass is well described by the leading order result but its
value is about factor 1.4 larger both in pure glue theory and 2+1 flavor QCD \cite{Kaczmarek:2007pb}
\footnote{The Debye mass obtained in the calculations with Wilson fermions is somewhat
larger \cite{Maezawa:2011aa}. More calculations are needed to resolve this discrepancy.}.
This means that the dependence of the Debye mass on quark flavors is well
described by perturbation theory. The Debye mass is non-perturbative beyond leading
order \cite{Rebhan:1993az,Rebhan:1994mx}.
The non-perturbative enhancement of the Debye mass over the leading order result 
is expected based on the gap equation
calculation of the Debye mass \cite{Patkos:1997cw}.

The qualitative features of the singlet free energy discussed above are not specific to the Coulomb gauge.
In fact, the singlet free energy defined in terms of Wilson loops show very similar behavior to the
one calculated in the Coulomb gauge \cite{Bazavov:2008rw}. Furthermore, the Debye mass can also be defined from the long
distance behavior of the electric gluon propagator 
\cite{Heller:1995qc,Heller:1997nqa,Karsch:1998tx,Cucchieri:2001tw,Nakamura:2003pu,Cucchieri:2012nx}.
Calculations in $SU(2)$ and $SU(3)$ gauge theories 
show no gauge dependence of the extracted Debye mass within statistical errors \cite{Cucchieri:2001tw,Nakamura:2003pu}.
The extracted screening masses  are in agreement
with the ones obtained from the singlet free energy 

\subsection{Color magnetic screening and dimensional reduction}

Contrary to ordinary plasmas that have no magnetic screening the static chromo-magnetic fields
are screened in QGP. This is due to the fact that unlike photons gluons interact with each other
(the stress tensor is non-linear in QCD). Magnetic screening is non-perturbative, i.e. it does not
appear at any finite order of pertubation theory and  is needed to cure the infrared
divergences of the pressure rendering it finite but also non-perturbative at order $g^6$ \cite{Linde:1980ts}.
In analogy with electric screening magnetic screening
can be studied using gauge fixed magnetic gluon propagators. However, the large distance behavior
of the magnetic propagators is more complex and much more susceptible to finite volume effects.
In Landau type gauges as well as in Coulomb gauge the magnetic propagator shows
oscillating behavior instead of decaying exponentially \cite{Cucchieri:2000cy}.
Thus no magnetic mass can be defined. However, one can define a common exponentially falling envelope
for different gauges that then gives a screening mass of about $0.5 g^2 T$ \cite{Cucchieri:2000cy}.
A magnetic mass of similar size was found in analytic approach of Refs. \cite{Karabali:1997wk,Nair:1998es,Nair:2002yg}.
Thus the corresponding
effective  magnetic screening length is larger than the electric screening length as one would expect
for weakly coupled QGP \footnote{Speculations on existence of a magnetic screening length of similar
size have been presented in Ref. \cite{Moore:2000ara} }. 
\begin{figure}
\includegraphics[width=8cm]{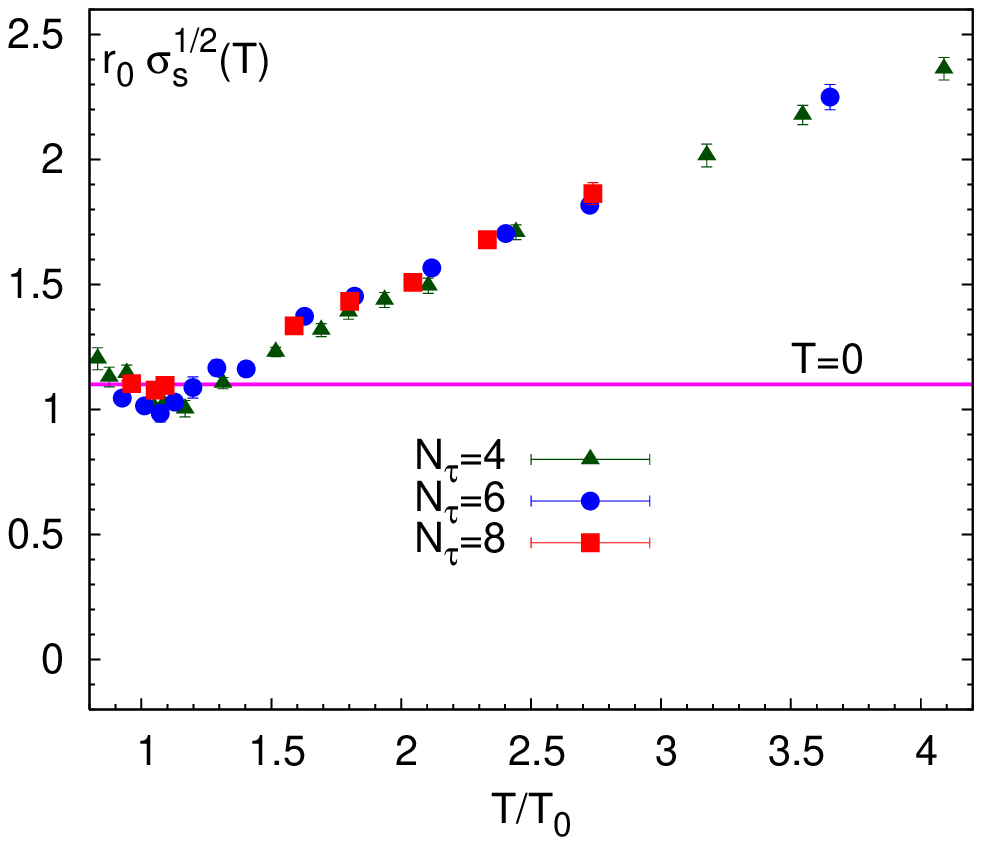}
\includegraphics[width=8cm]{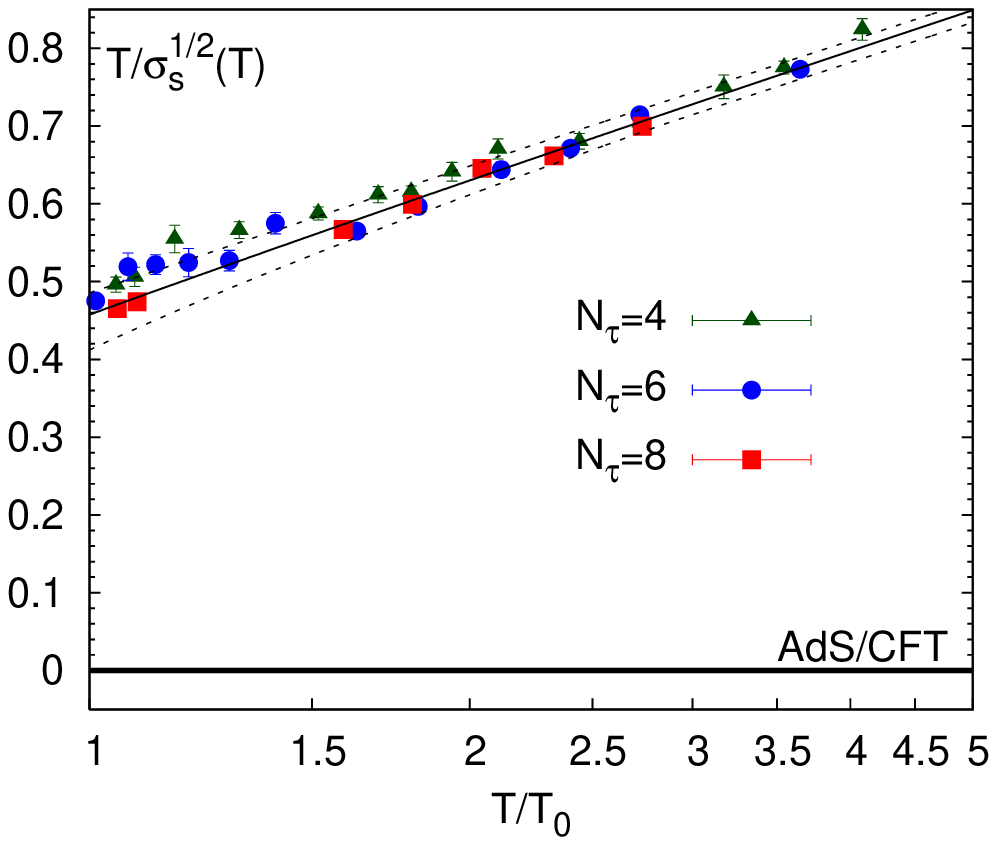}
\caption{(Left)The string tension in units of $r_0$ as function of the temperature
calculated on $N_{\tau}=4,~6$ and $8$ lattices. For better visualization we scale
the temperature axis by $T_0=200$MeV.
(Right) $T/\sqrt{\sigma_s(T)}$ calculated on $N_{\tau}=4,~6$ and $8$ lattices
compared with the prediction of dimensional reduction indicated by the line. 
The uncertainty in the prediction of the dimensionally reduced theory is show
 by dashed lines.
}
\label{fig:sigmas}
\end{figure}

Spatial Wilson loops can also be used to define a length scale associated
with static magnetic fields. They obey area law at any temperature governed by spatial string tension $\sigma_s(T)$.
The numerical results for the spatial string tension calculated in 2+1 flavor QCD with $p4$ action
and $m_l=m_s/10$ are shown in Fig. \ref{fig:sigmas}.
At low temperatures $\sigma_s(T)$ coincides with the zero temperature string tension, while at high temperatures
$\sqrt{\sigma_s(T)}\sim T$. This behavior can be understood in terms of dimensionally reduced effective theory.
At high temperatures there is a separation of different scales $2 \pi T \gg m_D \gg g^2 T$. If we integrate out the
modes associated with the scale $2 \pi T$ the thermodynamics and the screening lengths can be described in terms of an effective
field theory, which at the lowest order is the 3 dimensional adjoint Higgs model in the confined phase 
\cite{Appelquist:1981vg,Braaten:1995jr,Kajantie:1997tt}
\bea
S_3^E &=& \int d^3x \, {1 \over g_3^2} \,
\Tr \, F_{ij}({\bf x}) \, F_{ij}({\bf x}) + \, \Tr \, [D_i,A_0({\bf x})]^2
\nonumber \\
&{}&  
\, + \, m_D^2 \Tr A_0({\bf x})^2 + \lambda_A (\Tr A_0({\bf x})^2)^2 \; .
\label{eq.dm1} 
\eea
The static electric fields become adjoint scalars in the three dimensional effective theory.
The couplings of the effective theory can be calculated in perturbation theory at any order 
\cite{Braaten:1995jr,Kajantie:1997tt}.
The spatial string tension should be identified with the string tension of this effective three dimensional theory 
$\sigma_3$.
The gauge coupling constant in three dimensions $g_3^2$ has the dimension of the mass, and at the lowest order 
(tree level ) $g_3^2=g^2 T$. On dimensional grounds $\sigma_3=c g_3^2$ with $c$ being a constant. This explains
the linear rise of the string tension at high temperatures. 
The coefficient $c$ was calculated in the three dimensional adjoint Higgs model 
\cite{Hart:1999dj,Hart:2000ha}.
The three dimensional gauge $g_3$ coupling was calculated to 2-loop \cite{Laine:2005ai}. 
Combining this with the known value of $c$ we get
the weak coupling prediction for the spatial string tension which is compared with the 
lattice data in Fig. \ref{fig:sigmas}.
Quite surprisingly, the weak coupling result for the spatial string tension works well
down temperatures of $200$MeV, where the effective field theory is not supposed to work. 
Note that in the conformal strongly coupled gauge theory $T/\sqrt{\sigma_s(T)} \sim 1/\lambda^{1/4}$ 
(see e.g. Ref. \cite{Sin:2006yz} and references therein). Thus the ratio $T/\sqrt{\sigma_s(T)}$  is temperature
independent and vanishes in the limit of infinite t'Hooft coupling, $\lambda \rightarrow \infty$. This is
in sharp contrast with lattice calculations. 
To deal with this problem one can modify the AdS metric to break the conformal invariance and get a QCD-like theory. 
This approach
is known as AdS/QCD. Attempts to model the temperature dependence of the spatial 
string tension using  AdS/QCD were discussed in Ref. \cite{Andreev:2006eh,Andreev:2007rx,Alanen:2009ej}.

\subsection{The spectrum of gauge invariant screening masses}

\begin{figure}
\includegraphics[width=8cm]{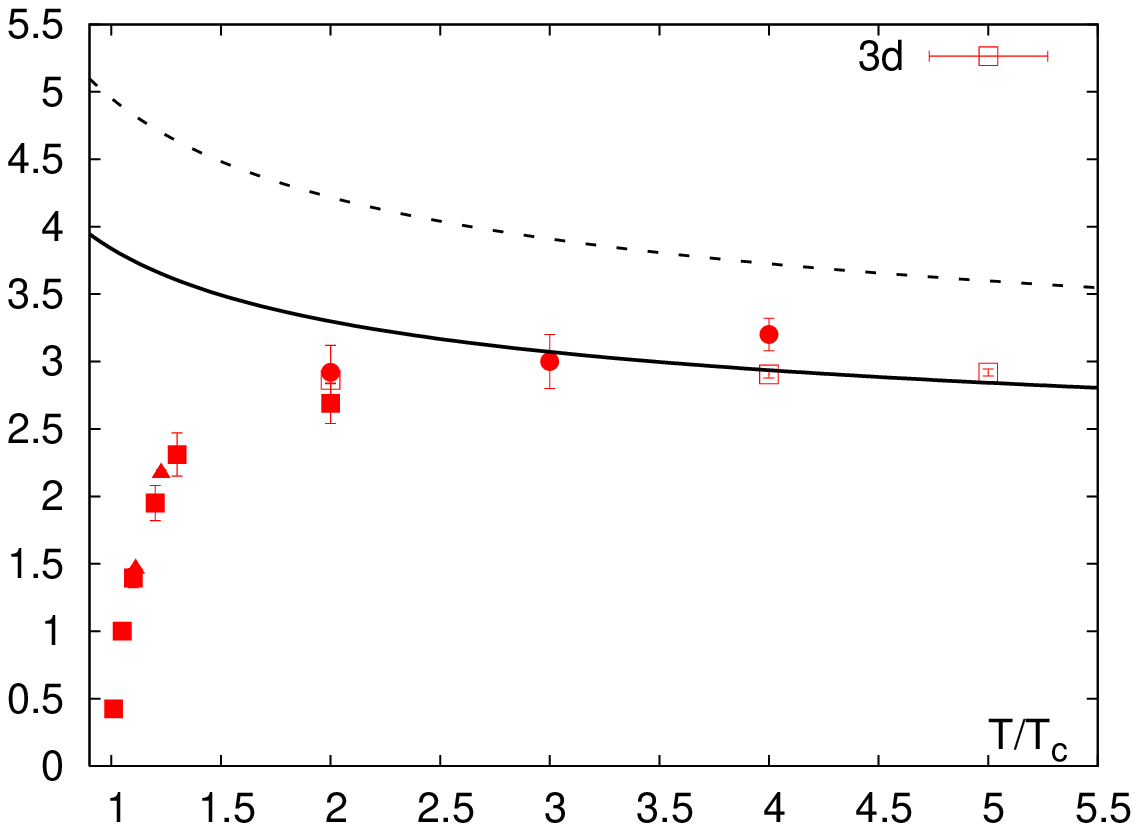}
\includegraphics[width=8cm]{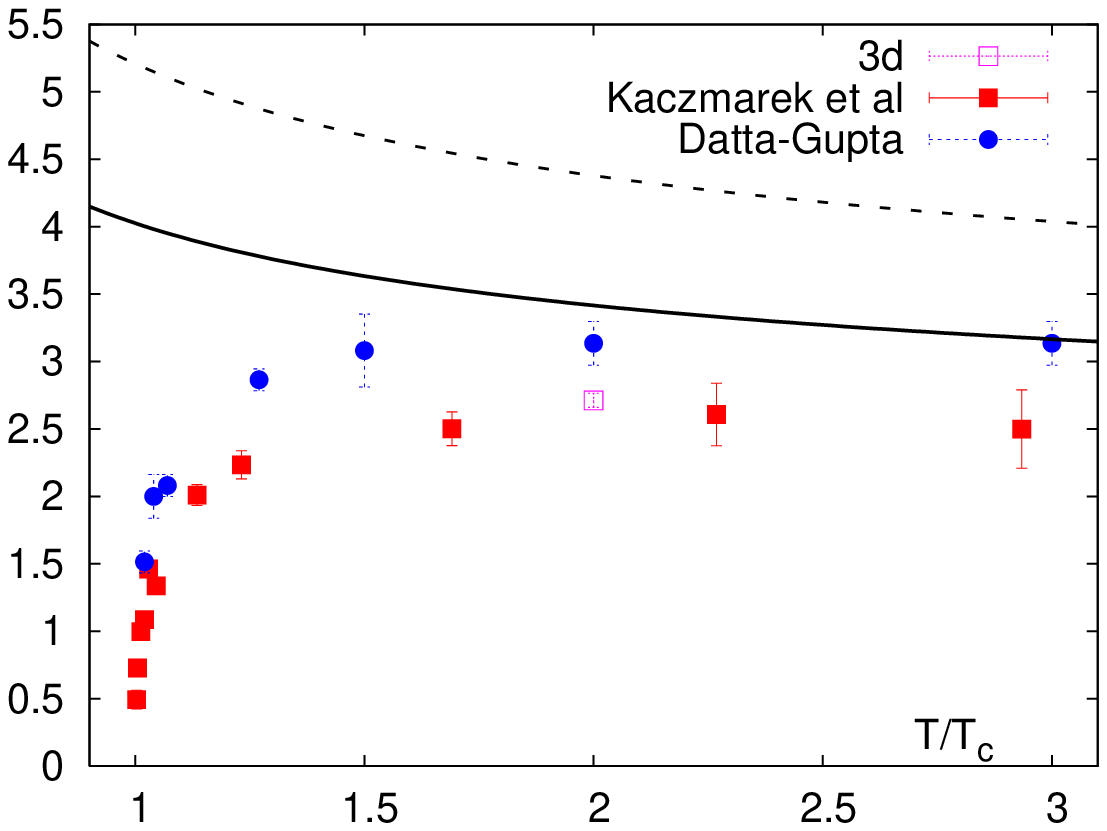}
\caption{The lightest screening mass in $0^{++}$ channel calculated from the Polyakov loop correlators
in $SU(2)$ gauge theory \cite{Digal:2003jc,Datta:1999yu,Fiore:2003yw} (left) and $SU(3)$ gauge theory \cite{Kaczmarek:1999mm,Datta:2002je} (right).
Also shown as open symbols are the results of the calculations 
in the 3d effective theory \cite{Hart:1999dj,Hart:2000ha}. The solid and dashed lines are the perturbative
predictions at leading order and next-to-leading order \cite{Laine:2009dh}. }
\label{fig:mplc}
\end{figure}

There are several 
screening masses that can be studied in the framework of the 3 dimensional effective theory 
\cite{Kajantie:1997pd,Laine:1997nq,Laine:1999hh,Hart:1999dj,Hart:2000ha}
as well as in the original four dimensional theory \cite{Datta:1998eb,Datta:1999yu,Datta:2002je,Digal:2003jc}.
Comparison of the screening masses
calculated in these two theories can provide information on the validity of the dimensionally reduced
effective theory. The screening masses can be classified according to $J^{PR}$ quantum numbers. 
Here $R$ is the analog of C-parity. It is related to the symmetry with respect to the following transformation
of the static electric field, $A_0 \rightarrow -A_0$.  All the operators transform trivially under charge
conjugation in the adjoint Higgs model \cite{Hart:1999dj}. 
In the $0^{++}$ channel there are two distinct screening masses: screening mass $m(A_0^2)$ corresponding
to the ``electric'' correlator of ${\rm Tr} A_0^2$, and the screening mass $m_G$ corresponding to the 
``magnetic'' correlator ${\rm Tr} F_{ij}^2$. These correlators can  mix but this mixing
is suppressed for weak coupling.  The lightest screening mass $m_E$ in the $0^{--}$ channel
was proposed as the non-perturbative definition of the Debye mass \cite{Arnold:1995bh} since the negative $R$
parity prevents mixing with the magnetic sector. 
For asymptotically small couplings we expect $m_G< m_E < m(A_0^2)$
and the large distance behavior of the Polyakov loop correlators is actually governed by the ``magnetic''
screening mass $m_G$ \cite{Arnold:1995bh,Braaten:1994qx}.
For the interesting temperature range, however, $m(A_0^2)$ turns out to be the smallest 
and $m_G$ turns out to be the largest of the above screening masses \cite{Hart:1999dj,Hart:2000ha}. 
Furtheremore,  there is almost
no mixing between the ``electric'' and ``magnetic'' correlators in the $0^{++}$ channel \cite{Hart:1999dj,Hart:2000ha}.
Thus the long distance behavior of the Polyakov
loop is determined by the screening mass corresponding to ${\rm Tr} A_0^2$ correlator (
see also discussion in Ref. \cite{Braaten:1994qx}).
In Fig. \ref{fig:mplc} I show the screening masses extracted from the Polyakov loop correlators in
$SU(2)$ and $SU(3)$ gauge theories and compared to the corresponding screening masses calculated in
the effective three dimensional theory. There is fairly good agreement with the screening masses 
calculated in the full theory and in the effective theory.

The lightest
screening mass in the $0^{++}$ channel was calculated in perturbation theory at next-to-leading order \cite{Laine:2009dh} and 
the corresponding result is also shown in the figure. 
The  next-to-leading order result significantly 
overestimates the screening masses, while the leading order result seems to work fine.
The next-to-leading order calculations also explains to some extent why the screening mass in the ${\rm Tr} A_0^2$
channel is the smallest \cite{Laine:2009dh}. An interesting observation has been made in Ref. \cite{Datta:2002je},
where screening masses corresponding to the correlators of the real and imaginary parts of the Polyakov
loop have been calculated. In terms of the effective 3d theory these correlators correspond
to the correlators of ${\rm Tr} A_0^2$ and ${\rm Tr} A_0^3$ respectively. The ratio of the screening
masses extracted from these correlators should be $2/3$ at leading order and the  lattice calculations
of Ref. \cite{Datta:2002je} confirm this result.

%% file: fluctuations.tex
\section{Deconfinement : Fluctuations of conserved charges}
\label{sec:fluct}
As mentioned before due 
to the infamous sign problem lattice QCD Monte-Carlo simulations are not possible at non-zero
quark chemical  potentials.
However, the pressure at non-zero chemical potentials can be evaluated using Taylor expansion. The 
Taylor expansion can be set up in terms of the quark chemical potentials $\mu_u$, $\mu_d$ and
$\mu_s$,  or in terms of the chemical
potentials corresponding to baryon number $B$, electric charge $Q$ and strangeness $S$ of hadrons
\bea
\frac{p}{T^4}&=&\frac{1}{VT^3}\ln Z(T,\mu_u,\mu_d,\mu_s)=\sum_{ijk} \frac{1}{i! j! k!} \chi_{ijk}^{uds} 
\left(\frac{\mu_u}{T}\right)^i \left(\frac{\mu_u}{T}\right)^i \left(\frac{\mu_d}{T}\right)^j \\
\chi_{ijk}^{uds}&=&\frac{\partial^{\,i+j+k}p/T^4}{\partial(\mu_{u}/T)^{i}\partial(\mu_{d}/T)^{j}\partial(\mu_{s}/T)^{k}}\\
\frac{p}{T^4}&=&\frac{1}{VT^3}\ln Z(T,\mu_B,\mu_S,\mu_Q)=\sum_{ijk} \frac{1}{i! j! k!} \chi_{ijk}^{uds} 
\left(\frac{\mu_u}{T}\right)^i \left(\frac{\mu_u}{T}\right)^i \left(\frac{\mu_d}{T}\right)^j \\
\chi_{ijk}^{BQS}&=&\frac{\partial^{\,i+j+k}p/T^4}{\partial(\mu_{B}/T)^{i}\partial(\mu_{Q}/T)^{j}\partial(\mu_{S}/T)^{k}}.
\eea
While Taylor expansion can be used to study the physics at non-zero baryon density the
expansion coefficients are interesting on 
their own right as they are related to the fluctuations and correlations of conserved charges.
The diagonal expansion coefficients are related to second and higher order fluctuations of conserved charges
\begin{eqnarray}
\chi_2^X &=& \frac{1}{VT^3}\langle N_X^2\rangle \nonumber \\
\chi_4^X &=& \frac{1}{VT^3}\left(\langle N_X^4\rangle - 
3 \langle N_X^2\rangle^2\right) \;\; \\
\chi_6^X &=& \frac{1}{VT^3}\left( \langle N_X^6\rangle - 
15 \langle N_X^4\rangle \langle N_X^2\rangle  +
30 \langle N_X^2\rangle^3 \right),
\label{fluc}
\end{eqnarray}
while off-diagonal expansion coefficients are related to correlations among conserved charges, e.g.
\bea
\chi_{11}^{XY}=\frac{1}{VT^3}\langle N_XN_Y\rangle.
\eea
The quark chemical potentials are related to the chemical potential of baryon number, electric charge and strangeness
\bea
\mu_u&=&\frac13\mu_B+\frac23\mu_Q,
\nonumber\\
\mu_d&=&\frac13\mu_B-\frac13\mu_Q,
\nonumber\\
\mu_s&=&\frac13\mu_B-\frac13\mu_Q-\mu_S.
\label{chem}
\eea
Therefore the expansion coefficients in quark chemical potential $\chi_{uds}^{jkl}$ are related to 
the hadronic ones $\chi_{BQS}^{jkl}$. In particular, for the second order expansion coefficients we
have 
\begin{eqnarray}
\chi_2^{B}&=& \frac19 \left(
\chi_2^{u}+\chi_2^{d}+\chi_2^{s}
+2\chi_{11}^{us}+2\chi_{11}^{ds}+2\chi_{11}^{ud}\right)\,,
\nonumber\\
\chi_2^{Q}&=& \frac19 \left(
4\chi_2^u+\chi_2^d+\chi_2^s
-4\chi_{11}^{us}+2\chi_{11}^{ds}-4\chi_{11}^{ud}
\right)\,,
\nonumber\\
\chi_2^{S}&=& \chi_2^s,\nonumber\\
\chi_{11}^{BQ}&=&\frac19 \left( 2 \chi_2^u-\chi_2^d-\chi_2^s+\chi_{11}^{us}-2 \chi_{11}^{ds}+\chi_{11}^{ud} \right),
\nonumber\\
\chi_{11}^{BS}&=&-\frac13 \left( \chi_2^s+\chi_{11}^{us}+\chi_{11}^{ds} \right),\nonumber\\
\chi_{11}^{QS}&=&\frac13 \left( \chi_2^s-2\chi_{11}^{us}+\chi_{11}^{ds} \right).
\end{eqnarray}

\begin{figure}[ht]
\includegraphics[width=8cm]{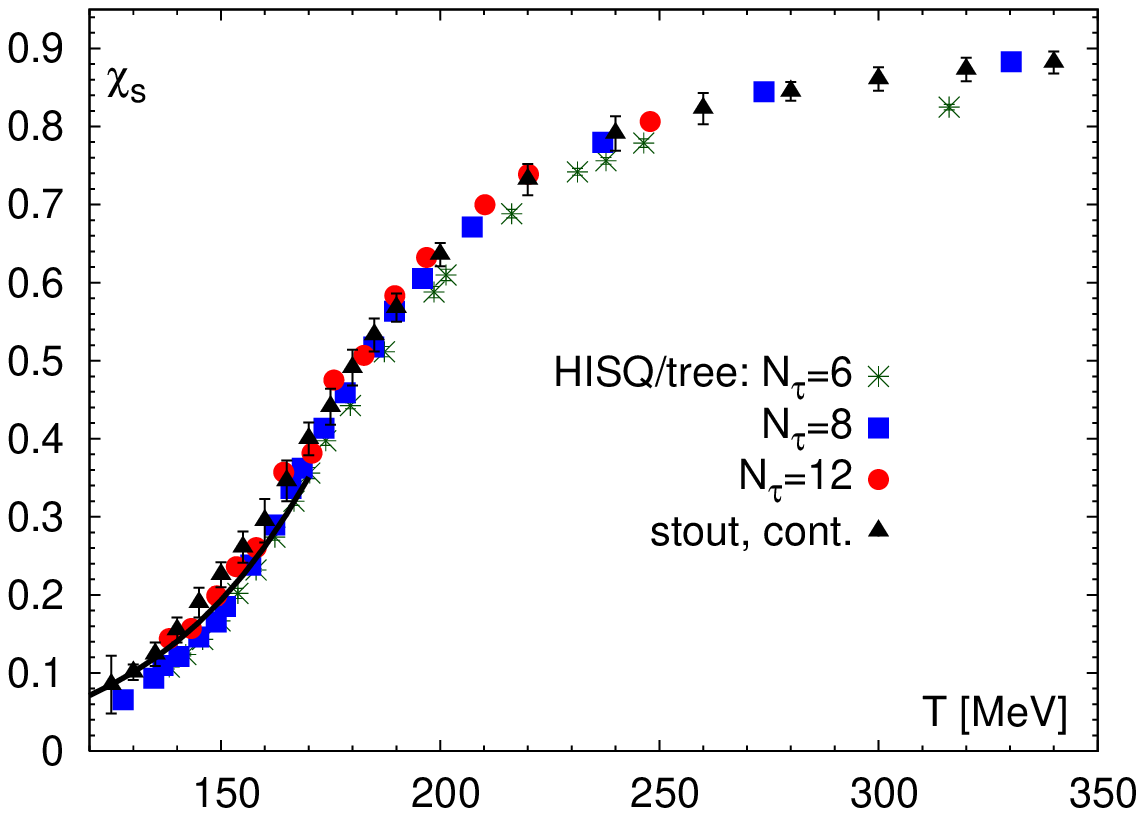}
\includegraphics[width=8cm]{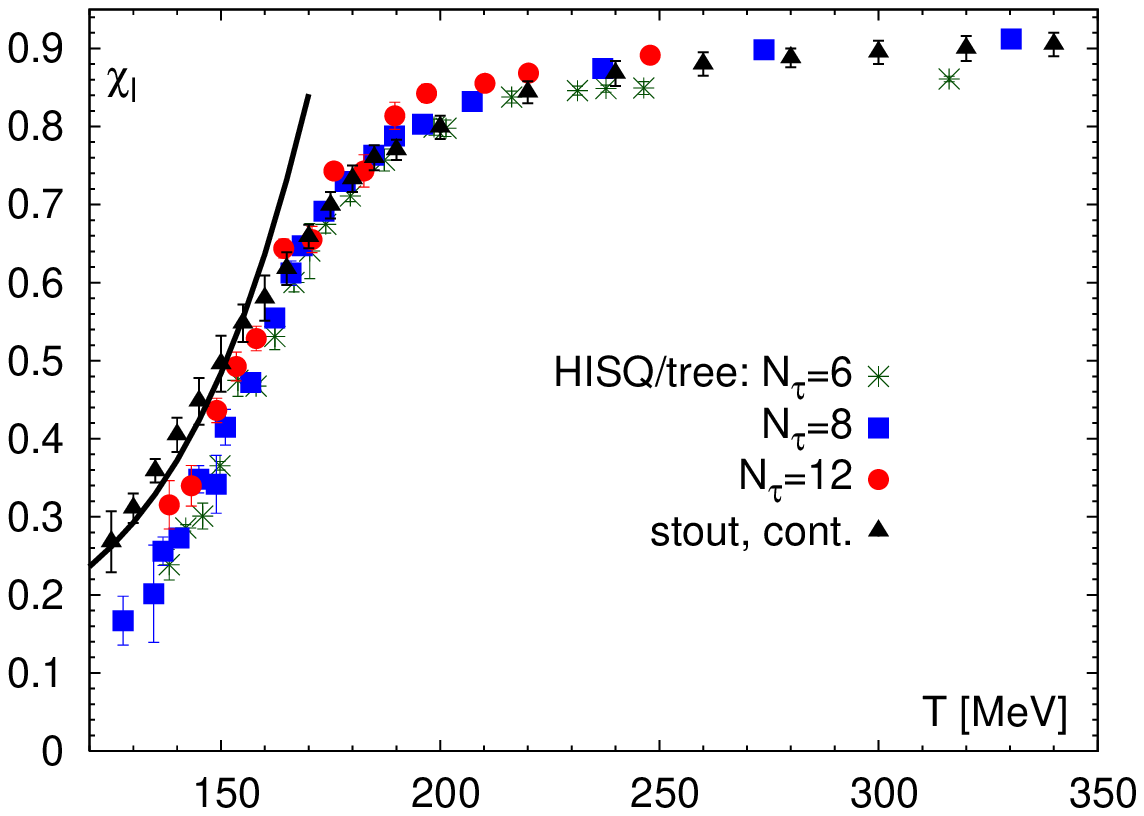}
\caption{Strange quark number (left) and light quark number fluctuations (right) 
calculated with $stout$ \cite{Borsanyi:2011sw} and $HISQ/tree$ actions
\cite{Bazavov:2011nk}. The lattice results are compared with the prediction of the HRG model shown as a black line.}
\label{fig:chis}
\vspace*{-0.2cm}
\end{figure}

Fluctuations and correlations of conserved charges are sensitive probes of deconfinement.
This is because fluctuation of conserved charges are sensitive to the underlying degrees of freedom
which could be hadronic or partonic. 
Fluctuations of conserved charges have been studied using different staggered 
actions \cite{Bernard:2004je,Bazavov:2011nk,Borsanyi:2010bp,Bazavov:2010sb,Cheng:2008zh,Bazavov:2009zn,Petreczky:2009cr,Mukherjee:2011td,Hegde:2011xe,Bazavov:lat11,Borsanyi:2011sw,Bazavov:2012jq}. 
As an example in  Fig. \ref{fig:chis}
I show the quadratic strangeness fluctuation as the function of the temperature
calculated with $HISQ/tree$ and $stout$ actions.  
Fluctuations are suppressed at low temperatures
because conserved charges are carried by massive strange hadrons (mostly kaons). 
They are well described
by Hadron Resonance Gas (HRG) model at low temperatures. 
Strangeness fluctuations rapidly grow in the transition region of
$T=(160-200)$ MeV as consequence of deconfinement. At temperatures $T>250$ MeV strangeness fluctuations are
close to unity that corresponds to the  massless ideal quark gas. Similar picture can seen in other fluctuations, 
in particular for light quark number fluctuation, $\chi_l$, which is also shown in Fig. \ref{fig:chis}. 
The more rapid rise of $\chi_l$ in the transition region is a quark mass effect. In fact, similar quark mass
dependence is seen in the HRG model. At high temperatures the strange quark mass plays little role. In
fact the difference between the  strange and the light quark number susceptibilities is zero within errors.
There is a good agreement between the calculations performed with
$stout$ and $HISQ/tree$ action as the continuum limit is approached. 
Let me finally note that strangeness 
fluctuations have been calculated very recently with Wilson action for the physical values of the quark masses,
and these calculations confirm the staggered result \cite{Borsanyi:2011kg}.
\begin{figure}[ht]
\includegraphics[width=8cm]{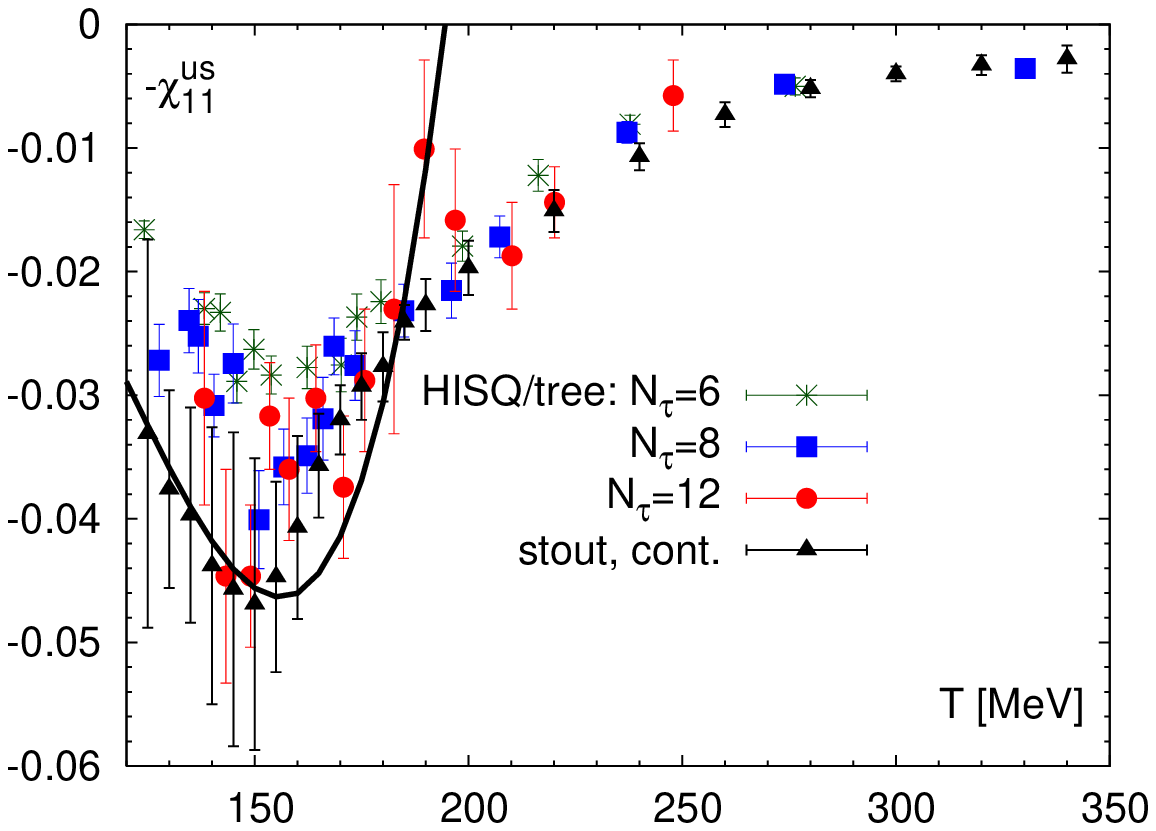}
\includegraphics[width=8cm]{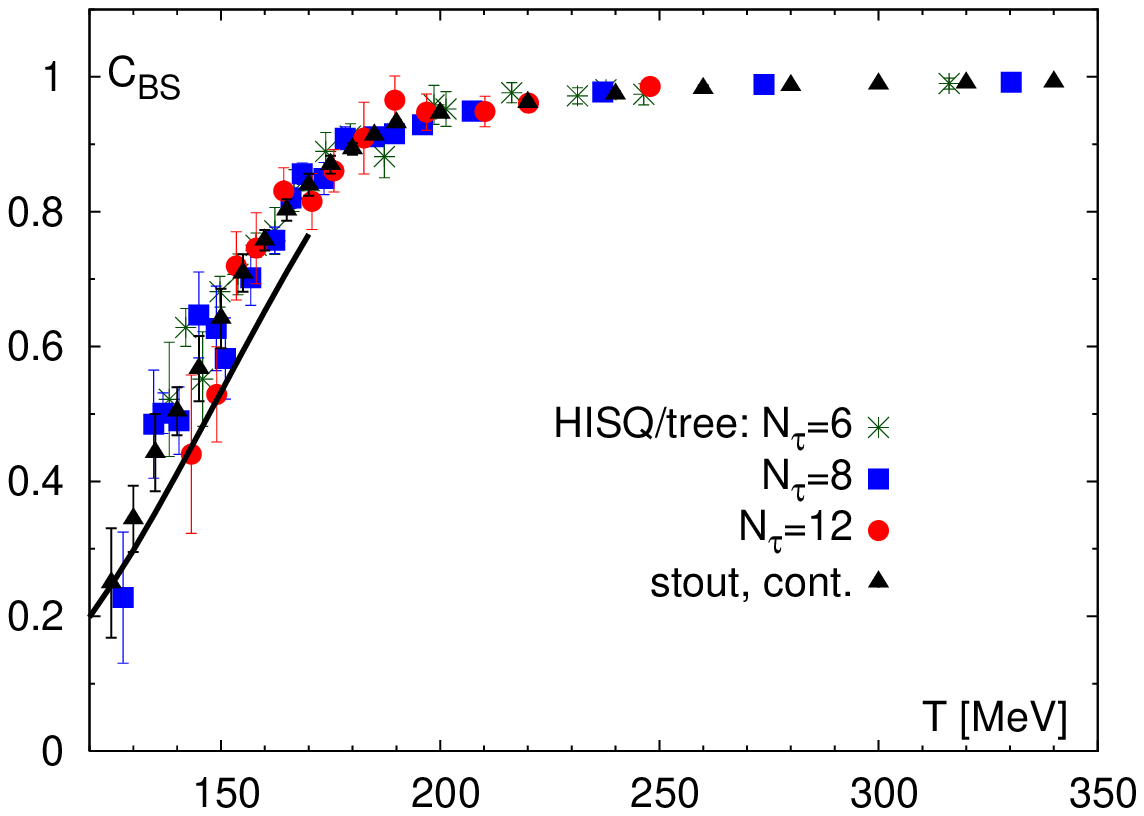}
\caption{Quark number correlation $\chi_{11}^{us}$ (left) and $C_{BS}$ (right)
calculated with $stout$ \cite{Borsanyi:2011sw} and $HISQ/tree$ actions
\cite{Bazavov:lat11,Bazavov:2012jq}. The lattice results are compared with the prediction of the HRG model shown as a black line.}
\label{fig:chi11}
\vspace*{-0.2cm}
\end{figure}

Correlations of conserved charges, in particular quark number correlations
are also sensitive to the relevant degrees of freedom. At infinitely high temperatures the correlations
between different quark numbers should vanish due to the weakly interacting nature of QGP. 
In the high temperature region quark number correlations have been calculated in perturbation theory 
\cite{Blaizot:2001vr,Rebhan:2003fj,Haque:2010rb} and in the framework
of dimensionally reduced effective theory \cite{Hietanen:2008xb}. The deviation from the ideal gas limit turned out
to be quite small.
At low 
temperatures, where hadrons are the relevant degrees of freedom such correlations naturally arise
as different quark flavors are bound in hadrons. In Fig. \ref{fig:chi11} I show the results of
the lattice calculations performed with $stout$ and $HISQ/tree$ action for $\chi_{11}^{us}$. These
results demonstrate the expected features: at high temperatures the correlations are very small while
at low temperatures they are compatible with HRG results. The fact that $\chi_{11}^{us}$ has a minimum
is due to interplay between meson and baryon contributions. Mesons and baryons contribute to $\chi_{11}^{us}$ 
with opposite signs. At low temperatures the negative contribution from mesons dominates $\chi_{11}^{us}$.
As the temperature increases baryon contribution becomes significant and eventually the dominant one because
of the larger density of states in the excited baryon sector. 
It is also interesting to consider the following combination 
\beq
C_{BS}=-3 \chi_{11}^{BS}/\chi_2^s=1+\frac{\chi_{11}^{us}+\chi_{11}^{ds}}{\chi_2^s},
\eeq
that was first introduced in Ref.\cite{Koch:2005vg}. The Boltzmann suppression at low temperature 
is canceled out in this 
combination and at high temperatures it should approach one. The lattice results obtained with
$stout$ and $HISQ/tree$ actions are also shown in Fig. \ref{fig:chi11}.
Again, at low temperatures the lattice results are well described by HRG model, while at high
temperatures $C_{BS}$ is close to one as expected for weakly interacting quarks.

\begin{figure}
\includegraphics[width=7cm]{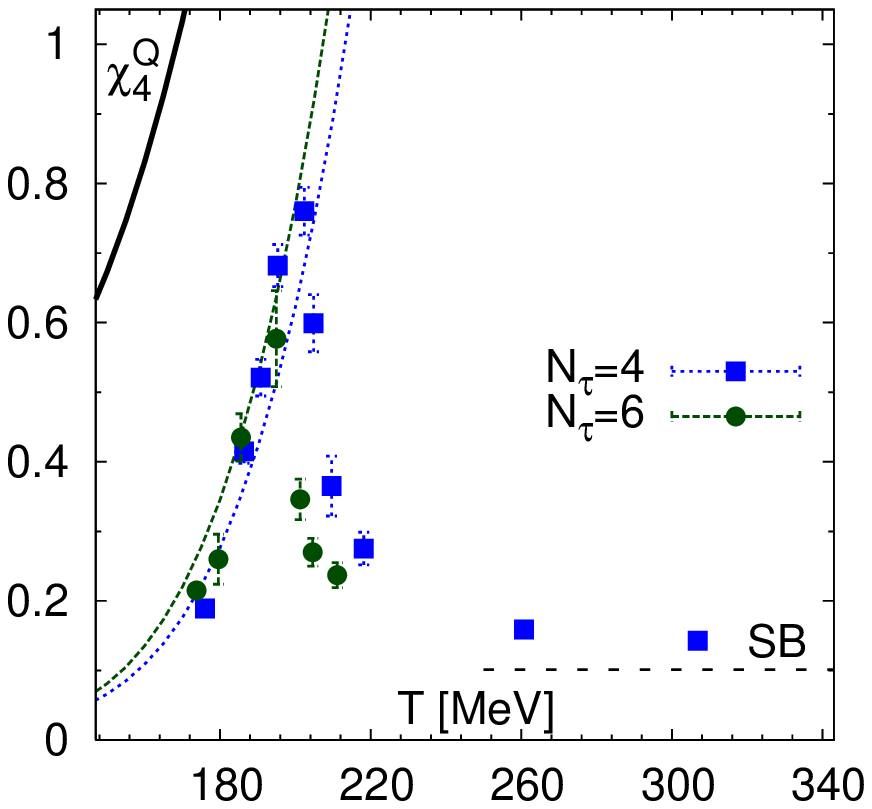}
\includegraphics[width=7cm]{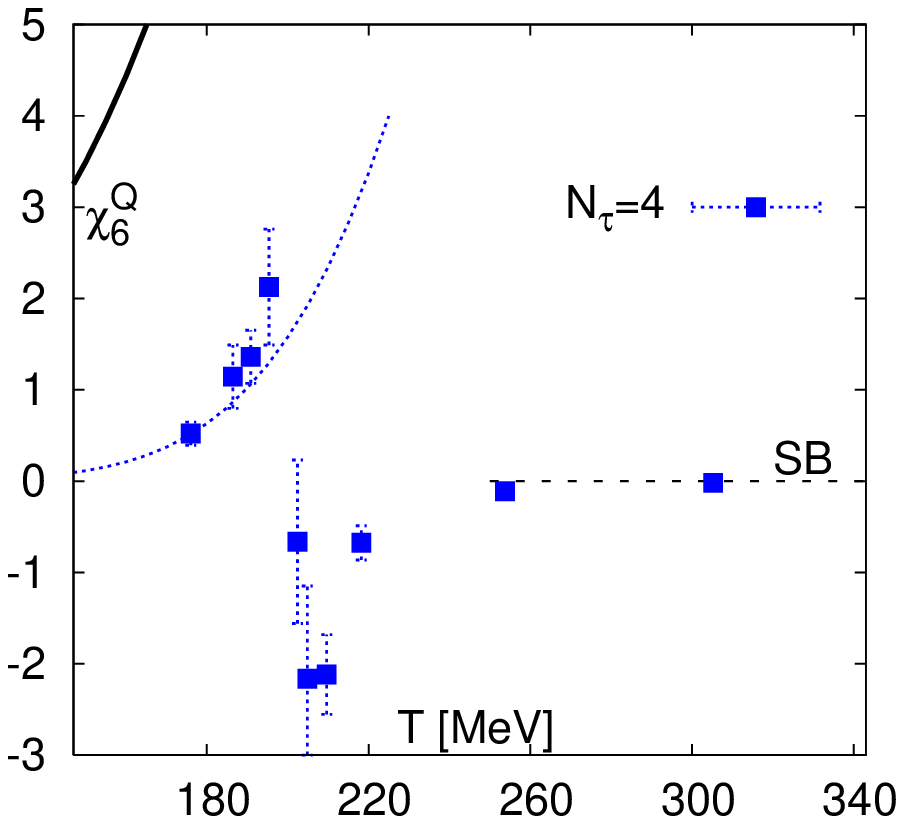}
\caption{The fourth (left) and sixth (right) Taylor expansion coefficients for the
electric charge chemical potential calculated for $p4$ action \cite{Cheng:2008zh}.
The prediction of the HRG model are shown as solid black line. The dotted and dahsed
lines correspond to the HRG model with the distorted hadron spectrum \cite{Huovinen:2010tv,Huovinen:2011xc}.}
\label{fig:chi6}
\end{figure}

Higher order expansion coefficients have been studied in detail for $p4$ action and light
quark mass $m_l=0.1m_s$ \cite{Cheng:2008zh}.  In Fig. \ref{fig:chi6} I show the fourth
and sixth order fluctuations of the electric charge. The forth order coefficient develops
a peak in the transition region and at high temperatures approaches the ideal quark gas value.
The sixth order expansion coefficient has a peak around the transition temperature and becomes
negative above the transition temperature. It approaches zero from below as the temperature 
increases as expected in the weakly interacting QGP. 
The six order expansion coefficient vanishes in the ideal gas limit as well
as in the next order of perturbation theory \cite{Blaizot:2001vr}.
Based on the discussion discussion of the second order expansion
coefficients we would expect that at low temperatures also the higher order expansion coefficients
should be reasonably well described by HRG. However, this is not the case. As one can see from
Fig. \ref{fig:chi6} the lattice results fall below the HRG prediction. This is due to the
large cutoff effects related to taste symmetry breaking in the $p4$ action. Taste symmetry breaking effects
lead to distortions of the hadron spectrum. If these distortions are taken into account in 
the HRG calculations a good agreement with the lattice data can be achieved 
\cite{Huovinen:2009yb,Huovinen:2010tv,Huovinen:2011xc}. The HRG model predictions with the distorted
spectrum are shown in Fig. \ref{fig:chi6} as the dashed and dotted lines and are in reasonably good
agreement with the lattice data. Preliminary results for the higher order expansion coefficients
have been recently obtained with $HISQ/tree$ action \cite{Mukherjee:2011td}. 
The qualitative features of these expansion coefficients remain the same, however, the location
of the peak is shifted to smaller temperatures. The observed shift in the peak position is
consistent with the shift in the chiral transition temperature observed in the calculations 
with $HISQ/tree$ action \cite{Bazavov:2011nk}.
The higher order expansion coefficients are sensitive to
the singular part of the free energy density in the vicinity of the chiral transition temperature.
Therefore I will come back to the discussion of these quantities in section \ref{sec:chiral}, which is
dedicated to the chiral aspects of the QCD transition.

In the high temperature region the fourth order expansion coefficient was also calculated with $p4$
action on $N_{\tau}=4$, $6$ and $8$ lattices \cite{Petreczky:2009cr}.
While no continuum extrapolation was performed the 
numerical data suggest that at high temperatures the fourth order expansion coefficient will approach
the ideal quark gas value from below \cite{Petreczky:2009cr}. In the same study charm quark number
fluctuations have been considered using partially quenched approximation, i.e. no charm quark loops.
It was found that the ideal quark gas is a good approximation also in this case  \cite{Petreczky:2009cr}.

\begin{figure}
\includegraphics[width=9cm]{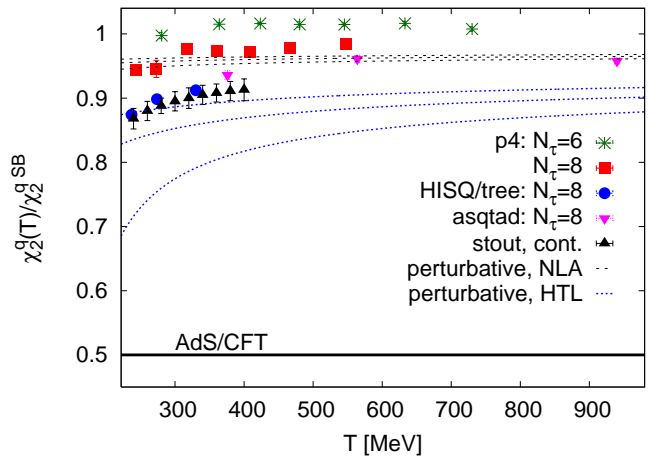}
\caption{Quark number fluctuations calculated on the lattice with different improved
staggered action and compared with the result of HTL resummed pertubation theory (dotted lines) 
\cite{Blaizot:2001vr} as well as the resummation that uses NLA approximation \cite{Rebhan:2003fj}.}
\label{fig:chiq_high}
\end{figure}

Finally let us compared the lattice results for quark number susceptibilities 
at high temperatures with the prediction
of resummed perturation theory \cite{Blaizot:2001vr,Rebhan:2003fj}. In Fig. \ref{fig:chiq_high}
I show the lattice results for light quark number susceptibilities 
obtained with $p4$, $stout$ and $HISQ/tree$ actions compared with the resummed perturbative results.
Since the difference between light and strange
quark number susceptibilities is negligible for $T>400$MeV in the figure I also show the strange
quark number susceptibilities obtained with asqtad action \cite{Bernard:2004je}.
The continuum extrapolated $stout$ results agree well with HTL resummed perturbative result \cite{Blaizot:2001vr}.
Whereas, the $N_{\tau}=8$ asqtad results, that are expected to be close to the continuum limit, agree with
the results obtained using the next-to-leading log (NLA) approximation \cite{Rebhan:2003fj}.
Clearly more lattice calculations in the high temperature limit are requited to settle this issue.
It is also interesting to compare the results 
of the lattice calculations with the results obtained in strongly coupled gauge theories using AdS/CFT 
correspondence. The result from AdS/CFT calculations shown in the figure as the solid black line is significantly
below the lattice results \footnote{The conserved charges considered in these calculations are not exactly the
quark numbers, see discussion in Ref. \cite{Petreczky:2009at}.}.

%% file: chiral.tex
\section{Chiral transition}
\label{sec:chiral}
The Lagrangian of QCD  has an approximate $SU_A(3)$ chiral symmetry. This symmetry is 
spontaneously broken in the vacuum. 
The chiral symmetry breaking is signaled by  non-zero expectation value of the quark or chiral condensate,
$\langle \bar \psi \psi \rangle \neq 0$ in the massless limit.
This symmetry is expected to be restored at high temperatures and the quark condensate vanishes.
There is an explicit breaking of the chiral symmetry by the small value of $u,d$ and $s$ quark masses. 
While due to the relatively large strange quark mass ($m_s \simeq 100$ MeV) $SU_A(3)$ may not
be a very good symmetry its subgroup $SU_A(2)$ remains a very good symmetry and is relevant
for the discussion of the finite temperature transition in QCD. If the relevant symmetry is 
$SU_A(2)$ the chiral transition is expected to be second order for massless light ($u$ and $d$) quarks
belonging to the $O(4)$ universality class \cite{Pisarski:1983ms}. 
Recent calculations with $p4$ action confirm this pictures \cite{Ejiri:2009ac}.
In other words $m_s^{phys}>m_s^{TCP}$, contrary to calculations in effective linear sigma model.
This also means that for non-zero light quark masses the transition must be a crossover. The later
fact seems to be supported by calculations in Ref. \cite{Aoki:2006we}.
The $U_A(1)$ symmetry is explicitly
broken in the vacuum by the anomaly but it is expected to be effectively restored at high temperatures
as the non-perturbative vacuum fluctuations responsible for its breaking are suppressed at high temperatures.
If the  $U_A(1)$ symmetry is restored at the same temperature as the  $SU_A(2)$ symmetry the transition
could be first order \cite{Pisarski:1983ms}. 
Recent calculations with staggered as well as with domain wall fermions
suggest that $U_A(1)$  symmetry gets restored at temperature that is significantly higher than the chiral
transition temperature \cite{Hegde:2011zg,Cheng:2010fe}.

For massless quarks the chiral condensate vanishes at the critical temperature $T_c^0$ and it is the order
parameter. Therefore in the lattice studies one calculates the chiral condensate and its derivative
with respect to the quark mass, called the chiral susceptibility. For the staggered fermion formulation
most commonly used in the lattice calculations these quantities can be written as follows:
\begin{eqnarray}
\langle \bar \psi \psi \rangle_{q,x}&=&\frac{1}{4} \frac{1}{N_{\sigma}^3 N_{\tau}} 
{\rm Tr} \langle D_q^{-1} \rangle ,\\
\chi_{m,q}(T)&=& 
n_f \frac{\partial \langle \bar\psi \psi \rangle_{q,\tau}}{\partial m_l}
=\chi_{q, disc} + \chi_{q, con} ~~q=l,s, \label{susc}
\end{eqnarray}
where the subscript $x=\tau$ and $x=0$ denote the expectation value
at finite and zero temperature, respectively.  
Furthermore, $D_q=m_q \cdot 1 + D$ is the fermion matrix in the canonical normalization and
$n_f=2$ and $1$ for light and strange quarks, respectively.
In Eq. (\ref{susc}) we made explicit that chiral susceptibility is the sum of connected 
and disconnected Feynman diagrams. The disconnected and connected contributions can be written
as
\begin{eqnarray}
\chi_{q, disc} &=&
{{n_f^2} \over 16 N_{\sigma}^3 N_{\tau}} \left\{
\langle\bigl( {\rm Tr} D_q^{-1}\bigr)^2  \rangle -
\langle {\rm Tr} D_q^{-1}\rangle^2 \right\}
\label{chi_dis} \; , \\
\chi_{q, con} &=&  -
{{n_f} \over 4} {\rm Tr} \sum_x \langle \,D_q^{-1}(x,0) D_q^{-1}(0,x) \,\rangle \; ,~~~q=l,s.
\label{chi_con}
\end{eqnarray}
The disconnected part of the light quark susceptibility describes the
fluctuations in the light quark condensate and is analogous to
the fluctuations of the order parameter of an $O(N)$ spin model. The
second term ($\chi_{q,con}$) arises from the explicit quark mass
dependence of the chiral condensate and is the expectation value of
the volume integral of the correlation function of the taste non-singlet
scalar operator $\bar{\psi}\psi$. Let me note that in the massless limit only
$\chi_{l,disc}$ diverges. In the next subsections I will discuss the temperature dependence of the
chiral condensate and the chiral susceptibility as well as the role of universal scaling in the transition
region. 

\subsection{The temperature dependence of the chiral condensate}
\begin{figure}
\includegraphics[width=7cm]{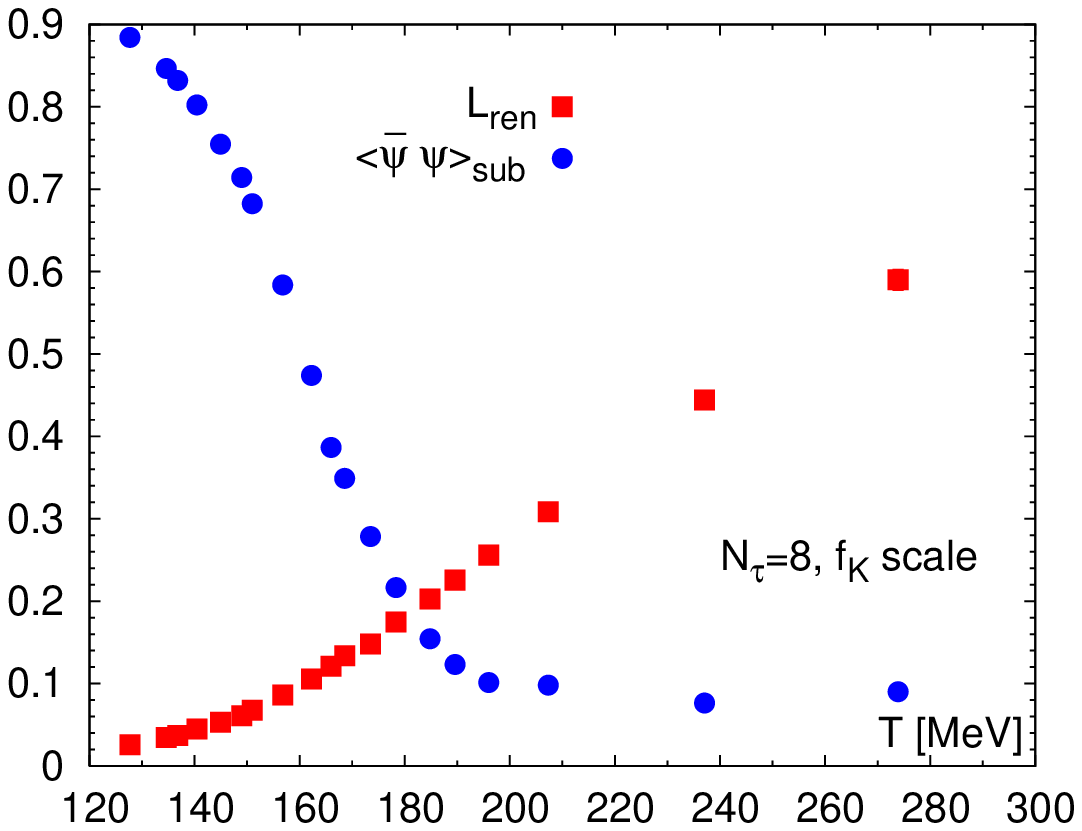}
\includegraphics[width=7cm]{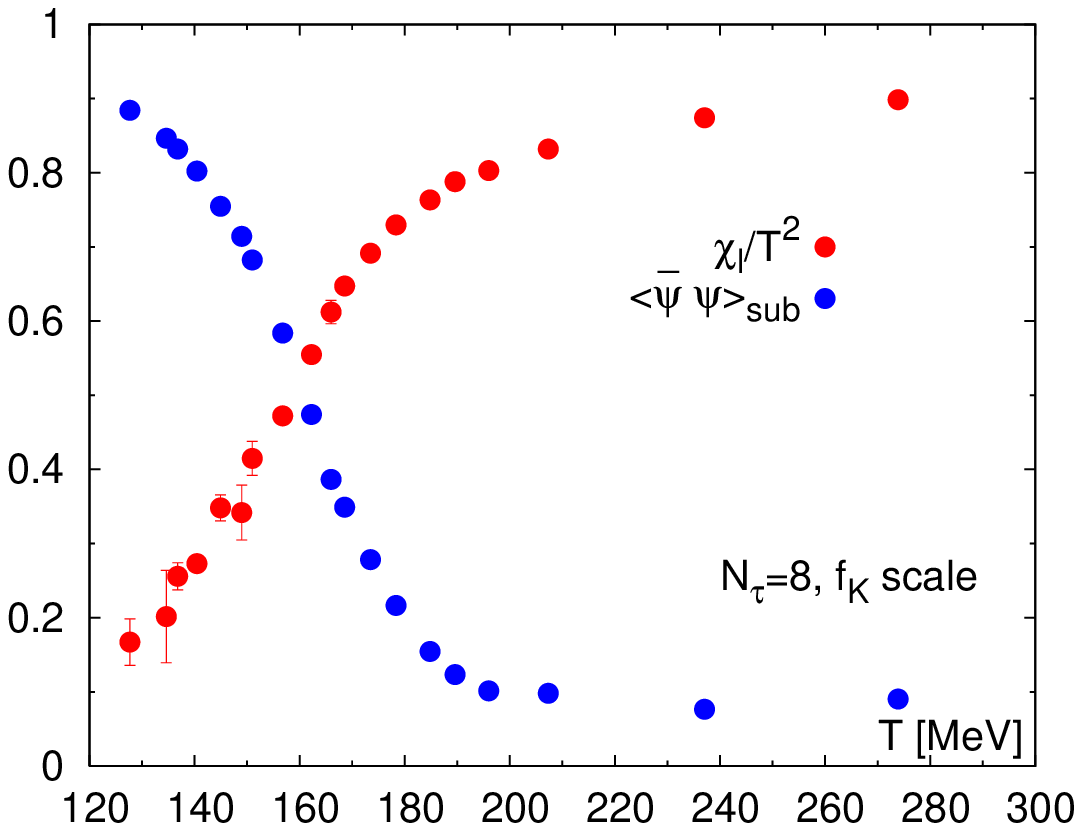}
\caption{The subtracted chiral condensate calculated with $HISQ/tree$ action on $N_{\tau}=8$ lattice
and compared to the renormalized Polyakov loop (left) and light quark number fluctuation (right).}
\label{fig:Delta}
\end{figure}

The chiral condensate needs a multiplicative, and also an additive renormalization if
the quark mass is non-zero. Therefore the subtracted chiral condensate is 
considered
\begin{equation}
\langle \bar \psi \psi \rangle_{sub}=\frac{\langle \bar\psi \psi \rangle_{l,\tau}-\frac{m_l}{m_s} \langle \bar \psi \psi \rangle_{s,\tau}}
{\langle \bar \psi \psi \rangle_{l,0}-\frac{m_l}{m_s} \langle \bar \psi \psi \rangle_{s,0}}.
\end{equation}
In Fig. \ref{fig:Delta} I show the results for $\langle \bar \psi \psi \rangle_{sub}$ 
calculated with $HISQ/tree$ action and compared to
the renormalized Polyakov loop and light quark number fluctuation discussed in relation to the deconfining
transition. Interestingly, the rapid decrease in the subtracted chiral condensate happens at temperatures
that are smaller than the temperatures where the Polyakov loop rises rapidly as first noticed in
Ref. \cite{Aoki:2006br}. On the other hand the rapid
change in $\chi_l$ and $\langle \bar \psi \psi \rangle_{sub}$ happen roughly in the same temperature region.
The above results for $\langle \bar \psi \psi \rangle_{sub}$ obtained with $HISQ/tree$ action agree well with
the continuum extrapolated $stout$ results.

\begin{figure}
\includegraphics[width=0.450\textwidth]{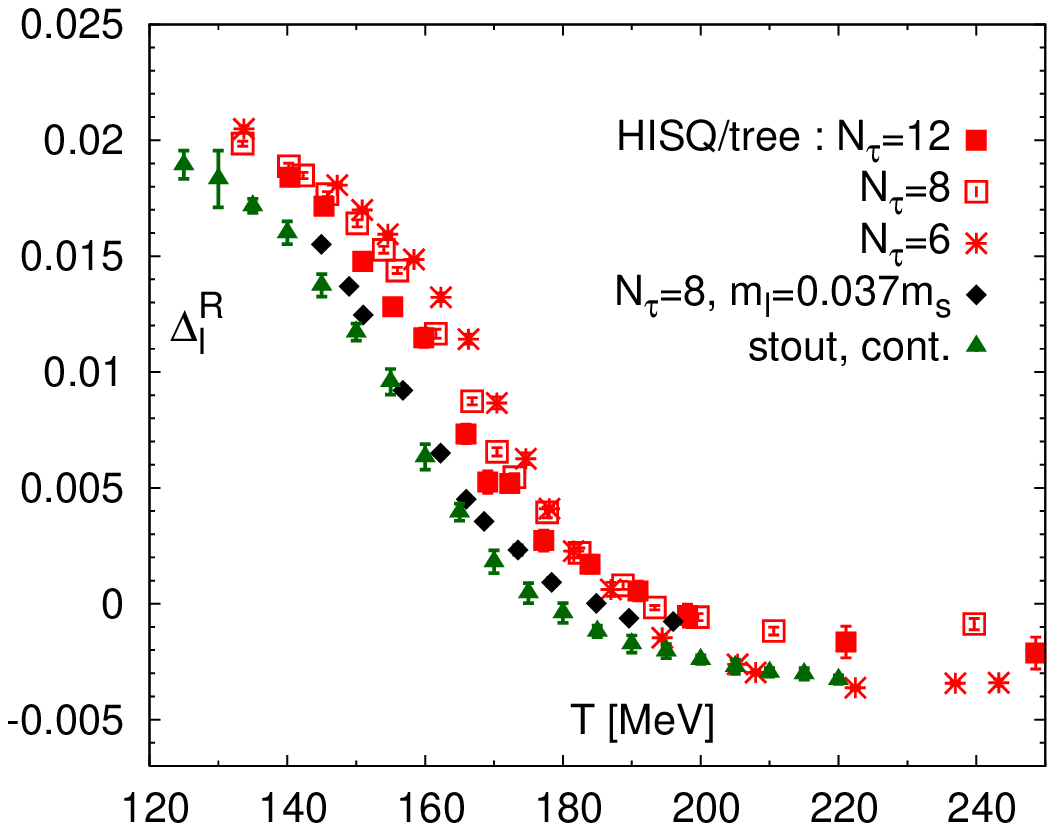}
\includegraphics[width=0.450\textwidth]{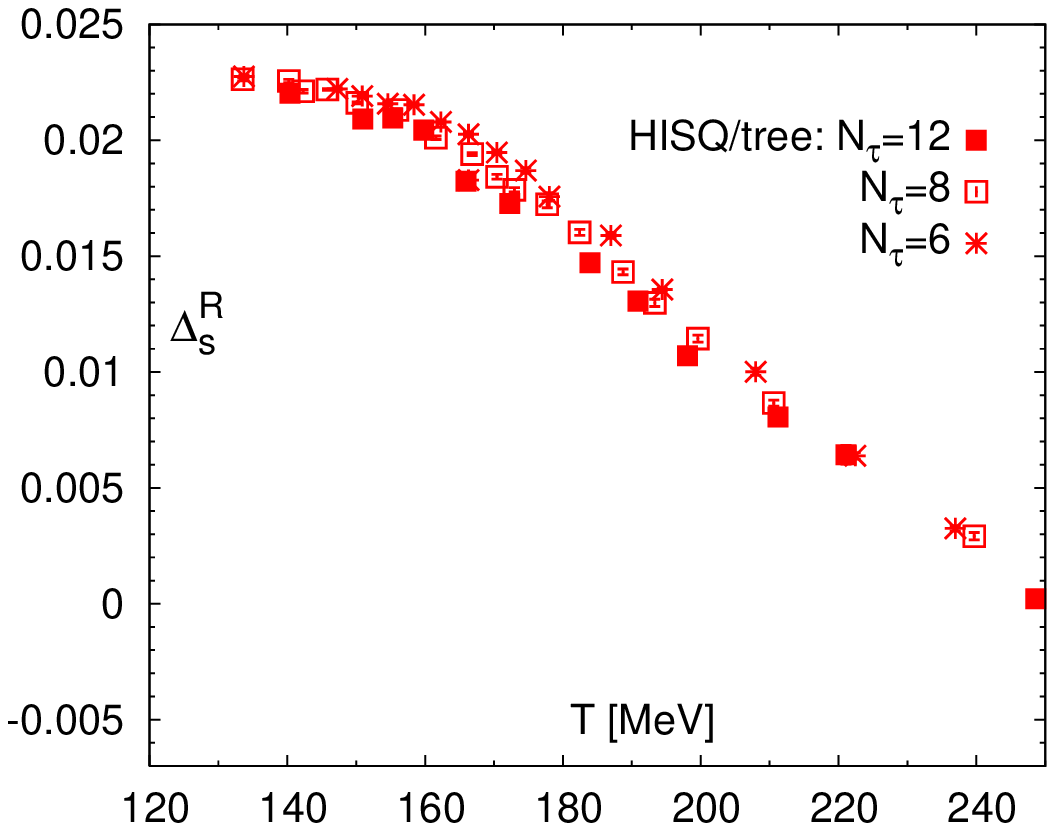}
\caption{The renormalized chiral condensate $\Delta_l^R$ for the
  $HISQ/tree$ action and $m_l/m_s = 0.05$ compared to the $stout$ data.
  In the right panel, we show the renormalized strange quark
  condensate $\Delta_s^R$ for the $HISQ/tree$ action.
}
\label{fig:pbpR}
\end{figure}

Another way to get rid of the multiplicative and additive renormalization is to subtract the zero
temperature condensate and multiply the difference by the strange quark mass, i.e. consider the following
quantity
\begin{equation}
\Delta_q^R=d+n_f m_s r_1^4 ( \langle \bar \psi \psi \rangle_{q,\tau}-
\langle \bar \psi \psi \rangle_{q,0} ), ~~~~ q=l,s .
\end{equation}
As before, $n_f=2$ for the light quarks and $n_f=1$ for the strange quark, while $d$ is a normalization constant.
The factor $r_1^4$ was introduced to make the combination dimensionless. 
It is convenient to choose the normalization constant to be the light quark condensate for $m_l=0$
multiplied by $m_s r_1^4$.  In Fig. \ref{fig:pbpR} the renormalized quark condensate is shown as function of
the temperature for $HISQ/tree$ and $stout$ actions. We see a crossover behavior for temperature of $(150-160)$MeV,
where $\Delta_l^R$ drops by $50\%$. The difference between the $stout$ and $HISQ/tree$ results is a quark mass
effect. Calculations for $HISQ/tree$ action were performed for $m_l=m_s/20$, while the $stout$ calculations
were done for the physical light quark masses, $m_l=m_s/27.3$. 
For a direct comparison with $stout$ results, we need to extrapolate the 
$HISQ/tree$ data in the light quark mass and also take care of the residual cutoff 
dependence in the $HISQ/tree$ data. 
This was done in Ref. \cite{Bazavov:2011nk} 
and the results are shown in the figure as black diamonds demonstrating
a good agreement between $HISQ/tree$ and $stout$ results.
Contrary to $\Delta_l^R$ the renormalized strange quark
condensate $\Delta_s^R$ shows only a gradual decrease over a wide temperature interval dropping by $50\%$
only at significantly higher temperatures of about $190$ MeV.

\subsection{The chiral susceptibility}
For a true chiral phase transition the chiral susceptibility diverges at the transition temperature.
For physical value of the quark masses we expect to see a peak in the chiral susceptibility at a certain
temperature that defines the crossover temperature. The chiral susceptibility also needs a multiplicative
and additive renormalization. Therefore the following quantity is considered
\begin{equation}
\frac{\chi_R(T)}{T^4}=\frac{m_s^2}{T^4} \left( \chi_{m,l}(T)-\chi_{m,l}(T=0) \right).
\label{chiR}
\end{equation} 
\begin{figure}
\includegraphics[width=8cm]{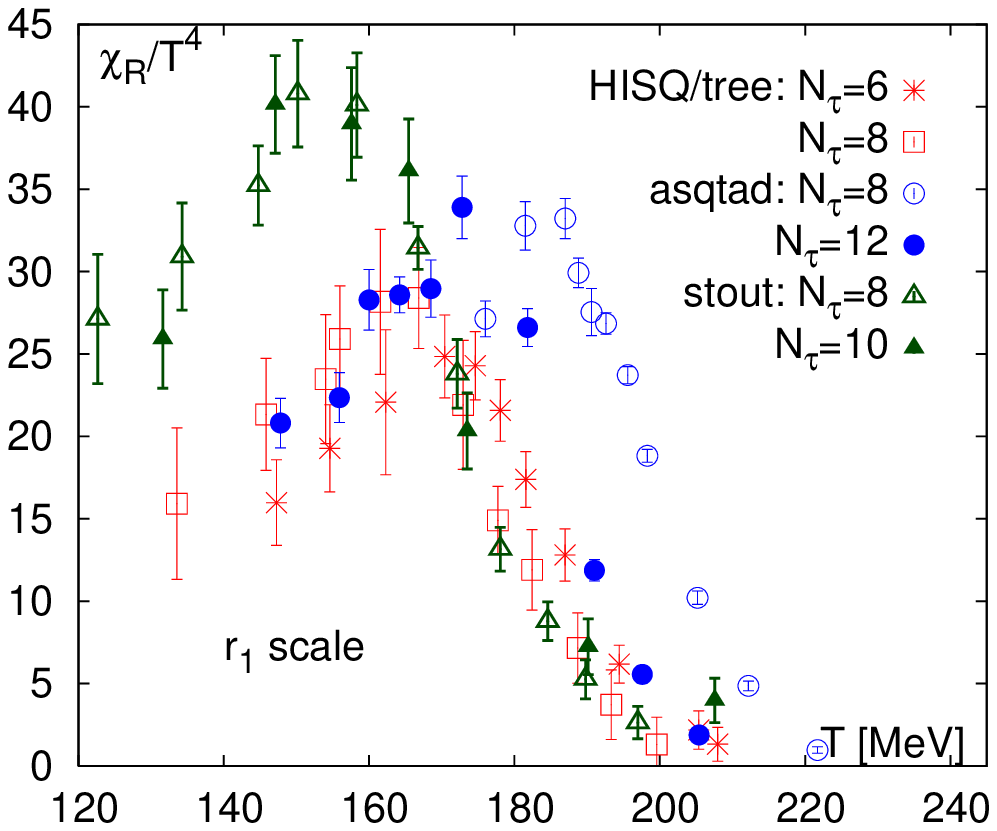}
\includegraphics[width=8cm]{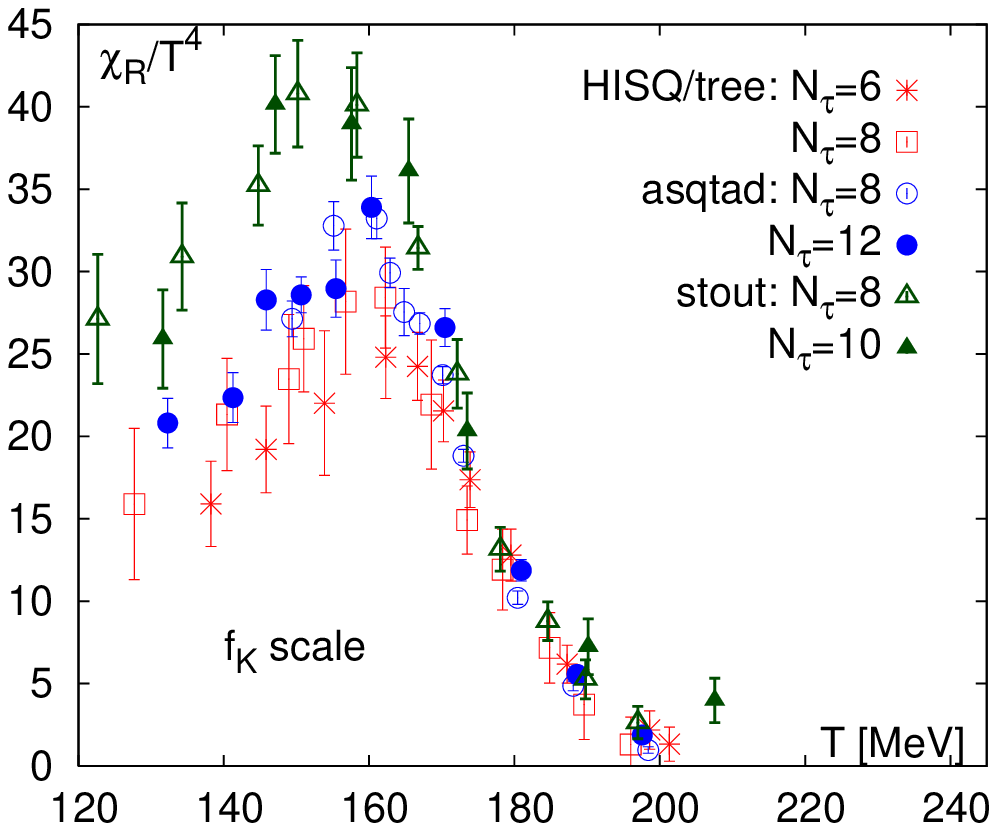}
\caption{The renormalized chiral susceptibility $\chi_R$
  for the asqtad and HISQ/tree actions obtained at $m_l=0.05m_s$ and
  compared with the stout action results \cite{Aoki:2006br}.  The
  temperature scale for $HISQ/tree$ and $asqtad$ actions is set using $r_1$ ($f_K$) in the left (right)
  panels.  }
\label{fig:chiR}
\end{figure}
The numerical results for this quantity are shown in Fig. \ref{fig:chiR} for $HISQ/tree$, $asqtad$ 
and $stout$ actions \footnote{In Ref. \cite{Aoki:2006br} the light quark mass
was used instead of $m_s$ in Eq. (\ref{chiR}). In the comparison this 
was taken into account \cite{Bazavov:2011nk}.}. There is a fairly good agreement between the results
obtained with different actions if $f_K$ scale is used. At low temperatures the $stout$ results
are above the $asqtad$ and $HISQ/tree$ results due to quark mass effects \cite{Bazavov:2011nk}.

\subsection{O(N) scaling and the transition temperature}

As discussed in the beginning of this section for physical $m_s$ and vanishing light quark masses
the transition is expected to be second order belonging to the $O(4)$ universality class 
in the continuum limit. At non-zero lattices spacing, however, we should expect $O(2)$ universality
class as only a part of the chiral symmetry is preserved in the staggered fermion formulation. 
Therefore in the following we will use the term $O(N)$ universality class as referring to either 
$O(4)$ or $O(2)$ universality class. 
In the vicinity of the chiral phase transition, the free energy
density may be expressed as a sum of a singular and a regular
part,
\begin{equation}
f = -\frac{T}{V} \ln Z\equiv f_{sing}(t,h)+ f_{reg}(T,m_l,m_s) \; .
\label{free_energy}
\end{equation}
Here $t$ and $h$ are dimensionless couplings that control deviations from
criticality. They are related to the temperature $T$ and the light quark mass $m_l$ as  
\begin{equation}
t = \frac{1}{t_0}\frac{T-T_c^0}{T_c^0} \quad , \quad 
h= \frac{1}{h_0} H \quad , \quad 
H= \frac{m_l}{m_s} \; ,
\label{reduced}
\end{equation}
where $T_c^0$ denotes the chiral phase transition temperature, 
{\it i.e.}, the transition temperature at $H=0$.  The scaling variables
$t$, $h$ are normalized by two parameters $t_0$ and $h_0$, which are
unique to QCD and similar to the low energy constants in the chiral
Lagrangian.  These need to be determined together with $T_c^0$. In the
continuum limit, all three parameters are uniquely defined, but depend
on the value of the strange quark mass. On the lattice they will depend on
$N_{\tau}$ as well. 

The singular contribution to the free energy density is a homogeneous
function of the two variables $t$ and $h$. Its invariance under scale
transformations can be used to express it in terms of a single
scaling variable
\begin{equation}
z=t/h^{1/\beta\delta} = \frac{1}{t_0}\frac{T-T_c^0}{T_c^0} \left( \frac{h_0}{H} \right)^{1/\beta\delta}
 = \frac{1}{z_0}\frac{T-T_c^0}{T_c^0} \left( \frac{1}{H} \right)^{1/\beta\delta}
\label{eq:defz}
\end{equation}
where $\beta$ and $\delta$ are the critical exponents of the $O(N)$
universality class and $z_0 = t_0/h_0^{1/\beta\delta}$.
Thus, the dimensionless free energy density
$\tilde{f}\equiv f/T^4$ can be written as
\begin{equation}
\tilde{f}(T,m_l,m_s) = h^{1+1/\delta} f_f(z) + f_{reg}(T,H,m_s) \; ,
\label{scaling}
\end{equation}
where $f_f$ is the universal scaling function and the regular term $f_{reg}$ 
gives rise to scaling violations. This regular term 
can be expanded in a Taylor series around $(t,h)=(0,0)$.

It should be noted that the reduced temperature $t$ may depend on other
parameters which do not explicitly break 
chiral symmetry. In particular, it depends on light and strange 
quark chemical potentials $\mu_q$, which in leading order enter only quadratically,
\begin{equation}
t = \frac{1}{t_0} \left( \frac{T-T_c^0}{T_c^0} + 
\sum_{q=l,s}\kappa_q\left(\frac{\mu_q}{T}\right)^2 +
\kappa_{ls} \frac{\mu_l}{T}\frac{\mu_s}{T} \right)
  \; .
\label{reduced2}
\end{equation}

The transition temperature can be defined as peaks in the susceptibilities (response functions) that are 
second derivatives of the free energy density with respect to relevant parameters. Since there are
two relevant parameters we can define three susceptibilities:
\begin{equation}
\chi_{m,l}=\frac{\partial^2 \tilde f}{\partial m_l^2},~~
\chi_{t,l}=\frac{\partial^2 \tilde f}{\partial t \partial m_l },~~
\chi_{t,t}=\frac{\partial^2 \tilde f}{\partial t^2}.
\end{equation}
Thus, three different  pseudo-critical temperatures $T_{m,l}$, $T_{t,l}$ and $T_{t,t}$ can be defined. 
In the vicinity of
the critical point the behavior of these susceptibilities is controlled by three universal
scaling functions that can be derived from $f_f$. In the chiral limit $T_{m,l}=T_{t,l}=T_{t,t}=T_c^0$. 
There is, however, an additional complication for $O(N)$ universality class:  while $\chi_{m,l}$ and $\chi_{t,l}$
diverge at the critical point for $m_l \rightarrow 0$
\begin{equation}
\chi_{m,l} \sim m_l^{1/\delta - 1},~~~
\chi_{t,l} \sim m_l^{(\beta -1)/\beta\delta}, 
\label{peaks}
\end{equation}
$\chi_{t,t}$ is finite because $\alpha<0$ for $O(N)$ models ( $\chi_{t,t} \sim |t|^{-\alpha}$  ).
Therefore, one has to consider the third derivative of $\tilde f$ with respect to $t$ :
\begin{equation}
\chi_{t,t,t}=\frac{\partial^3 \tilde f}{\partial t^3}.
\end{equation}

In the vicinity of the critical point the derivatives with respect to $t$ can be estimated 
by taking the derivatives with respect to $\mu_l^2$, i.e.
the response functions $\chi_{t,l}$ and $\chi_{t,t,t}$ are identical to the second Taylor expansion coefficient
of the quark condensate and the sixth order expansion coefficient to the pressure, respectively. The former
controls the curvature of the transition temperature as function of the quark chemical potential $\mu_q$ and
was studied for $p4$ action using $N_{\tau}=4$ and $8$ lattices \cite{Kaczmarek:2011zz}. The later corresponds to the sixth
order quark number fluctuation which is related to the deconfinement aspects of the transition.
The fact that this quantity is sensitive to the chiral dynamics points to a relation between  deconfining and chiral
aspects of the transition. 
In the following I discuss the determination of the transition temperature defined as
peak position of $\chi_{m,l}$, i.e. $T_c=T_{m,l}$. 
\begin{figure}[b]
\begin{center}
\includegraphics[width=7cm]{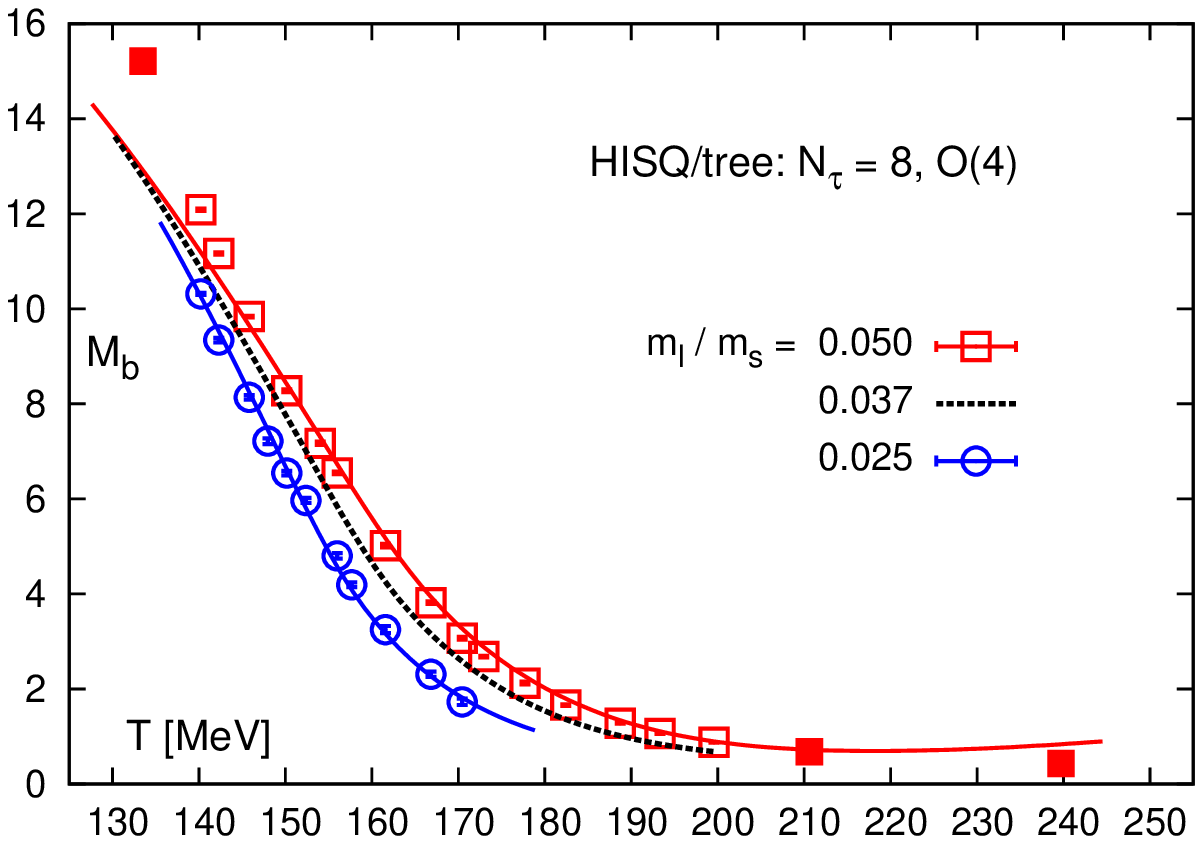}
\includegraphics[width=7cm]{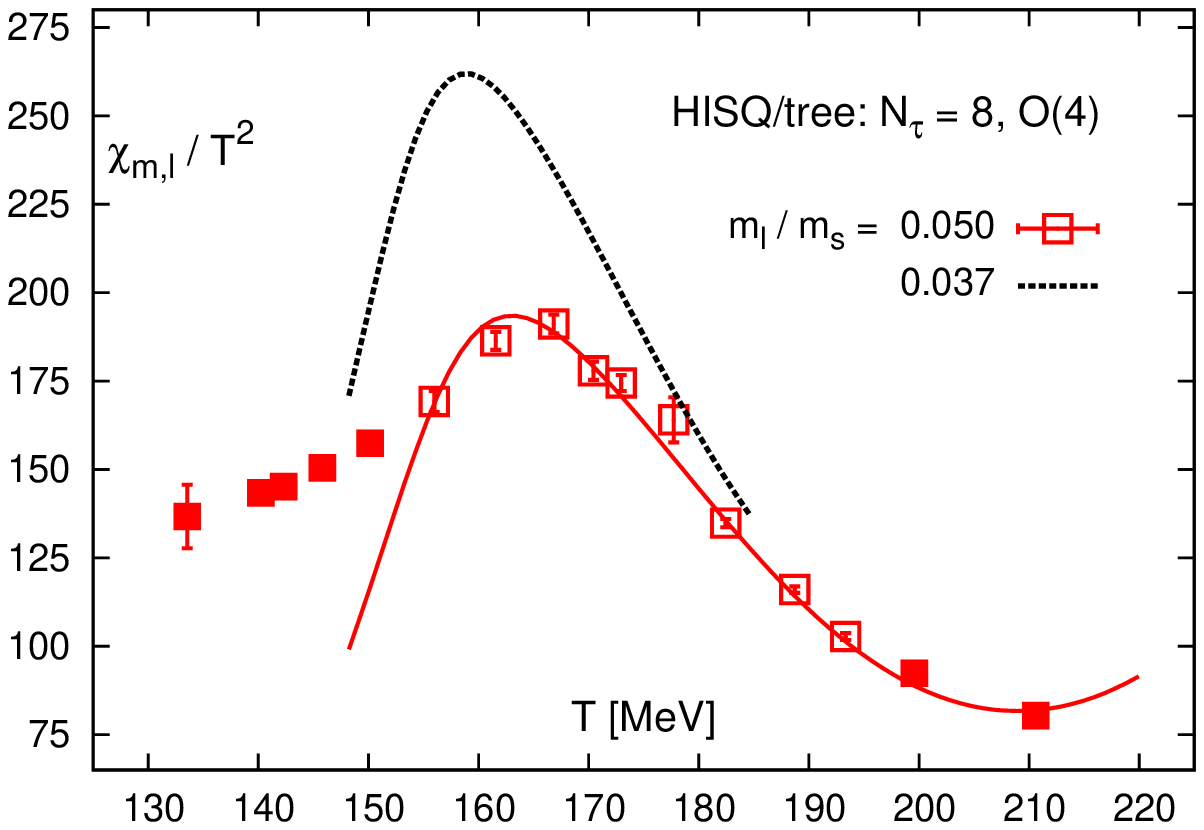}
\end{center}
\caption{
\label{fig:sfit_hisq8_O4} 
  Scaling fits and data for the chiral condensate $M_b$ calculated
  with the $HISQ/tree$ action on lattices with temporal extent
  $N_\tau=8$ (left) and the chiral susceptibility
  $\chi_{m,l}$(right). The data for $M_b$ at $m_l/m_s=0.025$ and for
  $M_b$ and $\chi_{m,l}$ at $m_l/m_s=0.05$ are fit simultaneously
  using the $O(4)$ scaling Ansatz. 
  The points used in the scaling fits are plotted using 
  open symbols.  The dotted lines give the data scaled to the physical
  quark masses.}
\end{figure}
\subsection{Determination of the transition temperature}
The $O(N)$ scaling described in the above subsection can be used to determine the pseudo-critical temperature of
the chiral transition. For the study of the $O(N)$ scaling it is convenient to consider the dimensionless 
order parameter
\begin{equation}
M_b=m_s \frac{\langle \bar \psi \psi\rangle_l}{T^4}.
\end{equation}
The subscript "b" refers to the fact that this is a bare quantity since the additive UV divergence is not removed.
From the point of view of the scaling analysis this divergent term is just a regular contribution.
For sufficiently small quark mass and in the vicinity of the transition region we can write 
\begin{equation}
M_b(T,H) = h^{1/\delta} f_G(t/h^{1/\beta\delta}) + f_{M,reg}(T,H). 
\label{order_scaling}
\end{equation}
Here $f_G(z)$ is the scaling function related to $f_f$ and was calculated for $O(2)$ and $O(4)$ spin models
\cite{Engels:2000xw,Toussaint:1996qr,Engels:1999wf,Engels:2001bq,Engels:2011km}.
The regular contribution can be parametrized as \cite{Bazavov:2011nk}
\begin{eqnarray}
f_{M,reg}(T,H) &=& a_t(T)  H   \nonumber \\
&=& \left( a_0 + a_1 \frac{T-T_c^0}{T_c^0} + a_2 \left(\frac{T-T_c^0}{T_c^0} \right)^2 \right) H.
\label{eq:freg}
\end{eqnarray}
Then we have the following behavior for the light chiral susceptibility
\begin{eqnarray}
\frac{\chi_{m,l}}{T^2} &=& \frac{T^2}{m_s^2}
\left( \frac{1}{h_0}
h^{1/\delta -1} f_\chi(z) + \frac{\partial f_{M,reg}(T,H)}{\partial H}
\right) \; , \nonumber \\
&&{\rm with}\;\; f_{\chi}(z)=\frac{1}{\delta} [f_G(z)-\frac{z}{\beta} f_G'(z)].
\label{eq:chiralsuscept}
\end{eqnarray}
One performs a simultaneous fit to the lattice data for $M_b$ and $\chi_{m,l}$ treating
$T_c^0, t_0, h_0, a_0, a_1$ and $a_2$ as fit parameters \cite{Bazavov:2011nk}. This gives a good description
of the quark mass and temperature dependence of $\chi_{m,l}$ and allows to determine
accurately the peak position in $\chi_{m,l}$.
As an example in Fig. I show the $O(4)$ scaling fits for $N_{\tau}=8$ lattice data obtained with
$HISQ/tree$ action. 
The scaling fit works quite well. Similar results have been obtained for $N_{\tau}=6$ and $12$ as
well as for $asqtad$ action on $N_{\tau}=8$ and $12$ lattices \cite{Bazavov:2011nk}. Furthermore, scaling fits have been
performed assuming $O(2)$ universality class. The quality of these fits were similar to the $O(4)$ 
ones and the resulting
transition temperatures turned out to be the same within statistical errors \cite{Bazavov:2011nk}. 
Having determined $T_c$ for 
$HISQ/tree$  and $asqtad$ action for each $N_{\tau}$ a combined continuum extrapolation was performed using different
assumption about the $N_{\tau}$ dependence of $T_c$. This analysis resulted in \cite{Bazavov:2011nk}:
\begin{equation}
T_c=(154 \pm 9) {\rm MeV}.
\end{equation}
The analysis also demonstrated that $HISQ/tree$ and $asqtad$ action give consistent results in the continuum limit.
The Budapest-Wuppertal collaboration found $T_c=147(2)(3)$MeV, $157(3)(3)$MeV and 
$155(3)(3)$MeV defined as peak position in $\chi_R$, inflection points in 
$\langle \bar \psi \psi \rangle_{sub}$  
and $\Delta_l^R$ respectively \cite{Borsanyi:2010bp}. These agree with the above value within errors.

\subsection{The chiral transition temperature for small chemical potential}
The scaling analysis described in the above subsection is also useful for
the determination of the chiral transition temperature for small quark chemical
potential. Also for non-zero quark chemical potential the phase transition happens
at $t=0$. Together with Eq. (\ref{reduced2}) this implies that the $\mu_q$ 
dependence of the chiral phase transition temperature is given by
\begin{equation}
T_c^0(\mu_q)=T_c^0(0)-\kappa_l \left(\frac{\mu_l}{T_c^0}\right)^2+{\cal O}\left(\left(\frac{\mu_l}{T_c^0}\right)^4\right).
\end{equation}
The value of $\kappa_l$ can be determined through lattice calculations of the mixed susceptibility $\chi_{t,l}$ 
\cite{Kaczmarek:2011zz}. The mixed susceptibility can be expressed in terms of the derivative of the 
scaling function of the order parameter $f_G(z)$
\begin{equation}
\frac{\chi_{t,l}}{T}=\frac{2 \kappa_l T}{t_0 m_s}h^{-(1-\beta)/(\beta \delta)} f_G'(z).
\end{equation}
Calculations with $p4$ action on $N_{\tau}=4$ and $8$ lattices of the mixed
susceptibility have been performed for  several quark masses and resulted in
the value  \cite{Kaczmarek:2011zz}
\begin{equation}
\kappa_l=0.059(2)(4).
\end{equation}
This result for the curvature of the critical line is about factor of two larger than the result
obtained in Ref. \cite{Fodor:2001pe,Fodor:2004nz} using multi-parameter re-weighting technique. 
On the other hand it is consistent
with the results obtained using the imaginary chemical potential 
technique \cite{deForcrand:2002ci,deForcrand:2006pv,Falcone:2010az}.

Finally let me mention that the study of the mixed susceptibility allows to estimate
the width of the chiral crossover for physical light quark masses. The pseudo-critical
temperatures $T_{t,l}$ and $T_{m,l}$ are determined by the peak positions $z_p$ of the
$f_G'(z)$ and $f_{\chi}(z)$ when scaling violations are neglected. 
Namely,
\begin{equation}
\frac{T_{p,l}-T_c^0}{T_c^0}=\frac{z_p}{z_0} H^{1/(\beta \delta)}, ~~~~p=m,t.
\end{equation}
The peak positions of the above scaling functions for $O(4)$ model have been determined to be 
$z_p=0.74(4)$ and $1.374(30)$ respectively \cite{Engels:2011km}. From these and
from the values of $z_0$ and $T_c^0$ published in Ref. \cite{Kaczmarek:2011zz} for $p4$ action
one can estimate $T_{m,l}$ to be higher than $T_{t,l}$ by $4MeV$ and $2MeV$
for physical value of the light quark mass
for $N_{\tau}=8$ and $4$ respectively. These estimates are reliable since the
scaling violations were found to be small for $p4$ action \cite{Kaczmarek:2011zz}.
In particular, the combination $-m_s t_0 h^{(1-\beta)/(\beta \delta)} \chi_{t,l}/T^2$
as function of $z$ has a peak at $z\simeq 0.74$ \cite{Kaczmarek:2011zz}. It remains to
be seen how this picture changes when $HISQ/tree$ action is used.

%% file: eos.tex
\section{Equation of State}
\begin{figure}[ht]
\includegraphics[width=7.3cm]{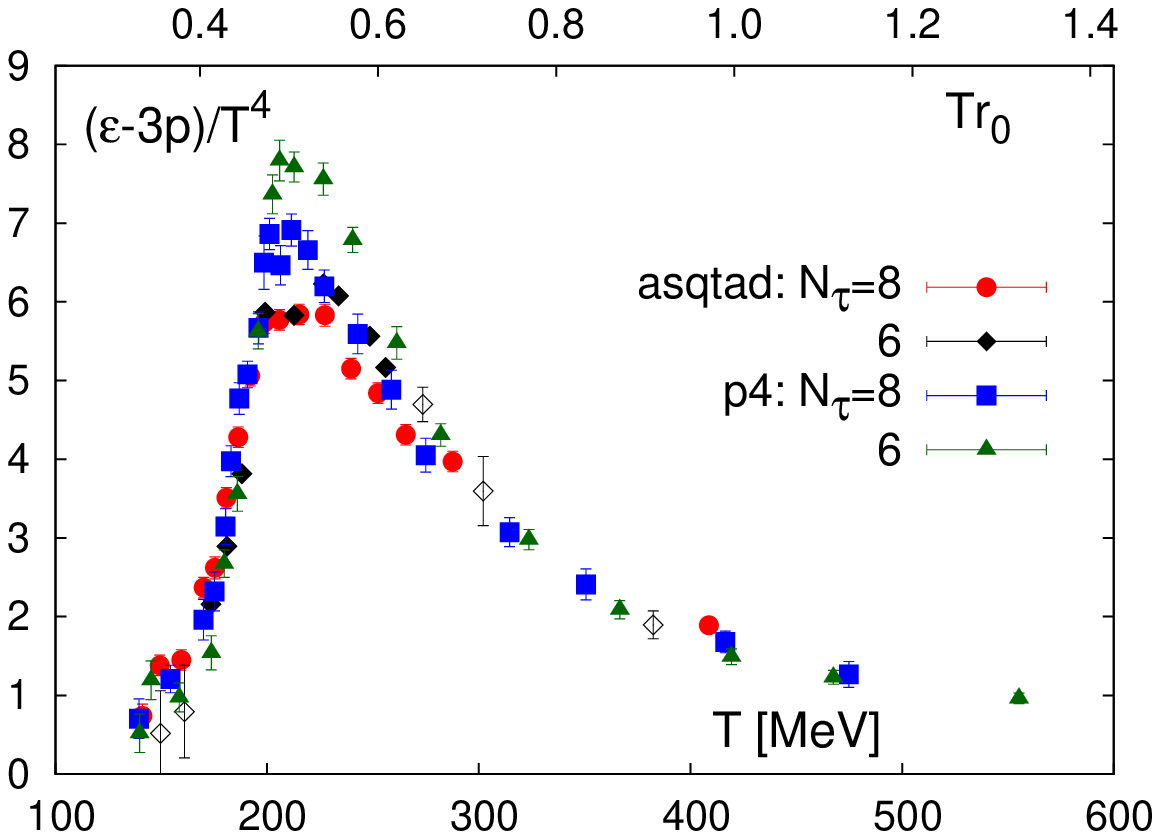} 
\includegraphics[width=7.3cm]{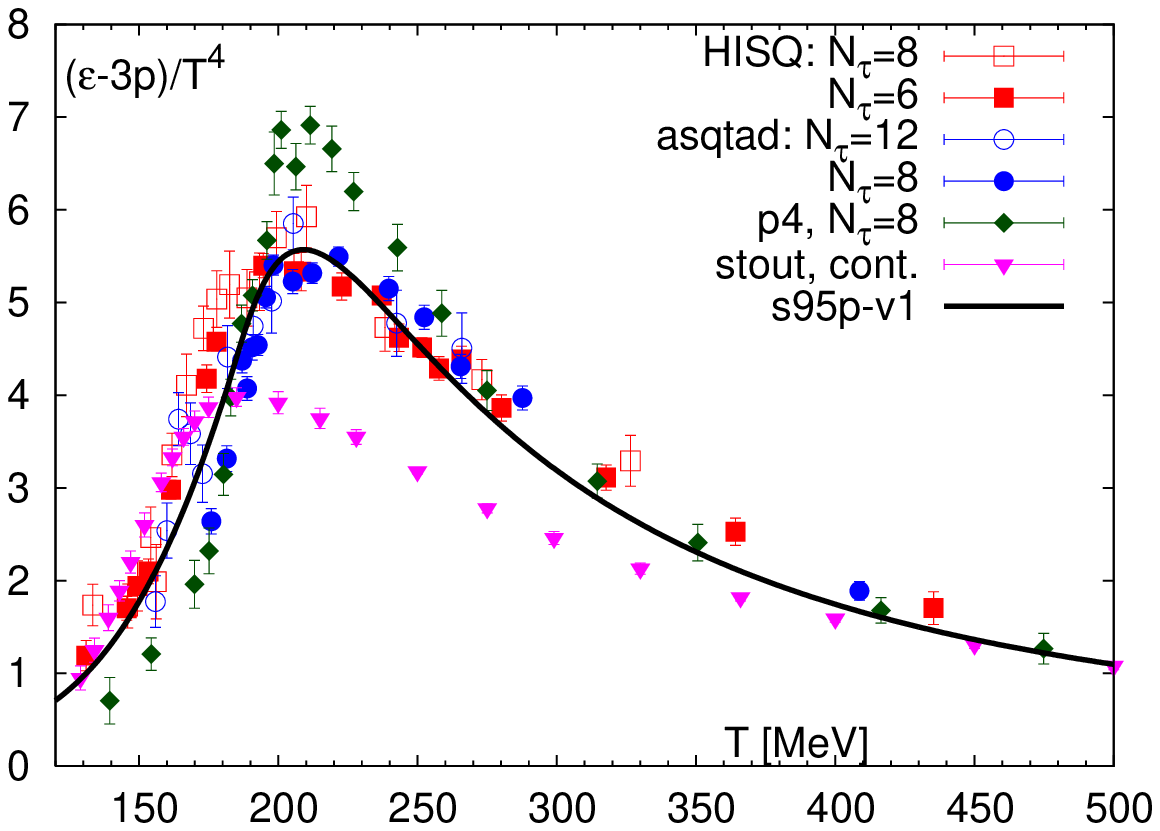}
\vspace*{-0.3cm}
\caption[]{
The interaction measure calculated with $m_l=m_s/10$ and $p4$ and $asqtad$ actions \cite{Bazavov:2009zn} (left)
and with $m_l=m_s/20$ for $p4$ action \cite{Cheng:2009zi} as well as with $HISQ/tree$ and $asqtad$ 
actions \cite{Bazavov:2010pg}.
Also shown in the figure are the continuum estimates obtained with $stout$ action and the parametrization
based on hadron resonance gas (HRG) model \cite{Huovinen:2009yb}.
}
\label{fig:e-3p}
\end{figure}

The equation of state has been calculated using different improved staggered fermion actions
$p4$, $asqtad$, $stout$ and $HISQ/tree$. 
The calculation
of thermodynamic observables proceeds through the calculation of the trace of the energy momentum tensor,
$\epsilon -3 p$, also
known as trace anomaly or interaction measure. This is due to the fact that this quantity can be expressed in
terms of expectation values of local gluonic and fermionic operators, (see e.g. Ref. \cite{Bazavov:2009zn}).  
Different thermodynamic observables can be obtained from the interaction measure through integration of
the trace anomaly
\footnote{A somewhat different approach was used in Ref. \cite{Borsanyi:2010cj}}. 
The pressure can be written as
\begin{equation}
\displaystyle
\frac{p(T)}{T^4}-\frac{p(T_0)}{T_0^4}=\int_{T_0}^T \frac{dT'}{T'^5} (\epsilon-3 p).
\end{equation}
The lower integration limit $T_0$ is chosen such that the pressure is exponentially small there.
Furthermore, the entropy density can be written as $s=(\epsilon+p)/T$. Since the interaction measure 
is the basic thermodynamic observable in the lattice calculations it is worth to discuss its properties
more in detail. In Fig. \ref{fig:e-3p} (left panel) I show 
the results of calculation with $p4$ and $asqtad$ actions using $N_{\tau}=6$ and $8$ lattices and light
quark masses $m_l=m_s/10$, where $m_s$ is the physical strange quark mass. These calculations correspond
in the continuum limit to the pion mass of $220$MeV and $260$MeV for $p4$ and $asqtad$ respectively.
The interaction measure shows a rapid rise in the transition region and after reaching a peak at temperatures
of about $200$MeV decreases. Cutoff effects (i.e. $N_{\tau}$ dependence) 
appears to be the strongest around the peak region and decrease at high temperatures. 
For temperatures $T<270$MeV calculations with
$p4$ and $asqtad$  actions have been extended to smaller quark masses, $m_l=m_s/20$, that correspond to
the pion mass of about $160$MeV in the continuum limit \cite{Cheng:2009zi,Bazavov:2010pg}.
It turns out that the quark mass dependence is negligible for $m_l < m_s/10$.  
Furthermore, for $astqad$  action calculations have been extended to $N_{\tau}=12$ lattices \cite{Bazavov:2010pg,Soldner:2010xk}.
The trace anomaly was calculated with $HISQ/tree$ action on lattices with temporal extent $N_{\tau}=6$ and
$8$ and $m_l=m_s/20$ \cite{Bazavov:2010pg,Soldner:2010xk} (corresponding to $m_{\pi}=160$MeV in the continuum limit). Finally, calculation
of the trace anomaly and the equation of state was performed with $stout$ action using $N_{\tau}=4,~6,~8,~10$ and
$12$ and physical light quark masses \cite{Borsanyi:2010cj}. Using the lattice data from $N_{\tau}=6,~8,~10$ 
a continuum estimate for different quantities was given \cite{Borsanyi:2010cj}. In  Fig. \ref{fig:e-3p} (right
panel) the results of different lattice calculations of $\epsilon-3p$  corresponding to the pion
masses close to the physical value are summarized. I also compare the lattice results with the parametrization s95p-v1 
of $\epsilon-3p$.
This parametrization combines lattice QCD results of Refs. \cite{Cheng:2007jq,Bazavov:2009zn} 
at high temperatures
with hadron resonance gas model (HRG) at low temperatures ($T<170$MeV) \cite{Huovinen:2009yb}.
At low temperatures there is a fair agreement between the results obtained with $stout$ action and the results
obtained with $HISQ/tree$ action as well as with $asqtad$ action for $N_{\tau}=12$. All these lattice results
are slightly above the HRG curve. 
For $N_{\tau}=8$  cutoff effects are significant for $asqtad$ and $p4$ action. As the result the
corresponding lattice data fall  below the HRG (s95p-v1) curve at low temperatures. 
When cutoff effects in the hadron spectrum are taken into account
in the HRG model a good agreement between the $p4$ data and HRG result can be achieved \cite{Huovinen:2009yb}.
The height of the peak in $\epsilon-3 p$ is the same for $asqtad$ and $HISQ/tree$ actions. At the same time
it is smaller for the $stout$ action. 
Since the dominant cutoff effects of order $a^2 T^2$  are eliminated, the $N_{\tau}$-dependence is 
expected to be small for $p4$, $asqtad$ and $HISQ/tree$ action at high temperatures.
We see that for $T>250$MeV all these lattice actions lead to similar results. At temperatures above
$350$MeV we also see a good agreement with the $stout$ results. 

\begin{figure}[ht]
\includegraphics[width=7.5cm]{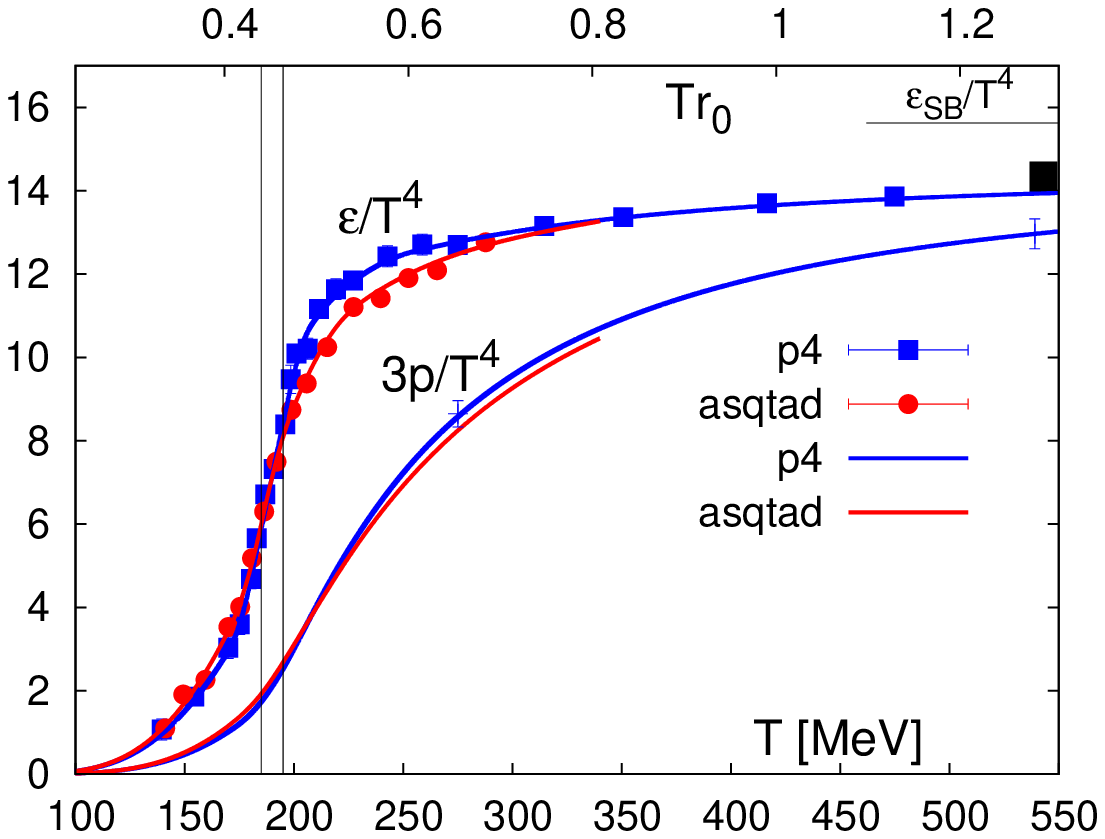}
\includegraphics[width=7.5cm]{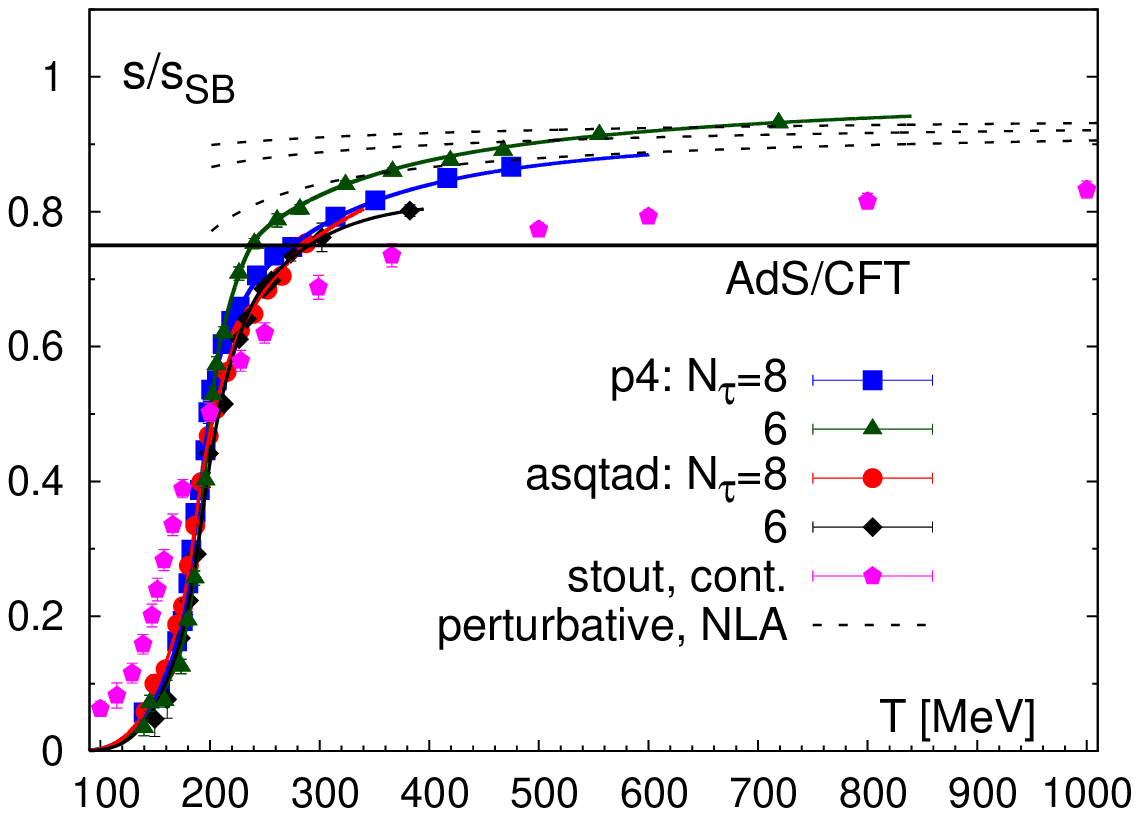}
\vspace*{-0.3cm}
\caption[]{
The energy density and the pressure as function of the temperature (left), and the entropy density divided
by the corresponding ideal gas value (right). The dashed lines in the right panel correspond to the resummed
perturbative calculations while the solid black line is the AdS/CFT result.
}
\label{fig:eos}
\end{figure}

The pressure, the  energy density and the entropy density are shown in Fig. \ref{fig:eos}. The energy density
shows a rapid rise in the temperature region $(170-200)$ MeV and quickly approaches about $90\%$ of the ideal gas
value. The pressure rises less rapidly but at the highest temperature it is also only about $15\%$ below the ideal
gas value. In the previous calculations with the $p4$ action it was found that the pressure and the energy density
are below the ideal gas value by about $25\%$ at high temperatures \cite{Karsch:2000ps}.
A possible reason for this larger deviation could
be the fact that the quark masses used in this calculation were fixed in units of temperature instead 
being tuned to give constant meson masses as lattice spacing is decreased. As discussed
in Ref. \cite{Csikor:2004ik} this could reduce the pressure by $10-15\%$ at high temperatures. 
In Fig. \ref{fig:eos} I also show the entropy density divided by the corresponding ideal gas value
and compare the results of lattice calculations with resummed perturbative calculation \cite{Blaizot:1999ip,Blaizot:2000fc}
as well as with the predictions from AdS/CFT correspondence for the strongly coupled supersymmetric
Yang-Mills theory \cite{Gubser:1998nz}.
The later is considerably below the lattice results. Note that the pressure, the energy density 
and the trace anomaly have also been recently discussed in the framework of 
resummed perturbative calculations which seem to agree with lattice data quite well at
high temperatures\cite{Andersen:2010wu}.

The difference between the $stout$ action and the $p4$ and $asqtad$ actions for the trace anomaly
translates into the differences in the pressure and the  energy density. 
In particular, the energy density is about 20\% below the ideal gas limit for the $stout$ action. 

%% file: spf.tex
\section{Meson correlators and spectral functions}
\label{sec:spf}

In-medium meson properties as well as some transport coefficients
are encoded in meson spectral functions. Medium modification
of meson spectral functions can serve as diagnostic tools of
the medium created in heavy ion collisions. For example, the excess
in the low mass dilepton rate can be related to the modification
of the light vector meson spectral function (see Ref. \cite{Rapp:2011zz} for
a recent review). The suppression of quarkonium yield was suggested
by Matsui and Satz as signal for QGP formation in heavy ion collisions \cite{Matsui:1986dk}.
Studying quarkonium melting in QGP and the consequent suppression of the quarkonium yield
is subject of a large experimental and theoretical effort 
(see Refs. \cite{Brambilla:2004wf,Brambilla:2010cs,Bazavov:2009us,Rapp:2009my} for reviews).

The spectral function $\sgh$ for a given 
meson channel $H$ in a system at temperature $T$ can be defined 
through the Fourier transform of the real time two point functions
$D^{>}$ and $D^{<}$ or equivalently as the imaginary part of 
the Fourier transformed retarded 
correlation function \cite{LeBellac:375551},
\ber
\sgh &=& \frac{1}{2 \pi} (D^{>}_H(\omega, \vec{p})-D^{<}_H(\omega, \vec{p}))
\nonumber\\
&&
=\frac{1}{\pi} Im D^R_H(\omega, \vec{p}) \nonumber \\
 D^{>(<)}_H(\omega, \vec{p}) &=& \int_{-\infty}^{\infty} dt \int d^3 x 
 e^{i \omega t -i \vec{p} \cdot \vec{x}} D^{>(<)}_H(t,\vec{x}) \label{eq.defspect} \\
D^{>}_H(t,\vec{x}) &=& \langle
J_H(t, \vec{x}), J_H(0, \vec{0}) \rangle \nonumber\\
D^{<}_H(t,\vec{x}) &=& 
\langle J_H(0, \vec{0}), J_H(t,\vec{x}) \rangle , t>0 \
\eer
In essence $\sigma_H$ is the Fourier transformation of the thermal average of the commutator $[J(x),J(0)]$.

In the present paper we study local meson operators of the form 
\beq
J_H(t,x)=\bar q(t,\vec{x}) \Gamma_H q(t,\vec{x})
\label{cont_current}
\eeq
with $q(t,\vec{x})$ is the quark field operator and
\beq
\Gamma_H=1,\gamma_5, \gamma_{\mu}, \gamma_5 \gamma_{\mu}, \gamma_{\mu} \gamma_{\nu}
\eeq
for scalar, pseudo-scalar, vector, axial-vector and tensor channels. 
The relation of these quantum number channels to different meson states is given
in Tab. \ref{tab.channels}. In the vector channel I will use the subscript $V$ to denote the sum over all four
components, while the subscript $ii$ will be used for the spatial components of
the vector correlators and spectral functions.

\begin{table*}
\begin{tabular}
[c]{||c|c|c||c|}\hline
$\Gamma$ & $^{2S+1}L_{J}$ & $J^{PC}$ & $u\overline{u}$\\\hline
$\gamma_{5}$ & $^{1}S_{0}$ & $0^{-+}$ & $\pi$\\
$\gamma_{s}$ & $^{3}S_{1}$ & $1^{--}$ & $\rho$\\
$\gamma_{s}\gamma_{s^{\prime}}$ & $^{1}P_{1}$ & $1^{+-}$ & $b_{1}$\\
$1$ & $^{3}P_{0}$ & $0^{++}$ & $a_{0}$\\
$\gamma_{5}\gamma_{s}$ & $^{3}P_{1}$ & $1^{++}$ & $a_{1}$\\
&&$2^{++}$&\\\hline
\end{tabular}%
\begin{tabular}
[c]{|cc|}\hline
$c\overline{c}(n=1)$ & $c\overline{c}(n=2)$\\\hline
$\eta_{c}$ & $\eta_{c}^{^{\prime}}$\\
$J/\psi$ & $\psi^{\prime}$\\
$h_{c}$ & \\
$\chi_{c0}$ & \\
$\chi_{c1}$ & \\
$\chi_{c2}$ & \\\hline
\end{tabular}
\begin{tabular}[c]{|cc|}\hline
$b\overline{b}(n=1)$ & $b\overline{b}(n=2)$\\
\hline
$\eta_b$ & $\eta_b'$ \\
$\Upsilon(1S)$ & $\Upsilon(2S)$\\
$h_b$ & \\
$\chi_{b0}(1P)$& $\chi_{b0}(2P)$\\
$\chi_{b1}(1P)$& $\chi_{b1}(2P)$\\
$\chi_{b2}(1P)$&  $\chi_{b2}(2P)$\\ 
\hline                                                                   
\end{tabular}
\caption{Meson states in different channels for light, charm and bottom quarks.}
\label{tab.channels}
\end{table*}

The correlators $D^{>(<)}_H(t,\vec{x})$ satisfy the 
well-known Kubo-Martin-Schwinger
(KMS) condition \cite{LeBellac:375551}
\beq
D^{>}_H(t,\vec{x})=D^{<}_H(t+i/T,\vec{x}).
\label{kms}
\eeq
Inserting a complete set of
states and using Eq. (\ref{kms}), one gets the expansion
\ber
&
\sgh = {1 \over Z} \sum_{m,n} e^{-E_n / T} \times \nonumber\\ 
&
\langle n | J_H(0) | m \rangle|^2 \left(\delta^4(p_\mu + k^n_\mu - k^m_\mu)- 
\delta^4(p_\mu + k^m_\mu - k^n_\mu)\right)
\label{eq.specdef}
\eer
where $Z$ is the partition function, 
$k^{n(m)}$ refers to the four-momenta of the state $| n (m) \rangle $ and $p_{\mu}=(\omega,\vec{p})$.

A stable meson state contributes a $\delta$ function-like
peak to the spectral function:
\beq
\sgh = | \langle 0 | J_H | H \rangle |^2 \epsilon(\omega)
\delta(p^2 - M_H^2),
\label{eq.stable}
\eeq
where $M_H$ is the mass of the state and $\epsilon(p_0)$ is the sign function. For 
a quasi-particle in the medium one gets a smeared peak, with the width
being the  thermal width. 
As one increases the temperature the width increases 
and at sufficiently high
temperatures, the contribution from the meson state in the spectral function may 
be sufficiently broad. At some point it is not very meaningful to speak of it
as a well defined state any more. 
The spectral function as defined in
Eq. (\ref{eq.specdef}) can be directly accessible by high energy
heavy ion experiments. More precisely, the spectral function for the vector 
current is directly related to the differential thermal cross section 
for the production of dilepton pairs \cite{McLerran:1984ay,Braaten:1990wp}:
\beq
\left.{dW \over d \omega d^3p}\right|_{\vec{p}=0} = {5 \alpha_{em}^2 \over 27 \pi^2} 
{1 \over \omega^2 (e^{\omega/T}-1)} \sigma_V(\omega, \vec{p}).
\label{eq.dilepton} \eeq
Then presence or absence of a bound state in the spectral function
will manifest itself in the peak structure of the differential 
dilepton rate.

In finite temperature lattice calculations, one calculates
Euclidean time propagators, usually
projected to a given spatial momentum:
\beq
G_H(\tau, \vec{p}) = \int d^3x e^{i \vec{p}.\vec{x}} 
\langle T_{\tau} J_H(\tau, \vec{x}) J_H(0,
\vec{0}) \rangle
\eeq
This quantity is an analytical continuation
of $D^{>}_H(x_0,\vec{p})$
\beq
G_H(\tau,\vec{p})=D^{>}_H(-i\tau,\vec{p}).
\label{cont}
\eeq
The KMS condition implies the following relation for the Fourier transforms
of $D_H^{<}$ and $D_H^{>}$
\beq
D_H^{<}(\omega,\vec{p})=D_H^{>}(-\omega,\vec{p})=\e^{-\omega/T} D_H^{>}(\omega,\vec{p}). 
\eeq
This leads to a relation between  $D^{>}(\omega,\vec{p})$ and the spectral
function
\beq
D_H^{>}(\omega,\vec{p})=2 \pi \sigma_H(\omega,\vec{p}) e^{\omega/T}/(e^{\omega/T}-1).
\eeq
Using this and Eq. (\ref{cont}) we get the following integral representation
for the Euclidean time correlator
\ber
G_H(\tau, \vec{p}) &=& \int_0^{\infty} d \omega
\sgh K(\omega, \tau), \label{eq.spect} \non\\
K(\omega, \tau) &=& \frac{\cosh(\omega(\tau-1/2
T))}{\sinh(\omega/2 T)}.
\label{eq.kernel}
\eer
This equation is the basic equation for extracting the spectral
function from meson correlators. 
Equation (\ref{eq.kernel})
is valid in the continuum. 
Formally the same spectral representation can be written for
the Euclidean correlator calculated on the lattice $G^{lat}_H(\tau,\vec{p})$.
The corresponding spectral function, however, will be distorted by the effect
of the finite lattice spacing, in particular, the spectral function is zero above 
certain energy $\omega>\omega_{max}$.
These distortions have been calculated in the
free theory \cite{Karsch:2003wy,Aarts:2005hg}. When discussing the numerical results in
following sections the subscript $H$ denoting different channels 
for meson correlators and spectral functions 
will be omitted unless stated otherwise.  

To get some information on the spectral functions from lattice QCD the
corresponding Euclidean time correlation functions have to be calculated
at as many separations in the time direction as possible, i.e. one should use 
large $N_{\tau}$. One way to accomplish this task with available computer
resources is to use anisotropic lattices, i.e. lattices with different spacings
$a_t$ and $a_s$ in time and space directions, such that $\xi=a_t/a_s>1$. In 
addition one also often uses the quenched approximation for studying spectral functions
on the lattice.

The low frequency limit of the vector spectral function gives information about the
transport coefficients of the medium, namely the electric conductivity and the heavy quark
diffusion constant. Other transport coefficients, such as the shear and bulk viscosity 
are related to the correlation functions of gluonic operators (see Ref. \cite{Meyer:2011gj} for a review).
The quark flavor diffusion constant $\D_q$ can be defined through
the time derivative of the quark number density in the rest frame of the thermal system
as follows
\beq
\partial_t n_q=\D_q \nabla^2 n_q+{\cal O}(\nabla^3).
\eeq
This equation holds for small deviation from thermal equilibrium. The quark flavor
diffusion constant can be determined from the spatial component of the vector 
spectral function as follows
\beq
\D_q=\frac{1}{3 \chi_q} \lim_{\omega\rightarrow 0} \frac{\sigma_{ii}(\omega)}{\omega},
\eeq
here $\chi_q$ is the quark number susceptibility. The electric conductivity is related
to the correlation function of electric currents
\beq
J_{\mu}^{em}=\sum_f Q_f \bar q_f(t,\vec{x}) \gamma_{\mu} q_f(t,\vec{x}),
\eeq
with $Q_f$ being the electric charge of quark flavor $f$. The correlator of
the electromagnetic current receives contribution from quark line connected
and quark line disconnected diagrams. At high temperatures the contribution
from the disconnected diagrams is small. For degenerate quark flavors the electric
conductivity can be written as \cite{Burnier:2012ts}
\beq
\zeta=\chi_q \left[ \left(\sum_f Q_f \right)^2 \D_{q,disc}+
  \left( \sum_f Q_f^2 \right) \D_{q,con} \right],~\D_q=\D_{q,con}+\D_{q,dis}.
\eeq 
For degenerate $u$, $d$ and $s$ quarks the disconnected contribution vanishes
since $\sum_{f=u,d,s} Q_f=0$.
The quark flavor diffusion coefficient for single heavy quark flavor ($m_q\gg T$)
is called the heavy quark diffusion constant $D$. It is related to the momentum
drag coefficient $\eta$ in the Langevin dynamics of heavy quarks \cite{Moore:2004tg}:
\beq
D=\frac{T}{m_q \eta}.
\eeq

In lattice QCD one also calculates meson correlation function in one of the spatial
directions, say $z$
\begin{equation}
G(z,T)=\int dx dy \int_0^{1/T} d \tau \langle J(x,y,z,\tau) J(0,0,0,0) \rangle .
\end{equation}
The spatial correlation function is related to the meson spectral function
at non-zero spatial momentum
\begin{equation}
G(z,T)=\int_{-\infty}^{\infty} d p_z e^{i p_z z} \int_{0}^{\infty} d \omega 
\frac{\sigma(\omega,p_z,T)}{\omega}.
\end{equation}
Thus the temperature dependence of the spatial correlation function 
also provides
information about the temperature dependence of the spectral function.
Medium effects are expected to be the largest at distances which are larger than $1/T$.
At these distances $G(z,T)$ decays exponentially and this exponential decay is governed
by a screening mass $M_{scr}$. If there is a lowest lying meson state of mass 
$M$, i.e. the spectral function can be well approximated by Eq. (\ref{eq.stable}),
then the long distance behavior of the spatial meson correlation 
function is determined by the meson mass, i.e. $M_{scr}=M$. At very high temperatures the quark and anti-quark 
are not bound and the meson screening mass is given by $2 \sqrt{(\pi T)^2+m_q^2}$, 
where $m_q$ is the quark mass and $\pi T$ is the lowest Matsubara frequency. 
Therefore, a detailed study of spatial meson correlators
and screening masses can provide some information about the melting of meson states at high
temperatures. 

One would like to obtain the spectral functions through lattice calculation of the temporal
meson correlation function.
The obvious difficulty in the reconstruction of the spectral function from
Eq. (\ref{eq.kernel}) is the fact that the Euclidean correlator is calculated
only at ${\cal O}(10)$ data points on the lattice, while for a reasonable discretization
of the integral in Eq. (\ref{eq.kernel}) we need ${\cal O}(100)$ degrees of freedom. 
The problem can be solved using Bayesian analysis
of the correlator, where one looks for a spectral function which maximizes the 
conditional probability $P[\sigma|DH]$ of having the spectral function $\sigma$ given
the data $D$ and some prior knowledge $H$  (for  reviews see \cite{Asakawa:2000tr,Lepage:2001ym}).
Different Bayesian  methods differ in the choice of the prior knowledge.
One version of this analysis which is extensively used in the literature is the 
{\em Maximum Entropy Method} (MEM) \cite{Bryan:1990,Nakahara:1999vy}.
It has been used to study different correlation functions in QCD
\cite{Nakahara:1999vy,Asakawa:2000tr,Karsch:2001uw,Yamazaki:2001er,Wetzorke:2001dk,Karsch:2002wv,Datta:2002ck,Asakawa:2002xj,Umeda:2002vr,Petreczky:2003iz,Asakawa:2003re,Datta:2003ww,Datta:2004im,Blum:2004zp,Sasaki:2005ap,Datta:2006ua}.
In this method the basic prior knowledge is the positivity of the spectral function and 
the prior knowledge is given by the Shannon - Janes entropy  
\beq
\displaystyle
S=\int d \omega \biggl [ \sigma(\omega)-m(\omega)-\sigma(\omega)
  \ln(\frac{\sigma(\omega)}{m(\omega)}) \biggr]. 
\eeq
The real function $m(\omega)$ is called the default model and parametrizes all additional prior knowledge about the
spectral functions, e.g. such as the asymptotic behavior at high energy  \cite{Nakahara:1999vy,Asakawa:2000tr}.
For MEM the conditional probability can be written as 
\beq
 P[\sigma|DH]=\exp(-\frac{1}{2} \chi^2 + \alpha S),
\label{eq:PDH}
\eeq
with $\chi^2$ being the standard likelihood function and $\alpha$ is a real parameter.

Early results on meson spectral functions obtained with MEM gave a number of unexpected results.
In particular, the MEM analysis of the charmonium correlators calculated in quenched QCD
suggested that 1S charmonium states can survive in QGP up
to temperatures as high as $1.6T_c$ (see e.g. \cite{Asakawa:2003re}).
Here $T_c$ is the deconfinement phase transition temperature of $SU(3)$ gauge theory.
This is in odds with the expected medium modification
of the heavy quark potential that in addition to the effects of color screening also includes
an imaginary part \cite{Brambilla:2008cx,Laine:2006ns}.
In fact, many potential model calculations indicate melting of 1S charmonium
state in the deconfined medium \cite{Mocsy:2005qw,Mocsy:2007jz,Mocsy:2007yj,Mocsy:2007py,Petreczky:2010tk,Riek:2010py}. 
Furthermore, even in the light meson spectral functions peak
structures have been observed \cite{Asakawa:2002xj}. It is not clear to what extent these correspond to physical effects
or are artifacts of the calculations.
Next we will consider the determination of the meson spectral function separately for the case of heavy
quarks and light quarks.

\subsection{Quarkonium spectral functions at zero temperature}
\begin{figure}
\includegraphics[width=7.5cm]{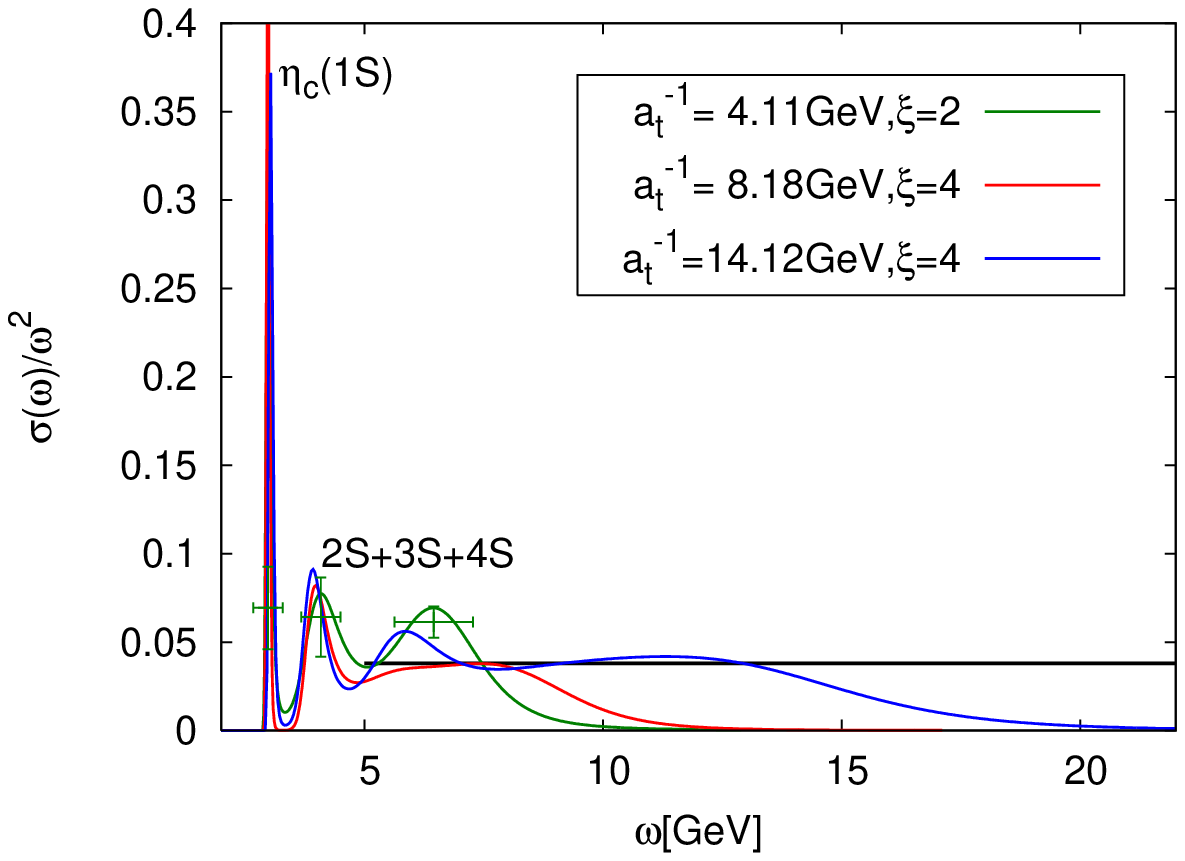}
\includegraphics[width=7.5cm]{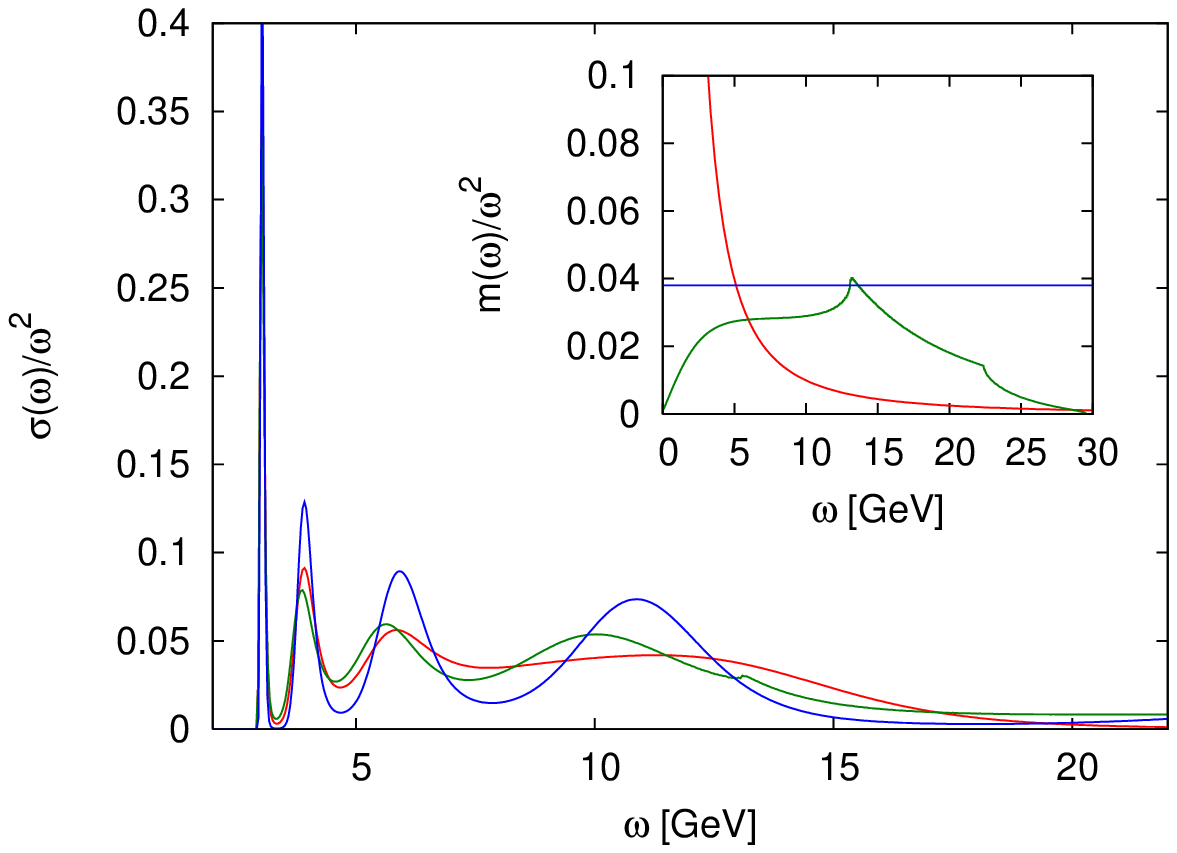}
\caption{The spectral function in the pseudo-scalar channel calculated for
several lattice spacings (left). The default model dependence of the spectral
function calculated on $24^3 \times 160$ lattice (right). The inset in the right
panel shows the different default models used in the analysis.}
\label{fig:spfcc}
\end{figure}

The best way to understand the structure of meson spectral functions and the issues
related to their determination on the lattice is to consider the case of heavy quarkonium
at zero temperature.
Charmonium correlation functions have been studied in detail 
in Ref. \cite{Jakovac:2006sf} using quenched anisotropic lattices and the so-called Fermilab
formulations for heavy quarks \cite{ElKhadra:1996mp}. Spectral functions have been extracted using MEM. The
spectral functions in the pseudo-scalar channel 
calculated at three lattice spacings are shown in Fig. \ref{fig:spfcc}.
The first peak corresponds to the ground state, $\eta_c(1S)$. The second peak is actually
the combination of several excited states, since MEM cannot resolve individual excited meson
states due to the small splitting between them \cite{Jakovac:2006sf}.
For $\omega>5$ GeV
we see the continuum. The spectral function becomes zero above certain energy $\omega$ due
to the presence of finite lattice as expected in the free theory. In Fig.  \ref{fig:spfcc}
I also show the spectral function calculated for different default models $m(\omega)$. 
The default model dependence is small for $\omega < 5$GeV, while above that energy it is
significant due to the fact that there are only very few data points that carry information
about the spectral function in that region. As the temperature increases the number of
data points available for the analysis as well as the maximal extent of
the time direction $\tau_{max}=1/(2 T)$ become smaller.
\begin{figure}
\includegraphics[width=7.5cm]{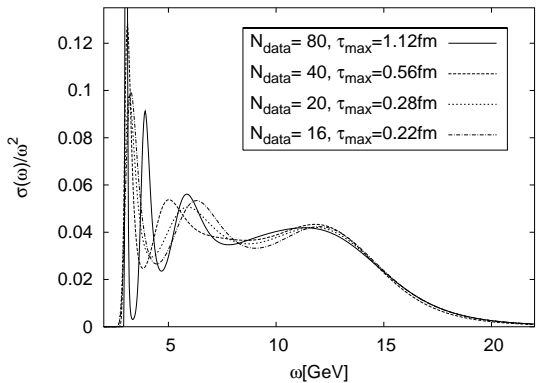}
\caption{The dependence of the pseudo-scalar spectral function on $\tau_{max}$.}
\label{fig:spf_ntdep}
\end{figure} 
To demonstrate this point in Fig. \ref{fig:spf_ntdep}
I show the zero temperature charmonium spectral function calculated with different number
of data points and $\tau_{max}$ that are characteristic for temperatures corresponding to QGP.
As $\tau_{max}$ decreases MEM looses the ability to reconstruct the peak corresponding the
excited states and the ground state peak is significantly broadens. For $\tau_{max}=0.22$fm
even the position of the ground state peak is not reproduced correctly. Furthermore, the shape
of the spectral functions, in particular, the presence of the ground state peak 
also becomes sensitive to the default model.

\begin{figure}
\includegraphics[width=8cm]{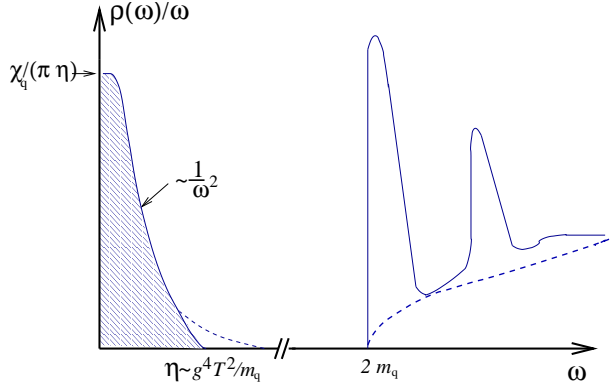}
\caption{The schematic structure of the vector spectral function for heavy quarks.} 
\label{fig:spf_demo}
\end{figure}

Since the analysis of the quarkonium spectral functions using MEM turns out to be
quite complicated already in the zero temperature limit, it is important to
understand the temperature dependence of the quarkonium correlation functions
and try to identify possible sources of the temperature dependence that are related
to melting of the bound states. From Eq. (\ref{eq.kernel}) it is clear that the temperature
dependence of the meson correlation functions comes from two sources: the trivial
temperature dependence of the integration kernel $K(\tau,\omega)$ and the temperature
dependence of the spectral function $\sigma(\omega,T)$. To get rid of the first trivial
temperature dependence one can consider the reconstructed correlation function 
\beq
G_{rec}(\tau,T) = \int_0^{\infty} d \omega \sigma(\omega,T=0) K(\omega, \tau).
\eeq
If the spectral function does not change across the deconfinement transition 
$G(\tau,T)/G_{rec}(\tau,T)$ should be unity. Deviations of this ratio from unity indicate
temperature dependence of the spectral functions. The ratio $G(\tau,T)/G_{rec}(\tau,T)$
was first studied in Ref. \cite{Datta:2003ww} and subsequently in Refs. \cite{Datta:2004im,Datta:2006ua,Jakovac:2006sf,Datta:2004js}. It was found that in the pseudo-scalar
channel this ratio stays close to one and shows only small temperature dependence.
This seemed to support the conclusion based on the spectral functions extracted from MEM that
ground state charmonium survives in the deconfined phase up to temperatures $1.6T_c$.
In the scalar and axial-vector channels large temperature dependence in $G(\tau,T)/G_{rec}(\tau,T)$
was observed. This was interpreted as melting of 1P charmonium states and fitted well into
the expected picture of sequential melting. However, this conclusion was premature. 
In the deconfined phase there is an additional contribution to the spectral functions
at very low frequency. This can be easily seen by calculating
the spectral function in the free theory. In addition to the contribution to the spectral
function that starts at twice the heavy quark mass there is a contribution proportional to
$\omega \delta(\omega)$ in all but the pseudo-scalar channel \cite{Aarts:2005hg}.
The delta function is smeared once 
the interactions of the heavy quarks with the medium are taken into account. In particular,
in the vector channel, where this low $\omega$ structure is related to the heavy
quark diffusion we have \cite{Petreczky:2005nh}
\beq
\omega \delta(\omega) \rightarrow \frac{1}{\pi} \frac{\eta \omega}{\omega^2+\eta^2}, ~\eta=T/(D m_q),
\eeq
with $m_q$ being the heavy quark mass and $D$ being the heavy quark diffusion constant.
The schematic structure of the spectral functions is shown in Fig. \ref{fig:spf_demo}.
The area under the transport peak is given by the quark number susceptibility $\chi_q$.
For large quark mass the transport peak is narrow and is well separated 
from the high energy part of the spectral function 
that corresponds to bound states and/or unbound heavy quark anti-quark
pairs. We expect similar structure in the spectral function in other channels as well.
Thus we can write
\ber
\sigma(\omega,T)&=&\sigma_{low}(\omega,T)+\sigma_{high}(\omega,T)\\
G(\tau,T)&=&G_{low}(\tau,T)+G_{high}(\tau,T).
\eer
\begin{figure}
\includegraphics[width=7.5cm]{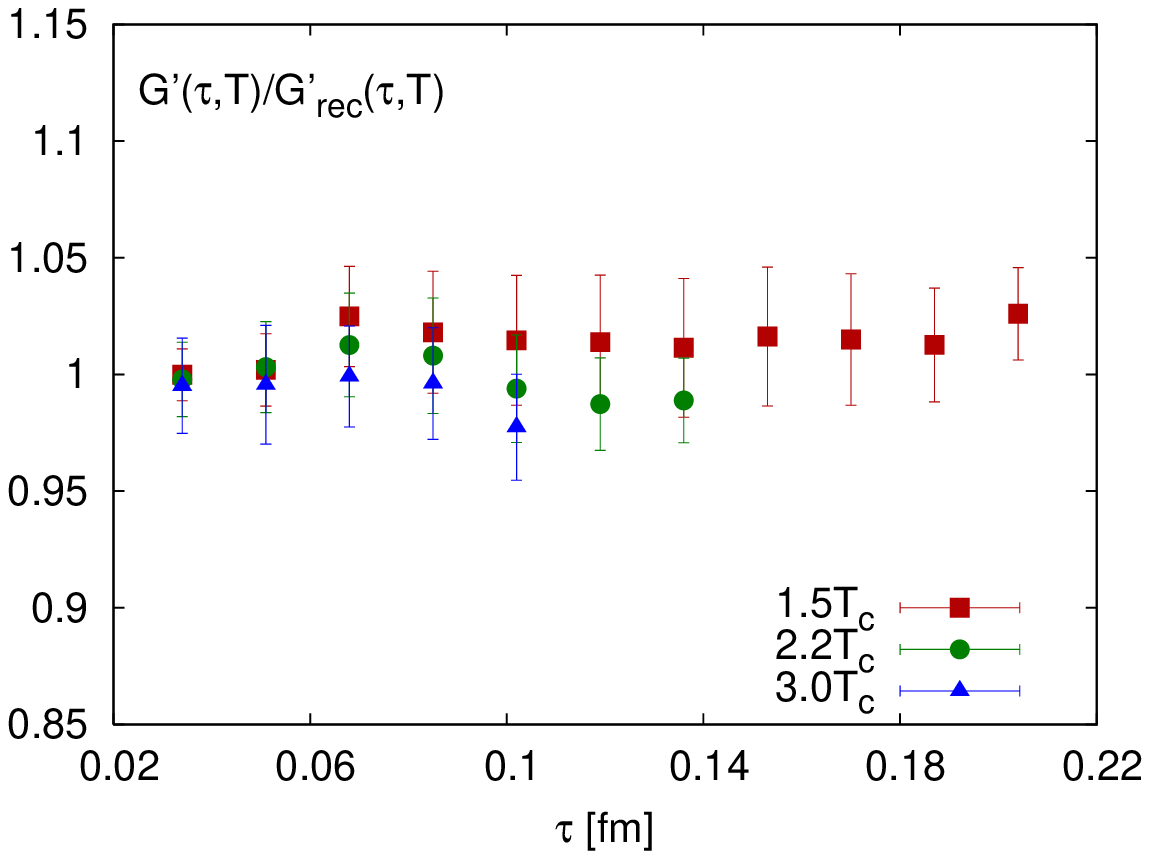}
\includegraphics[width=7.5cm]{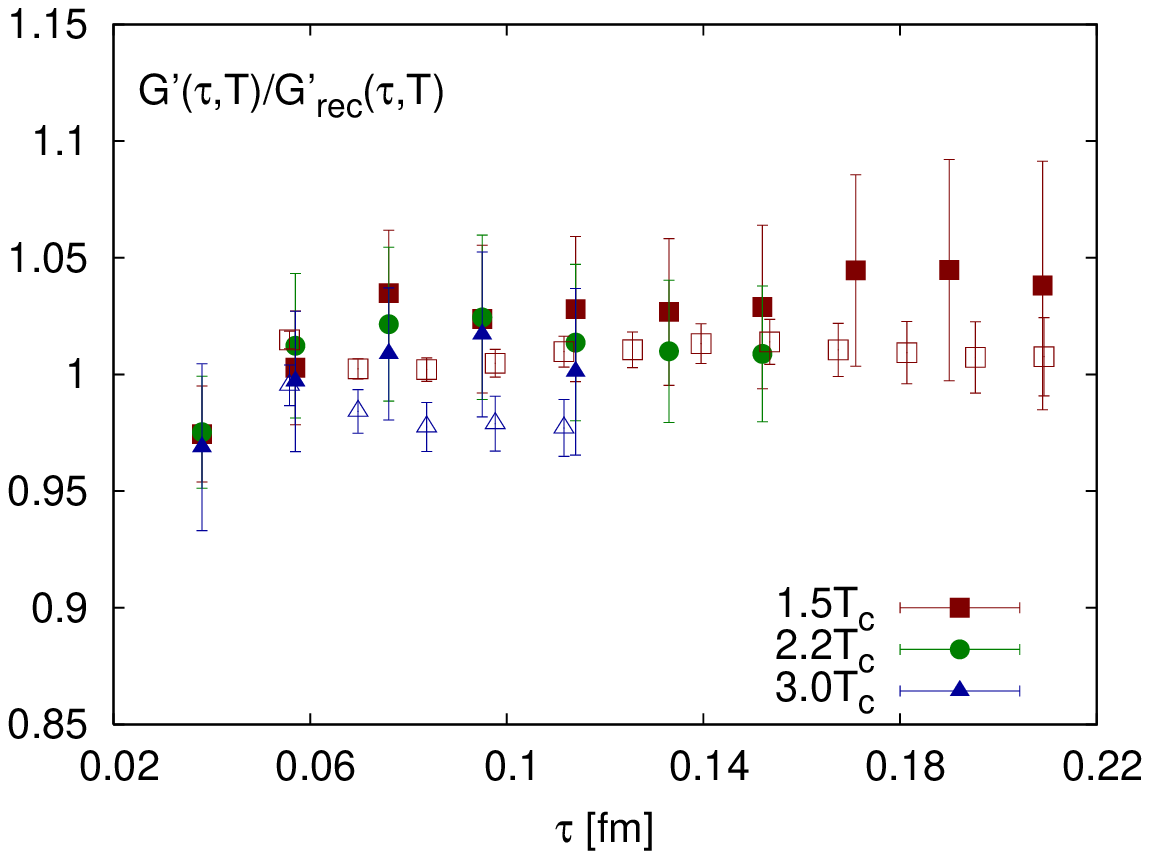}
\caption{The ratio of the derivatives of the charmonium correlators to the corresponding
reconstructed correlators for the pseudo-scalar (left) and scalar (right) channels \cite{Petreczky:2008px}.}
\label{fig:gp}
\end{figure}
Since the peak at low $\omega$ is narrow we expect that the derivative of
$G_{low}(\tau,T)$ with respect of $\tau$ is small, $G_{low}'(\tau,T) \simeq 0$.
In Fig. \ref{fig:gp} I show the ratio of the derivatives $G'(\tau,T)/G_{rec}'(\tau,T)$
in the pseudo-scalar and scalar channels in the deconfined phase. As once can
see from the figure these ratios show no strong temperature dependence 
and are close to one for all temperatures, including the highest temperature of $3T_c$.
This means that the large temperature dependence
seen in $G/G_{rec}$ for the scalar and axial-vector channels 
comes mostly from the temperature dependence of $G_{\rm low}(\tau,T)$ and is
not related to melting of P-wave  charmonium states. A question arises how the observed 
temperature (in)dependence of $G_{high}(\tau,T)$ is related to the expected in-medium
modification of the heavy quark potential and melting of quarkonium states at sufficiently
high temperatures. This question was addressed in Refs. \cite{Mocsy:2007yj}, 
where the quarkonium spectral
functions have been calculated using potential model with screened potential.
It was found that despite the significant
change in the spectral functions the Euclidean correlator obtained from them do not
change significantly in the deconfined medium and are compatible with the lattice results.
Including the imaginary part in the analysis did not change this conclusion \cite{Petreczky:2010tk}.
\begin{figure}
\includegraphics[width=7.5cm]{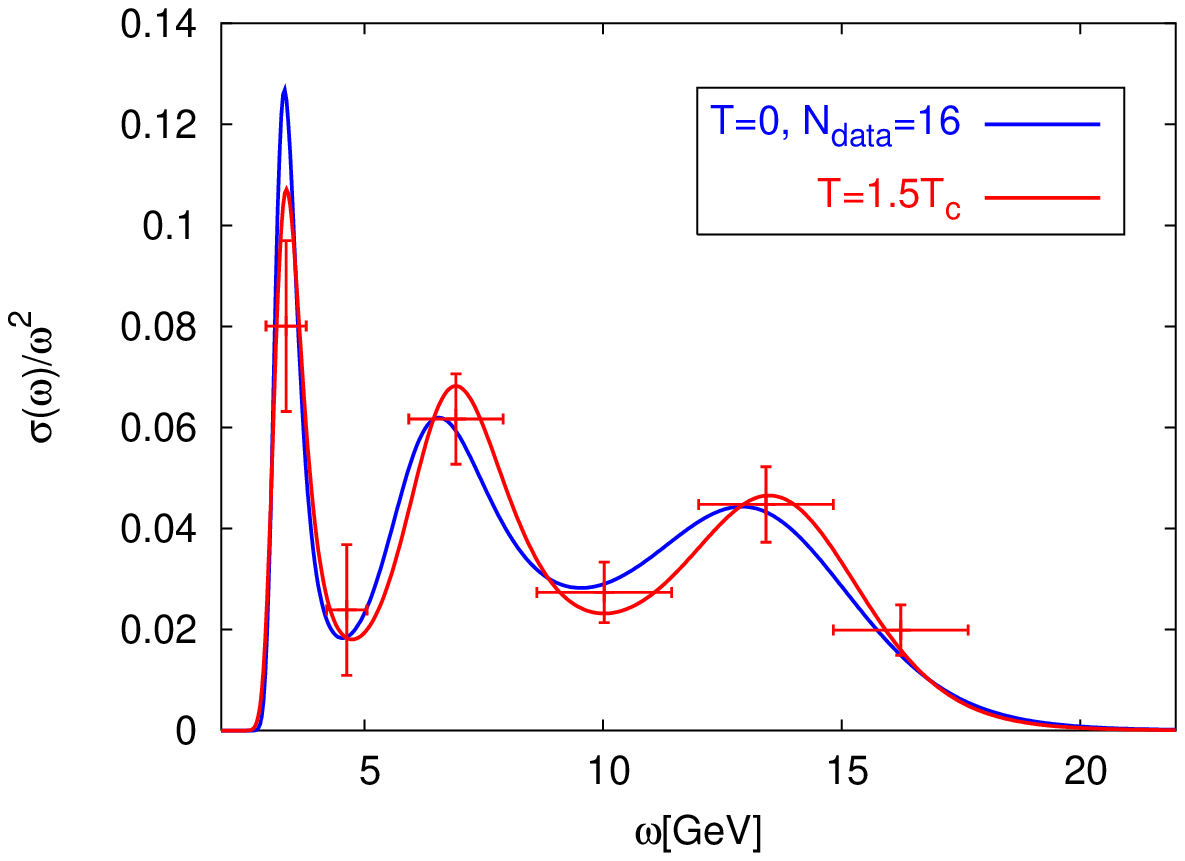}
\includegraphics[width=7.5cm]{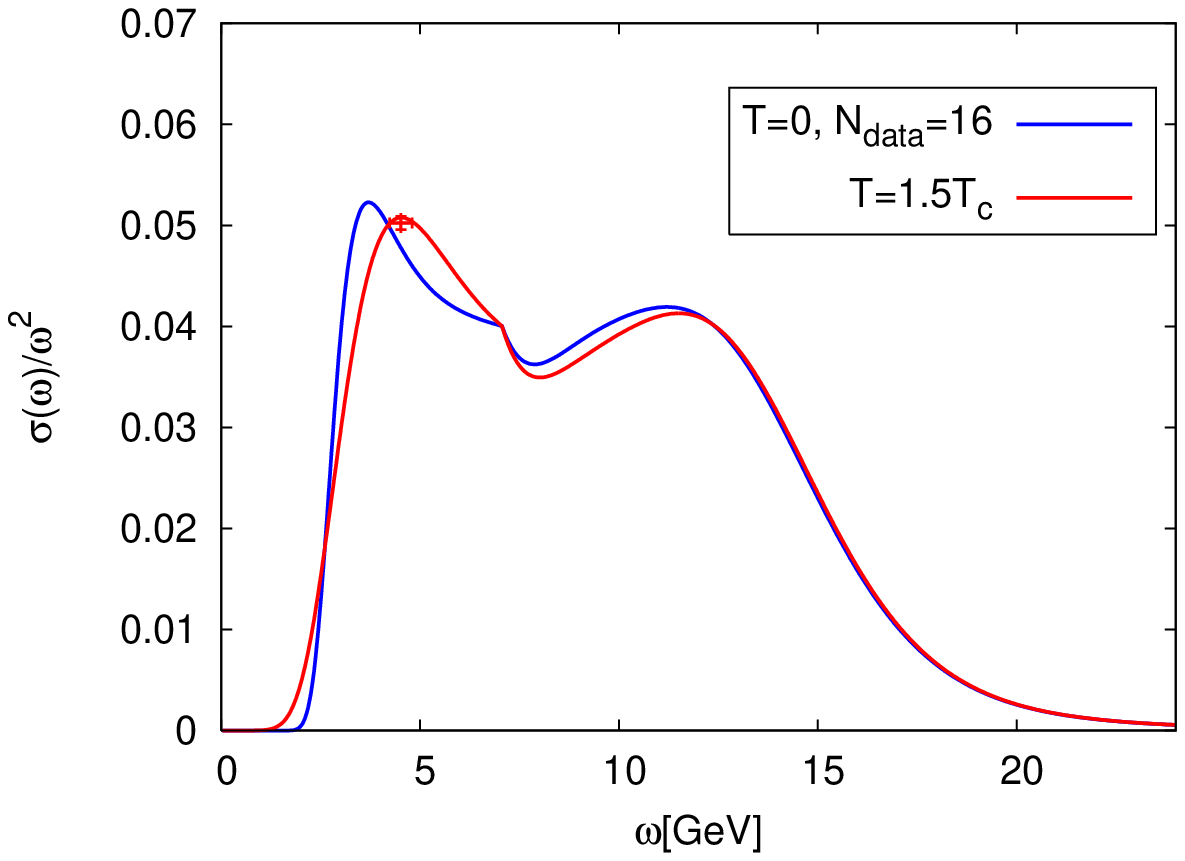}
\caption{The spectral functions in pseudo-scalar channel calculated at $T=1.5T_c$
and compared to the zero  temperature spectral function obtained using $\tau_{max}=1/(2T)$.
In the left panel the spectral functions obtained with $m(\omega)=const$ are shown, while
in the right panel the default model was obtained from the $T=0$ spectral function.}
\label{fig:spf1.5Tc}
\end{figure}
Attempts to reconstruct quarkonium spectral functions using MEM have been presented
in Refs. \cite{Umeda:2002vr,Asakawa:2003re,Datta:2003ww,Datta:2004im,Jakovac:2006sf} 
and no evidence for melting of quarkonium states has been found. However, in 
view of the difficulties related to the MEM analysis for small $\tau_{max}$ and limited 
number of data points one should be careful drawing conclusions. In particular the dependence
of the results on the default model should be examined. In Fig. \ref{fig:spf1.5Tc} I show
charmonium spectral functions in the pseudo-scalar channel calculated at $1.5T_c$ and at
zero temperature for two default models: $m(\omega)=const$ and $m(\omega)$ that equals
to the zero temperature spectral function for $\omega>5$GeV and smoothly matched to a constant
below that energy \cite{Jakovac:2006sf}. In both cases the difference between the spectral function at $1.5T_c$ and
$T=0$ is small. However, the shape of the spectral function depends on the default model. While
for $m(\omega)=const$ we see a peak structure for more realistic choice of the default
model this peak structure is absent even at zero temperature.
Recent analysis of charmonium spectral function based
on isotropic lattices that uses free lattice spectral function also does not find peak
structure that can be associated with 1S state in the deconfined phase \cite{Ding:2010yz} 
It is not clear
to what extent the absence of bound state peaks is consequence of quarkonium melting in QGP
or due to the limited data set.
It is clear, however, that existing MEM calculations of the spectral function do not provide evidence
for existence of quarkonium bound states in QGP. 

The analysis described so far was performed in the quenched approximation.
Calculation of the charmonium correlators and spectral functions was also performed in two flavor QCD
\cite{Aarts:2007pk}.
The findings of this analysis are similar to the ones described above. 
Furthermore, in the case of the bottomonium correlators and spectral functions have been studied
using NRQCD \cite{Aarts:2010ek,Aarts:2011sm}.
The main advantage of the NRQCD approach is that it allows to study correlators
at larger Euclidean time separations, namely $\tau_{max}=1/T$. Furthermore, there is no low
energy contribution to the spectral functions and thus the temperature dependence of the corresponding
Euclidean time correlation function is directly related to the properties and melting of the bound
states. A significant temperature dependence has been seen in the scalar bottomonium correlators
that may indicate the melting of the P-wave bottomonium at temperatures slightly above the transition
temperature. 

\subsection{Meson spectral functions at non-zero temperature in the light quark sector}
Early attempts to calculate light meson spectral functions in QGP using MEM 
were presented in Refs. \cite{Karsch:2001uw,Karsch:2002wv,Asakawa:2002xj,Petreczky:2003iz}. 
The results were inconclusive and to some extent confusing as peak
structures in the spectral functions have been found up to $3T_c$ and no continuum
was observed  at high energies. It was pointed out, however, that even without using
MEM Euclidean correlation functions can put stringent constraints on the spectral
functions \cite{Karsch:2001uw,Petreczky:2003iz}. In the vector
channel the correlation function calculated on $N_{\tau}=12$ and $N_{\tau}=16$ lattices
was found to deviate from the free value by less than $10\%$ \cite{Karsch:2001uw}. 
In the pseudo-scalar channel on the other hand a large enhancement over the free theory result was
observed \cite{Petreczky:2003iz} possibly indicating non-perturbative effects in QGP.

\begin{figure}
\includegraphics[width=8cm]{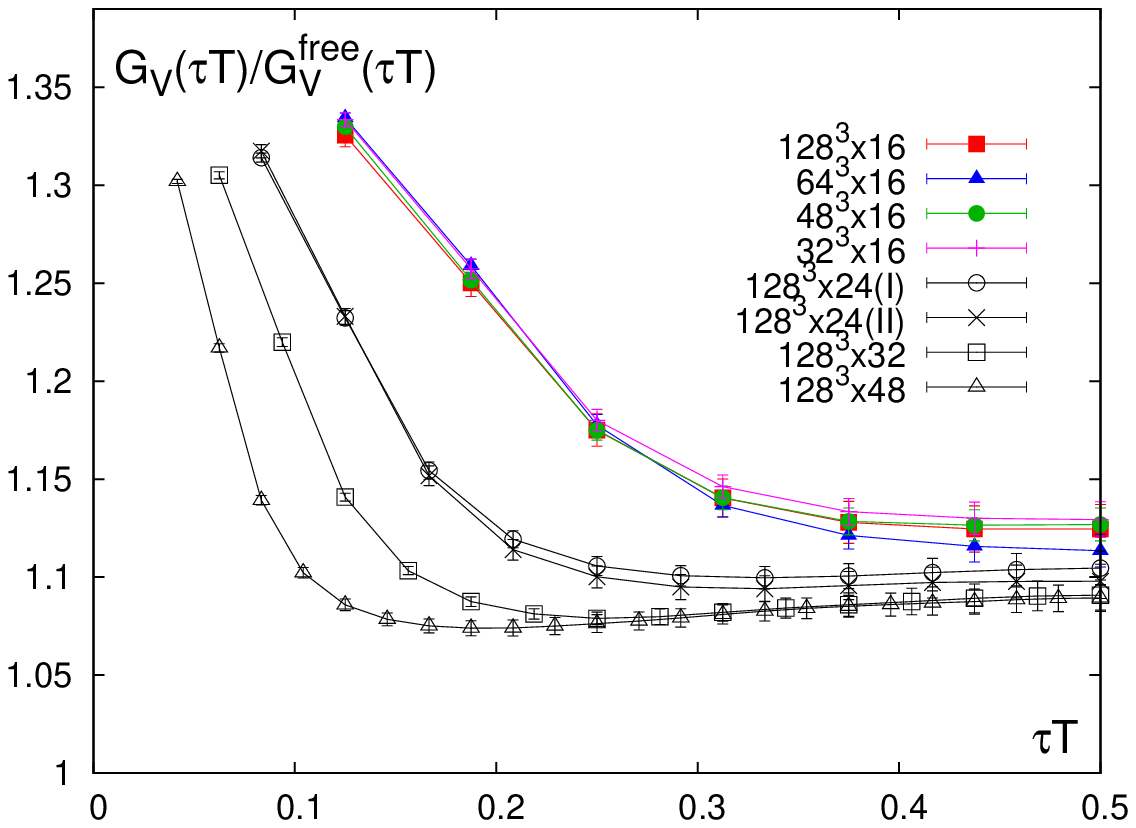}
\includegraphics[width=8cm]{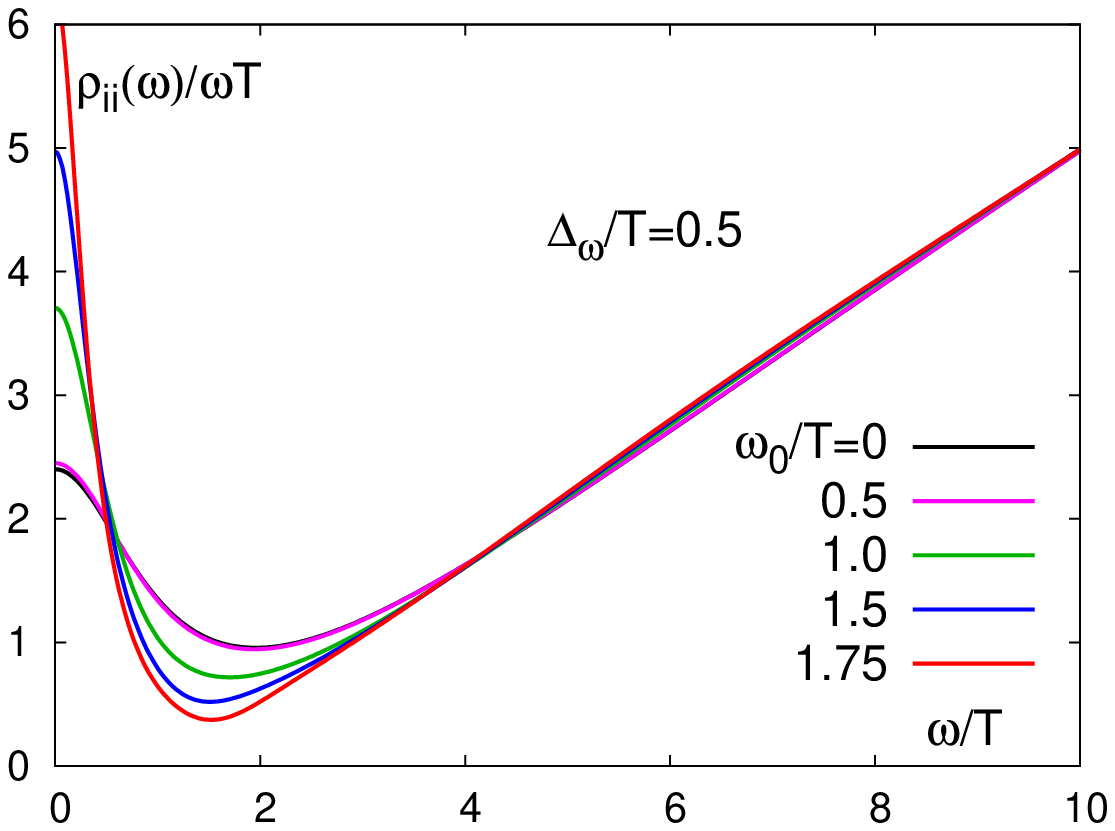}
\caption{The vector correlation function, $G_V(\tau,T)$ calculated on different lattices
for $T=1.45T_c$ (left) and the vector spectral function (right) obtained using the fit form
given by Eq. (\ref{fit}).}
\label{fig:vect}  
\end{figure}

More recently a detailed calculation of the vector correlation function on large
quenched lattices was reported for $T\simeq 1.45T_c$ \cite{Ding:2010ga}. Calculations were 
performed using several lattice spacings. This 
enabled a reliable extrapolation to the continuum limit and it was found that
the correlator never exceeds the free theory value by more than $9\%$ \cite{Ding:2010ga}.
The corresponding results are shown in Fig. \ref{fig:vect}.
Furthermore, the spectral functions reconstructed with MEM showed no evidence for  peak like structures
\cite{Ding:2010ga}.
Since vector correlator is so close to the free limit the authors of Ref. \cite{Ding:2010ga} obtained the spectral function
using the following model
\ber
\sigma_{ii}(\omega)&=&c_{BW} \chi_q \frac1\pi \frac{\omega \Gamma/2}{\omega^2+(\Gamma/2)^2}+
\frac{3}{2 \pi} (1+k) \omega^2 \tanh(\omega/4T) \Theta(\omega_0,\Delta_{\omega}),\nonumber\\
\Theta(\omega_0,\Delta_{\omega})&=&(1+e^{(\omega_0-\omega)/\omega \Delta_{\omega}})^{-1},
\label{fit}
\eer
and treating $c_{BW}$, $\Gamma$ and $k$ as fit parameters. Furthermore, several choices
for the parameters $\omega_0$ and  $\Delta_{\omega}$ have been considered, including 
$\omega_0=\Delta_{\omega}=0$. These fits are shown in Fig. \ref{fit} and gave the following constrains for
the electric conductivity\cite{Ding:2010ga}  
\beq
1/3 < \frac{1}{C_{em}} \frac{\zeta}{T} < 1, ~C_{em}=\sum_f Q_f^2
\eeq
The disconnected part was neglected in this calculation. 
One can see from Fig. \ref{fig:vect} that while 
somewhat broad the transport peak is clearly visible in the vector spectral function contrary
to the expectation based in strongly coupled supersymmetric Yang-Mills theory 
where the transport peak is absent \cite{Teaney:2006nc}. On the other hand the analysis of the same
lattice data using a different approach lead to much smaller value of the electric conductivity
and no clear transport peak \cite{Burnier:2012ts}. This may indicate that much more work is needed till a reliable
result for the electric conductivity can be quoted.

\subsection{Spatial meson correlation functions}
\begin{figure} 
\includegraphics[width=8.5cm]{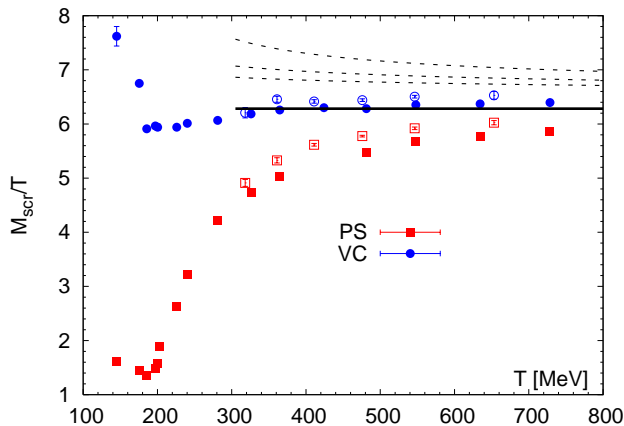}
\caption{Meson screening masses for light quarks in the pseudo-scalar and vector channels
as function of the temperature calculated in 2+1 flavor QCD with $p4$ action on $N_{\tau}=6$ (filled symbols)
and $N_{\tau}=8$ (open  symbols) lattices \cite{Cheng:2010fe}. The solid black line corresponds to the limit
of free quarks $2 \pi T$. The dashed lines are the NLO results \cite{Laine:2003bd} evaluated with a 2-loop
running coupling constant in $\overline{MS}$-scheme for $\mu=\pi T,~2 \pi T$ and $4 \pi T$.}
\label{fig:mscr}
\end{figure}

As discussed above spatial meson correlators are also sensitive to in-medium modification of the
meson spectral functions. In particular, the change in the meson spectral functions
is reflected in the meson screening masses, which should be close to the vacuum masses
at low temperatures. At high temperatures, on the other hand, the meson screening masses should approach $2 \pi T$
corresponding to the free quark limit. In the past meson screening masses have been
studied in quenched approximation \cite{Laermann:2001vg,Gavai:2002jt} and the sharp
change in their behavior was observed at the transition temperature.
More recently spatial  meson correlators and screening masses have been studied in 2+1 flavor
QCD for physical strange quark mass and light quark masses $m_l=m_s/10$ with $p4$ action \cite{Cheng:2010fe}.
In the case of full QCD rapid change of the spatial correlators and screening masses is expected
in the vicinity of the chiral crossover. Numerical calculations show that this is indeed the case 
\cite{Cheng:2010fe}.
In Fig. \ref{fig:mscr} the meson screening masses in the pseudo-scalar and vector channels are shown as 
function of the temperature and the rapid change in the behavior of the screening masses takes place around $200$MeV,
which is close to the chiral crossover temperature $T_c$ for those lattices. The effective restoration of the chiral
symmetry above the crossover temperature should manifest in approximate degeneracy of the vector and axial-vector
correlation functions. Such degeneracy is observed at temperature $200$MeV (c.f. Fig. 5 of Ref. \cite{Cheng:2010fe}).
As discussed in section \ref{sec:chiral} the $U_A(1)$ symmetry is expected to be restored at sufficiently high 
temperatures. The effective restoration of the $U_A(1)$ should manifest itself in the approximate degeneracy
of flavor non-singlet pseudo-scalar and scalar correlators. The difference between the scalar and pseudo-scalar screening masses
rapidly decreases above the transition temperature but becomes compatible with zero only temperatures above $240$MeV.
Thus the effective restoration of the axial symmetry happens at $T \simeq 1.2 T_c$, with $T_c$ being 
the chiral transition temperature.
Calculations using DWF formulations
confirm this result \cite{Hegde:2011zg}.

At high temperatures the screening masses can be calculated using weak coupling methods. The calculations
of the screening masses up to next-to-leading order was performed in Ref. \cite{Laine:2003bd} and the first
perturbative correction turned out to be positive, namely
\begin{equation}
M_{scr}=2 \pi T + \frac{4}{3} \frac{T}{2 \pi} g^2 (\frac{1}{2}+\hat E_0),
\end{equation}
with $\hat E_0 \simeq 0.46939139$ for $N_f=3$ \cite{Laine:2003bd}. Furthermore, all the screening masses are 
expected to be degenerate at next-to-leading order \cite{Laine:2003bd}.
In Fig. \ref{fig:mscr} we compare the lattice data in 2+1 flavor case with the perturbative result
of Ref. \cite{Laine:2003bd}. At high temperatures we still see significant differences between the 
pseudo-scalar and vector screening masses. Moreover the pseudo-scalar screening masses are smaller than
$2 \pi T$. Note, however, that the lattice spacing ($N_{\tau}$) dependence of the screening masses is
non-negligible. Thus it will be important in the future to perform calculations on finer lattices to
get control over the continuum limit. 

Spatial meson correlators and  screening masses have been calculated also for charmonium  \cite{Karsch:2012na}.
The changes in the correlators and screening masses around the transition
temperature are much smaller than for the light mesons. This fits into the picture that ground state
charmonium does not melt at the crossover temperature. However, for $T> 300$MeV we see large change also
for the charmonium screening masses. The behavior of the screening masses and their dependence on
the spatial boundary conditions is compatible with the picture of unbound quarks \cite{Karsch:2012na}.
This corroborates
the discussion in the previous subsection that there is no evidence for survival of charmonium states
at temperatures of $1.5T_c$.

\subsection{Heavy quark diffusion constant}

Since for sufficiently heavy quarks the transport peak is very narrow it is very difficult
if not impossible to determine the heavy quark diffusion using the procedure that was used
for the electric conductivity \footnote{attempts along these lines have been presented in 
Ref. \cite{Ding:2011hr}}. Even for the smallest possible heavy quark diffusion constant  $D \simeq 1/(2 \pi T)$
the low energy part of the vector correlator $G_{\rm low}(\tau,T)$ shows very little $\tau$-dependence 
\cite{Petreczky:2005nh}. An alternative way to estimate the heavy quark diffusion constant is integrate
out the heavy quark fields and relate the drag coefficient to the correlation function of
chromo-electric fields at time $t$ and time $0$ connected by Wilson lines
\cite{CasalderreySolana:2006rq,CaronHuot:2009uh}. In particular, at leading order in
the expansion in the inverse of heavy quark mass we have
\ber
\kappa &=& \int dt G_E(t)\nonumber\\
G_E(t) &=& \frac{1}{3T \chi_q} \int d^3x \langle O_E^i(t,x) O_E^i(0,0) \rangle\nonumber\\
O_E^i(t,x)&=&\phi^{\dagger}(t,x)g E^i(t,x) \phi(t,x)-\theta^{\dagger}(t,x) g E^i(t,x)\theta(t,x).
\eer
Here $\phi$ and $\theta$ are the static quark and anti-quark fields, and $E^i$ is the chromo-electric field.
It is easy to generalize the above expression to Euclidean time and define the corresponding correlator
$G_E(\tau)$ which after integrating the heavy quark fields becomes \cite{CaronHuot:2009uh}
\beq
G_E(\tau)= -\frac{1}{3} 
\frac{\langle {\rm Re Tr} \left( W(\beta,\tau) g E_i(\tau) W(\tau,0) E_i(\tau) \right) \rangle}{
{\rm Re Tr} \langle W(\beta,0) \rangle}.
\eeq
The above correlation function was also calculated in perturbation theory to next-to-leading order
\cite{CaronHuot:2009uh}. 
The key difference between this correlation function and meson correlation function is that
the transport contribution does not appear as a narrow peak in the corresponding spectral function
This is principle should make it easier to estimate the corresponding transport coefficient.
Numerical calculations of $G_E(\tau)$  have been performed in $SU(3)$ gauge theory on lattices
with temporal extent $N_{\tau}=12-24$ \cite{Meyer:2010tt,Banerjee:2011ra,Francis:2011gc}. 
The lattice data are significantly larger than the next-to-leading order  result. 
Comparing the numerical results with a model correlation
function based on perturbative calculations of Ref. \cite{CaronHuot:2009uh} and strong coupling
calculations in $N=4$ Super-Yang-Mills theory \cite{CasalderreySolana:2006rq} it was concluded
that the heavy quark diffusion constant should be in the range 
\beq
D=(0.5-1.0)/T.
\eeq
The above value is considerably smaller than the perturbative estimate \cite{Moore:2004tg} and lies in
the range used in phenomenological models \cite{Rapp:2009my}.

%% file: conclusion.tex
\section{Conclusions}
In recent years significant progress has been made in studying chiral and deconfining aspects
of the QCD transition in finite temperature QCD. For several quantities that are relevant
for the discussion of these aspects continuum extrapolation has been performed.
Calculations by two groups that use different
staggered quark formulation give results for the chiral transition temperature that agree in
the continuum limit. It is now established that for physical value of chiral transition temperature
is $T_c \simeq 150$MeV. Calculations using Wilson fermion or DWF formulation have been performed for
physical or nearly physical quark masses and seem to confirm the staggered fermion results. 
The universal behavior of the chiral transition in the limit of vanishing light quark masses plays an important
role also for the physical values of the light quark masses and allows to define the chiral crossover 
temperature in a meaningful way.

The interplay
between chiral and deconfinement aspects of the transition appears to be more complicated
than earlier lattice studies suggested. There is no transition temperature that can be associated
with the deconfining aspects of the transition for physical values of the light quark masses. Furthermore,
the behavior of the Polyakov loop suggests that color screening sets in at temperatures that are
higher than the chiral transition temperature. The deconfinement aspects of the QCD transition have
been also studied in terms of fluctuations of conserved charges. In principle these fluctuations are 
are also sensitive to universal aspects of the chiral transition. However, for the lowest order, i.e.
quadratic fluctuations the contribution coming from the regular part of the free energy density dominates.
This implies in particular, that inflection points of the quark number susceptibilities cannot be used
to define the transition temperature. Higher order fluctuations of conserved charges are more sensitive
to the singular contribution and therefore are more suitable to study the interplay between chiral and
deconfining aspects of the QCD transition at finite temperature. The restoration of the $U_A(1)$ symmetry
at high temperature has been also discussed and it was found that it is effectively restored at temperatures
$T \simeq 1.2T_c$.

In the high temperature region ( $T>300$ MeV) lattice results have been compared with the results obtained
in weak coupling approaches, which seem to capture the qualitative features of the lattice data.
In some cases we see a good agreement at quantitative level. 

Finally I discussed lattice results on meson correlation functions. These are useful to study the fate
of quarkonium states in QGP as well as for determination of some transport coefficients. While a lot of progress
has been made in studying meson correlation functions and extracting the corresponding spectral functions
more work is needed to reach definitive conclusions.